\documentclass[pra, onecolumn, superscriptaddress, nofootinbib, aps, 10pt]{revtex4-2}

\usepackage{laura}

\newcommand{\flip}{\mathbb{F}}
\newcommand{\I}{\mathbb{I}}
\newcommand{\ntr}{\overline{\tr}}
\newcommand{\pauli}{\mathcal{P}}
\newcommand{\supp}{\mathrm{supp}}
\newcommand{\dlow}{D^{\mathrm{low}}}
\newcommand{\dlowd}{D^{\mathrm{low}\dagger}}
\newcommand{\dhigh}{D^{\mathrm{high}}}
\newcommand{\dhighd}{D^{\mathrm{high}\dagger}}
\newcommand{\flow}{F^{\mathrm{low}}}
\newcommand{\fhigh}{F^{\mathrm{high}}}
\newcommand{\err}{\mathrm{err}}

\SetKwInput{KwData}{Input}
\SetKwInput{KwResult}{Output}

\begin{document}

\title{A provably efficient classical algorithm for simulating OTOC$^{(1)}$ in 1D}

\author{Laura Lewis}
\email{lllewis@berkeley.edu}
\affiliation{Google Quantum AI, Venice, CA, USA}
\affiliation{University of California, Berkeley, CA, USA}

\author{Antonio A. Mele}
\affiliation{Google Quantum AI, Venice, CA, USA}
\affiliation{Dahlem Center for Complex Quantum Systems, Freie Universit\"at Berlin, 14195 Berlin, Germany}

\author{Ronald de Wolf}
\affiliation{Google Quantum AI, Venice, CA, USA}
\affiliation{QuSoft, CWI, and University of Amsterdam, Netherlands}

\author{Jarrod R. McClean}
\affiliation{Google Quantum AI, Venice, CA, USA}

\author{Amira Abbas}
\email{amiraabbas@google.com}
\affiliation{Google Quantum AI, Venice, CA, USA}

\begin{abstract}
Out-of-time-order correlators (OTOCs) are quantities of broad interest in many-body physics due to their connection with information scrambling and quantum chaos.
Recently, estimating higher-order OTOCs in random quantum circuits was proposed as a promising avenue for quantum advantage, yet the classical hardness of this task remains poorly understood.
Even for first-order OTOCs, where growing evidence suggests classical tractability, no efficient classical algorithm with rigorous guarantees has been established to date.
In this work, we introduce \emph{window propagation}, the first provably polynomial-time classical algorithm for approximating first-order OTOCs in 1D Haar-random brickwork circuits, at arbitrary depths and with inverse-polynomial mean-squared error.
We also conduct numerical experiments comparing our simulation results to exact computation on up to $512$ qubits for Clifford circuits with $768$ $T$-gates and up to $22$ qubits for Haar-random circuits.
We emphasize that our results do not directly apply to second-order OTOCs in 2D, which is the subject of recent quantum advantage proposals.
Nevertheless, our work provides a rigorous step towards understanding the classical complexity of OTOC estimation and represents a family of non-trivial quantum circuits for which a provable, polynomial time algorithm is now known, establishing foundations for understanding these widely used diagnostics of quantum chaos.
\end{abstract}

\maketitle

\stoptoc
\section{Introduction}

Simulating quantum dynamics and identifying complex many-body correlations are central problems in physics and quantum computation.
However, traditional approaches, which estimate dynamical two-point correlation functions (e.g., \emph{time-ordered correlators}), quickly lose meaningful signal about the dynamics at long evolution times due to ergodicity, especially when studying quickly thermalizing or random quantum systems~\cite{srednicki1994chaos, hayden2007black, d2016quantum, hosur2016chaos,maldacena2016bound, schuster2023learning, swingle2018unscrambling}.
To remedy this, one can instead consider \emph{out-of-time-order correlators (OTOCs)}~\cite{shelankov2000quasiclassical,mi2021information, xu2024scrambling, yunger2018quasiprobability, huang2017out, von2018operator,nahum2018operator,shenker2014black}, which repeatedly interleave forward and backward time evolution, and have the remarkable property of maintaining a strong signal even when used with fully random quantum circuits~\cite{mele2026towards}.
OTOCs have also been used to study operator spreading~\cite{von2018operator,nahum2018operator,khemani2018operator}, quantify information scrambling~\cite{swingle2018unscrambling,swingle2016measuring,landsman2019verified,mi2021information,shenker2014black,maldacena2016bound,shenker2014multiple}, and characterize chaotic systems~\cite{aleiner2016microscopic,roberts2016lieb,roberts2017chaos} with further applications in quantum sensing~\cite{davis2016approaching,li2023improving}.

Recently, Ref.~\cite{google2025observation} proposed the task of estimating high-order OTOCs as a demonstration of quantum advantage and showed experimentally that they indeed remain sensitive to quantum dynamics at large timescales.
Their experiments consider \emph{random quantum circuit dynamics}, which are often used as minimal models for Hamiltonian dynamics~\cite{hayden2007black,lashkari2013towards,sekino2008fast,brandao2016local}.
Even more recently, Ref.~\cite{mele2026towards} proved that OTOCs in random quantum circuits exhibit such sensitivity at all orders, confirming rigorously that OTOCs can capture meaningful signal in chaotic systems, in the sense that the final output depends significantly on details of the specific circuit that was run.

However, the presence of sensitivity does not necessarily imply classical hardness.
In fact, surprisingly little is known about the classical computational complexity of OTOCs.
This leads to the central question of our work:
\begin{center}
    \emph{Are OTOCs classically hard to simulate?}
\end{center}
For first-order OTOCs, where the forward and backward time evolution are interleaved only twice, numerical methods have performed well on finite-size instances of up to $40$ qubits~\cite{google2025observation,liu2026efficient}, but no rigorous efficient algorithm for large system sizes is known.
Meanwhile, higher-order OTOCs exhibit a form of many-body constructive interference that is widely believed to be intractable for known classical simulation algorithms~\cite{google2025observation}.
However, a formal proof of such classical hardness is still unknown.
If classical hardness indeed holds, then OTOCs would combine two properties rarely found together: robust signal in chaotic regimes and classical intractability, offering a clear route to quantum advantage with relatively minimal hardware requirements~\cite{king2025simplified}.

In this work, we establish the first polynomial-time classical simulation algorithm for first-order OTOCs whose dynamics are modeled as $1$D brickwork random quantum circuits, which we call \emph{window propagation}. While this result has been conjectured and empirically explored in previous work~\cite{google2025observation,liu2026efficient}, no efficient classical algorithm with rigorous guarantees was known, even in $1$D where the \textit{average} first-order OTOC value can be computed exactly on a classical computer in polynomial-time~\cite{nahum2018operator}. 
We emphasize that our results do not apply to the seemingly more challenging problem of estimating second-order OTOCs in 2D, which is the subject of the recent quantum advantage proposals~\cite{google2025observation}.

In addition to our theoretical guarantees, we conduct numerical experiments to assess the performance of our algorithm in practice.
We consider first-order OTOCs with dynamics modeled as 1D brickwork random quantum circuits (up to $22$ qubits) and Clifford circuits with $T$-gates placed uniformly at random ($768$ $T$-gates on $512$ qubits).  
In both cases, the results exhibit excellent agreement with exact simulation, demonstrating the practical viability of our algorithm.
Additionally, in the case of 1D brickwork random quantum circuits, we observe an empirical wall-clock scaling of roughly~$\sim n^{2.2}$ to achieve a target accuracy on the instance circuits, where $n$ is the number of qubits.

\section{Problem statement}
\label{sec:problem-main}

To define the problem precisely, we follow the presentation of~\cite{king2025simplified}.
Consider a distribution of brickwork random quantum circuits with Haar-random two-qubit gates of depth $d\geq 2$ on a 1D chain of $n$ qubits with open boundaries, denoted by $\mathcal{U}_{n,d}$ (see \Cref{def:brickwork}).
Take $n,d$ to be even-numbered for notational simplicity.
For a unitary $U \sim \mathcal{U}_{n,d}$, we write $U = U_d \cdots U_1$, where each $U_j$ is a layer of the circuit.
We also write $U = L_{d/2} \cdots L_1$ when grouping the layers in groups of two, i.e., $L_j = U_{2j} U_{2j-1}$.

Let $B$ and $M$ denote the \emph{butterfly} and \emph{measurement} operators, respectively, which are local Pauli operators acting on distant qubits.
For concreteness, we take $B \triangleq X_n$ and $M \triangleq Z_1$.
Then, for an initial state $\rho$, the $K^\text{th}$ order OTOC, sometimes denoted $\mathrm{OTOC}_\rho^{(K)}$, is defined as
\begin{equation}
	f^{(K)}_\rho(U) \triangleq \tr((U^\dagger BUM)^{2K}\rho).
\end{equation}
In this work, we mainly consider the \emph{infinite-temperature} OTOC, which takes the initial state $\rho = I/2^n$ to be the maximally mixed state.
When using $\rho = I/2^n$, we drop the subscript~$\rho$. 
The simulation task is as follows: given $U \sim \mathcal{U}_{n,d}$ and a fixed initial state, estimate $f^{(K)}_\rho(U)$ to inverse-polynomial error $\epsilon = 1/\mathrm{poly}(n)$.
Our error metric is good mean-squared error, i.e., we want to output an estimate $h^{(K)}_\rho(U)$ such that
\begin{equation}
	\E_{U \sim \mathcal{U}_{n,d}}[|h^{(K)}_\rho(U) - f^{(K)}_\rho(U)|^2] \leq \epsilon.
\end{equation}

We highlight that this task is only interesting for the intermediate depth regime $d = \Theta(n)$, where OTOCs exhibit instance-to-instance fluctuations~\cite{mele2026towards}.
In particular, Ref.~\cite{mele2026towards} shows that in 1D, the most interesting regime is when $|d - 5n/3| \leq c\sqrt{n}$, for a constant $c > 0$.
Outside of this linear depth regime, estimating OTOCs is trivial and can be achieved by outputting a constant independent of $U$.
For depths that are too small, the lightcone of $U^\dagger B U$ does not reach $M$, so these operators commute and $f_\rho^{(K)}(U) = 1$.
Moreover, for depths that are too large, brickwork random circuits become unitary designs~\cite{brandao2016local,harrow2023approximate,mittal2023local,haferkamp2022random}, so $f_\rho^{(K)}(U)$ is close to $0$.

Note that the simulation task can be achieved efficiently on a quantum computer by a simple algorithm.
For example, for $\rho = \ketbra{0^n}$, one can prepare the state $(U^\dagger B UM)^K \ket{0^n}$, measure $M$, and repeat sufficiently many times to estimate the expectation value with high confidence.
However, no rigorous efficient classical algorithm was known for any order $K$ prior to our work.

\section{Algorithm and rigorous guarantee}
\label{sec:algo-main}

\begin{figure}
    \centering
    \includegraphics[width=\linewidth]{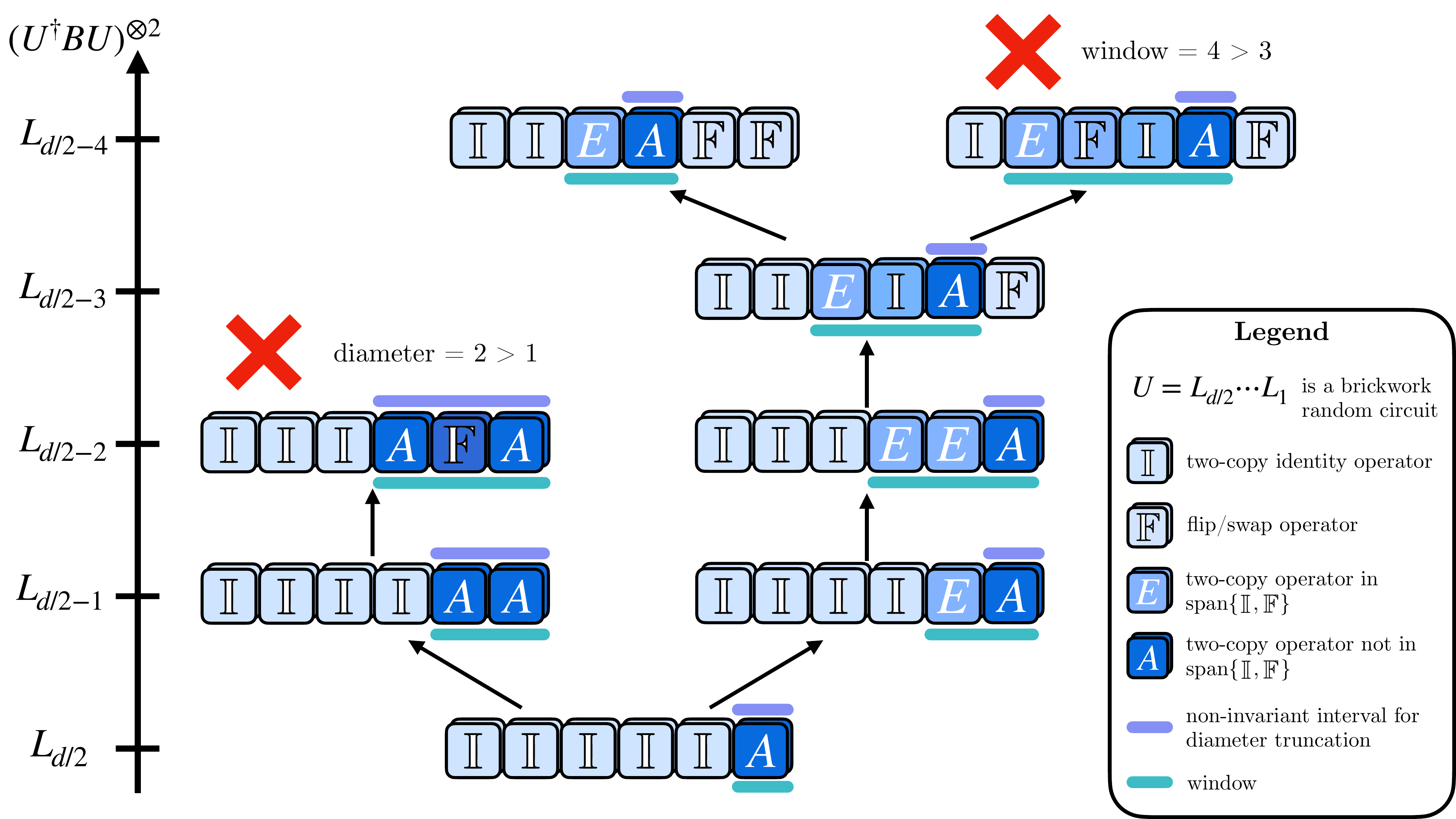}
    \caption{\textbf{Overview of our window propagation algorithm.} We depict steps of our algorithm for system size $n = 6$, diameter cutoff $k = 1$, and window width cutoff $w = 3k = 3$. The algorithm starts with the two-copy butterfly operator $B \otimes B = X_6 \otimes X_6$ and Heisenberg evolves it through each layer of a circuit $U= L_{d/2} \cdots L_1$, indicated by the arrows. Here, $B\otimes B$ can be written as a sum of a non-invariant basis element and a basis element in $\mathrm{span}\{\I,\flip\}$. In the figure, non-invariant basis elements are labeled by $A$, and elements in $\mathrm{span}\{\I,\flip\}$ that are not $\I,\flip$ themselves are labeled by $E$. We track the evolution of the non-invariant element, which is why the figure begins with $\I^{\otimes 5}\otimes A$. The darkness of the block coloring indicates the ``level of invariance,'' where the regions in the lightest blue are invariant under two-qubit gates, the middle blue are invariant under single-qubit gates, and the darkest blue are non-invariant. The teal underline indicates the window, and the purple overline indicates the non-invariant interval which is used in the diameter truncation. The leftmost branch is truncated by the diameter truncation rule because its non-invariant interval has length $2$, which is greater than $k = 1$. The rightmost branch is truncated by the window truncation rule because its window has size $4$, which is greater than $w = 3$. The middle branch survives both truncation rules and is propagated through the rest of the circuit.}
    \label{fig:algo}
\end{figure}

Our main result is a polynomial-time classical simulation algorithm for $\mathrm{OTOC}^{(1)}$ in 1D random brickwork circuits of arbitrary depths.
In contrast, we note that recent quantum advantage experiments~\cite{google2025observation} focus on the problem of estimating $\mathrm{OTOC}^{(2)}$ in $2$D random circuits.
To state our result precisely, we prove the following theorem.

\begin{theorem}
	\label{thm:sim-main}
	Let $0 < \epsilon < 1$.
    Let $n, d \geq 2$.
    Let $\mathcal{U}_{n,d}$ denote the distribution of 1D brickwork random quantum circuits of depth $d$ on $n$ qubits.
    Then, there exists a deterministic classical algorithm that, when given an efficient description of a quantum circuit $U$ sampled from $\mathcal{U}_{n,d}$, outputs a real number $h(U)$ such that
	\begin{equation}
		\E_{U \sim \mathcal{U}_{n,d}}[|h(U) - f^{(1)}(U)|^2] \leq \epsilon.
	\end{equation}
	The runtime of this algorithm is $\mathrm{poly}(n,d,1/\epsilon)$.
\end{theorem}

As a simple corollary, via Markov's inequality, we can also transform this mean-squared-error statement into an additive error guarantee with high probability (see \Cref{coro:sim-whp}).
We highlight that our result also holds for 1D brickwork circuits with two-qubit gates that form a $4$-design, as $|h(U)-f^{(1)}(U)|^2$ is a polynomial of degree $(4,4)$ in each gate and its conjugate.

Moreover, with a small modification to our algorithm, it is also possible to extend our result to zero-state $\mathrm{OTOC}^{(1)}$ (where the initial state is $\rho = \ketbra{0^n}$) and ``bulk'' $\mathrm{OTOC}^{(1)}$ (where the butterfly and measurement operators can be placed anywhere in the 1D chain).
The extension to zero-state OTOCs is particularly noteworthy.
In the worst case, estimating infinite-temperature $\mathrm{OTOC}^{(1)}$ to inverse-polynomial additive accuracy is $\mathsf{DQC1}$-complete~\cite{roymoulik2024dqc1hardness}, whereas the zero-state version is $\mathsf{BQP}$-complete.
Furthermore, infinite-temperature $\mathrm{OTOC}^{(1)}$ depends only on the squared Pauli coefficients of the evolved butterfly operator, so efficient sampling from this distribution suffices for efficient estimation.
In contrast, zero-state $\mathrm{OTOC}^{(1)}$ additionally contains off-diagonal contributions between distinct Pauli strings and are not, in general, determined by this distribution alone~\cite[Supplementary Information,
Sec.~III.B.1]{google2025observation}.
Overall, this suggests that zero-state $\mathrm{OTOC}^{(1)}$ may be more challenging to simulate classically than infinite-temperature $\mathrm{OTOC}^{(1)}$.
Nevertheless, a modification of our algorithm also suffices to simulate zero-state $\mathrm{OTOC}^{(1)}$ efficiently.   
We will include the proof of these two extensions in a future work.

We now describe our algorithm, which we call \emph{window propagation}, achieving \Cref{thm:sim-main}.
Using the swap trick, we can rewrite the first-order infinite-temperature OTOC as
\begin{equation}
    \label{eq:two-copy-otoc-main}
	f^{(1)}(U) = \frac{1}{2^n}\tr((U^\dagger BUM)^2) = \frac{1}{2^n}\tr((U^\dagger BUM)^{\otimes 2} \flip_{\mathrm{all}}),
\end{equation}
where $\flip_{\mathrm{all}}$ denotes the swap operator on all $2n$ qubits.
At a high level, our algorithm starts with the two-copy butterfly operator $B \otimes B$ and evolves it backwards through each layer of the circuit in the Heisenberg picture.
After evolving by each layer, two separate truncation rules are applied to ensure that the simulation is both accurate and efficient to compute.

To describe the algorithm more precisely, consider a basis $\mathcal{A}$ of the $16^n$-dimensional space of two-copy operators (see~\Cref{def:basis}).
On each pair of corresponding qubits from the two copies, the basis $\mathcal{A}$ consists of $\I, \flip$ (the identity and swap operators on two copies of a single qubit, respectively), and fourteen operators spanning the orthogonal complement of $\mathrm{span}\{\I,\flip\}$.
Importantly, this basis distinguishes $\I$ and $\flip$, which are invariant under two-copy rotations. Moreover, let the diameter of a finite set be defined as $\mathrm{diam}(S) = \max S - \min S$.
With this, we can describe our algorithm in \Cref{alg:sim-main}.

\begin{algorithm}
    \caption{Window propagation algorithm (Simplified version)}
    \label{alg:sim-main}
    \KwData{A 1D brickwork, depth-$d$, $n$-qubit quantum circuit $U = L_{d/2} \cdots L_1$; butterfly operator $B$; measurement operator $M$; error parameter $\epsilon > 0$.}
    \KwResult{An estimate $h(U)$ such that
    \begin{equation}
    	\E_{U \sim \mathcal{U}_{n,d}}[|h(U) - f^{(1)}(U)|^2] \leq \epsilon.
    \end{equation}}
    Set truncation parameters $k \triangleq \mathcal{O}(\log (d/\epsilon))$ and $w \triangleq 3k$.\\
    Start with the observable $O_{d/2} = B \otimes B$.\\
    \For{$j=d/2,d/2-1,\dots,1$}{
        Compute the Heisenberg evolved operator $L_j^{\dagger \otimes 2} O_j L_j^{\otimes 2}$ in the basis $\mathcal{A}$.\\
        Apply a \emph{diameter truncation}: discard terms $A \in \mathcal{A}$ for which $\mathrm{diam}(S_A) > k$, where $S_A = \{i \in [n]: A_i \notin \mathrm{span}\{\I,\flip\}\}$ is the set of sites that are non-invariant under two-copy single-qubit Haar-random unitaries.\label{step:diameter}\\
        Apply a \emph{window truncation}: orthogonally project onto the span of terms of the form\label{step:window}
	 	\begin{equation}
            \label{eq:window-main}
	 		\I^{\otimes(\ell-1)} \otimes W_{[\ell,r]} \otimes \flip^{\otimes(n-r)},\quad r-\ell + 1 \leq w,
	 	\end{equation}
	 	where $W_{[\ell,r]}$ denotes an arbitrary two-copy operator on sites $[\ell,r]$.\\
        Denote the operator after the two truncations as $O_{j-1}$.\\
    }
    Output $h(U) = h_k(U) = \frac{1}{2^n}\tr(O_0(M\otimes M)\flip_{\mathrm{all}})$.
\end{algorithm}
A more detailed description of the algorithm is given in~\Cref{alg:sim}.
We also illustrate the algorithm in \Cref{fig:algo}.

\subsection{Intuition behind our window propagation algorithm}

In this section, we explain the ideas behind the window propagation algorithm presented in \Cref{alg:sim-main}.
First, recall from prior work that first-order OTOCs can be approximated well by their ensemble mean, with additional circuit-dependent corrections~\cite{google2025observation,mele2026towards}.
Ref.~\cite{nahum2018operator} establishes that the mean can be estimated efficiently via Monte Carlo methods, even in higher spatial dimensions. 
Thus, the difficult part is keeping track of the corrections that depend on~$U$. 
A natural starting point is to then understand the algorithm for estimating the mean and attempt to extend it to efficiently capture individual circuit information.
At a high-level, \Cref{alg:sim-main} achieves this using its two truncation rules. 

\subsubsection{The window truncation}
Recall that the $\mathrm{OTOC}^{(1)}$ function, $f^{(1)}(U)$, is a linear functional of the two-copy Heisenberg-evolved butterfly operator $(U^\dagger B U)^{\otimes 2}$ (see~\Cref{eq:two-copy-otoc-main}).
Its mean can be computed by averaging over the random circuit ensemble $U \sim \mathcal{U}_{n, d}$.
Remarkably, the averaged evolution has a particularly simple structure. 
For example, consider $n=6$ qubits and $B=X_6$.
Successively averaging over two-qubit Haar-random gates acting on qubits $(5,6)$ and $(4,5)$ gives
\begin{align}
\E_{G_{56}}
[(G_{56}^\dagger B G_{56})^{\otimes 2}]
&=
-\frac{1}{15}\I^{\otimes 6}
+\frac{4}{15}\I^{\otimes 4}\otimes\flip^{\otimes 2},
\\
\E_{G_{45},G_{56}}
[(G_{45}^\dagger G_{56}^\dagger B
G_{56}G_{45})^{\otimes 2}]
&=
-\frac{1}{15}\I^{\otimes 6}
+\frac{8}{75}\I^{\otimes 5}\otimes\flip
+\frac{8}{75}\I^{\otimes 3}\otimes\flip^{\otimes 3}.
\end{align}
This structure persists at arbitrary depths:
\begin{equation}
\E_{U\sim\mathcal{U}_{n,d}}
[(U^\dagger B U)^{\otimes 2}]
=
\sum_{i=0}^{n}c_i\,
\I^{\otimes(n-i)}\otimes\flip^{\otimes i},
\end{equation}
for some coefficients $c_i$. 
These averaged dynamics, with identities on the left and flips on the right, suggest a natural extension for adding circuit-dependent corrections: allow an arbitrary two-copy operator $W$ within a small window between the identity and flip regions.
More precisely, we retain the span of operators of the form
\begin{equation}
    \label{eq:trunc_ops}
    \I^{\otimes(\ell-1)}
    \otimes W_{[\ell,r]}
    \otimes\flip^{\otimes(n-r)},
    \qquad r-\ell+1\leq w,
\end{equation}
as in \Cref{eq:window-main}.
Our algorithm starts with the two-copy butterfly operator $B \otimes B$, Heisenberg-evolves through each two-layer block of the circuit, and orthogonally projects back onto this span.
At each iteration, our approximation of the Heisenberg-evolved butterfly operator is a linear combination of terms of the form of \Cref{eq:trunc_ops} for different choices of operators $W$ inside the window and positions $\ell,r$.
This is the window truncation in Step~\ref{step:window} of our algorithm (and the motivation for our algorithm's name: window propagation).

This representation is efficient because gates acting entirely within either the identity region or the flip region leave that region unchanged.
Indeed, for any two-qubit gate $G$,
\begin{align}
    (G^\dagger)^{\otimes 2}
    (\I\otimes\I)
    G^{\otimes 2}
    = \I\otimes\I, \qquad(G^\dagger)^{\otimes 2}
    (\flip\otimes\flip)
    G^{\otimes 2}
    = \flip\otimes\flip.
\end{align}
Thus, only gates intersecting the window require nontrivial updates.
Moreover, each two-layer block enlarges a window by at most four sites before truncation. 
Since the single-qubit two-copy operator space has dimension $16$, the number of coefficients to keep track of is at most $\mathcal{O}(n16^w)$ at each iteration.
Taking the window width $w$ to scale logarithmically then allows the simulation algorithm to run in polynomial-time.

Interestingly, our numerical experiments in~\Cref{sec:numerics-main} suggest that the window truncation alone is already sufficient for accurate simulation.
However, proving a rigorous guarantee for our algorithm with only this truncation appears challenging and remains open.
To prove~\Cref{thm:sim}, we perform another independent truncation, namely, the \emph{diameter truncation}.

\subsubsection{The diameter truncation}
The window truncation only considers three regions: an identity region on the left, a flip region on the right, and a central window of width at most $w$ containing arbitrary two-copy operators. While the identity and flip regions are invariant under two-copy two-qubit gates acting entirely within them, the central region has no such restriction.

The idea behind our additional diameter truncation is to make this transition less abrupt, by introducing intermediate regions with a weaker form of invariance. To do so, we introduce a second parameter~$k\leq w$, which bounds the maximum distance between sites carrying operators outside $\mathrm{span}\{\I,\flip\}$.
More precisely, for each basis configuration $A$, let $S_A$ denote the set of such non-invariant sites.
The diameter truncation enforces that all non-invariant sites must fit within an interval of at most $k+1$ consecutive sites (see Step~\ref{step:diameter}). 
Outside this interval, but still within the window of width $w$, only operators in $\mathrm{span}\{\I,\flip\}$ are allowed, which are invariant under two-copy single-qubit rotations, but not necessarily under two-qubit gates.
We also comment on how the diameter truncation can be viewed as an extension of Pauli propagation to the two-copy setting in \Cref{sec:pauli-prop}.

The two truncations together therefore give rise to five regions: an exact identity tail, a single-qubit invariant region, a fully arbitrary region, another single-qubit invariant region, and an exact flip tail. Thus, moving from either tail towards the center, the invariance becomes progressively weaker.
For example, a configuration retained after performing \emph{both} truncations can have the form:
\begin{equation}
\underbrace{\I^{\otimes 4}}_{\substack{\text{Two-qubit}\\\text{invariant}}}
\otimes
\underbrace{\flip\otimes\I\otimes\flip}_{\substack{\text{Single-qubit}\\\text{invariant}}}
\otimes
\underbrace{W}_{\text{Arbitrary}}
\otimes
\underbrace{\flip\otimes\I}_{\substack{\text{Single-qubit}\\\text{invariant}}}
\otimes
\underbrace{\flip^{\otimes 3}}_{\substack{\text{Two-qubit}\\\text{invariant}}},
\end{equation}
where $W$ acts on at most $k+1$ sites, and the entire transition region has width at most $w$.
We note that the single-qubit invariant region can also contain elements in $\mathrm{span}\{\I, \flip\}$, rather than only $\I$ or $\flip$ individually.

Our algorithm efficiently tracks linear combinations of such configurations.
The remaining challenge is to prove that these truncations give an accurate approximation of $\mathrm{OTOC}^{(1)}$, which we address in the next section.

\section{Proof ideas}
\label{sec:proof-ideas}

We describe the key ideas behind the proof of \Cref{thm:sim-main}.
Recall that the goal is to prove that the mean-squared error between $f^{(1)}$ and the output $h(U)$ of \Cref{alg:sim-main} is small, i.e.,
\begin{equation}
	\E_{U\sim \mathcal{U}_{n,d}}[|h(U) - f^{(1)}(U)|^2] \leq \epsilon.
\end{equation}
The key intermediate step is to consider the estimate obtained when \emph{only} the diameter truncation rule is applied.
Call the resulting estimate $g_k(U)$.
We emphasize that the diameter-truncation-only algorithm is not one that we execute; it is only used in the proof.
The proof is then separated into two parts: (1) bounding the mean-squared error between $f^{(1)}$ and $g_k$; (2) bounding the mean-squared error between $g_k$ and $h_k$.
By Cauchy-Schwarz, showing both of these suffices to prove \Cref{thm:sim-main}.
We discuss each of these steps below.

In both steps, we find it useful to first insert layers of single-qubit independent Haar-random gates into our brickwork random circuit.
In other words, we consider $U = L_{d/2} V_{d/2} \cdots L_1 V_1$, where $V_j = \bigotimes_{i=1}^n V_{j,i}$ for single-qubit Haar-random gates $V_{j,i}$.
By invariance of the Haar measure, these~$U$ have the same distribution as our original circuits, so it suffices to bound the mean-squared error with respect to them.

\begin{figure}
    \includegraphics[width=0.6\linewidth]{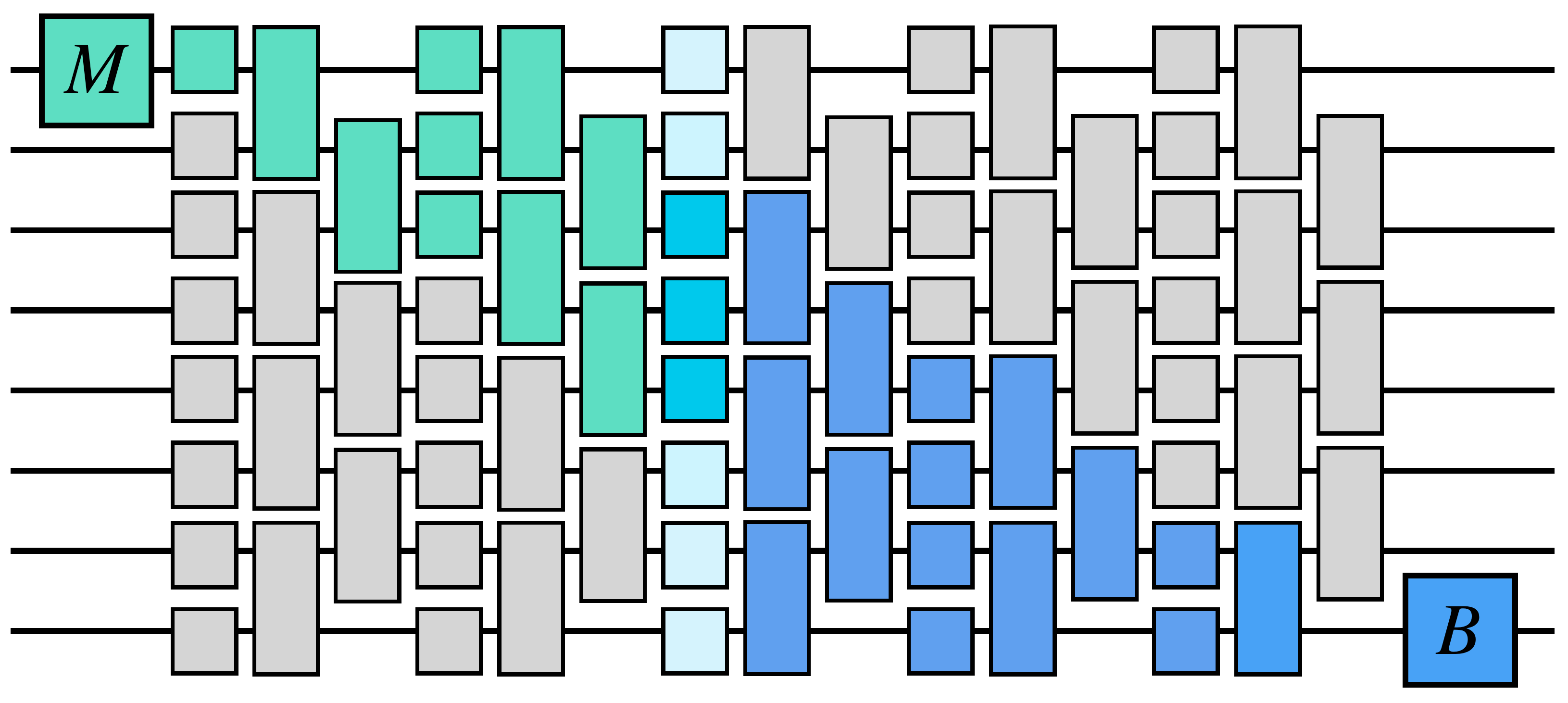}
    \caption{\textbf{Overlap between the butterfly and measurement operators' lightcones.} The gates shown constitute $U$, after adding layers of single-qubit Haar-random gates. The green/blue gates show the measurement/butterfly operator's forward/backward lightcone up until the center layer of single-qubit gates, respectively. The gray gates are not in the geometric lightcone of either operator. The dark teal gates show the single-qubit gates for which the lightcones overlap. These are the gates which can affect the $\mathrm{OTOC}^{(1)}$ value in the proof. The light teal gates are single-qubit gates which are not in the intersection of the lightcones.}
    \label{fig:lightcone-overlap}
\end{figure}

\subsection{Error of the diameter-truncation-only algorithm}
In this step, we prove that
\begin{equation}
	\label{eq:k-only-error-main}
	\E_{U\sim \mathcal{U}_{n,d}}[|g_k(U) - f^{(1)}(U)|^2] \leq \mathcal{O}(d\alpha^k),
\end{equation}
for some universal constant $0 < \alpha < 1$.
First, by using that the diameter truncation implements an orthogonal projection at each layer, we show that it suffices to consider the error accumulated one iteration at a time.
In other words, it suffices to bound the mean-squared error between $f^{(1)}$ and the output of the algorithm when the diameter truncation is applied \emph{only} at that iteration.
The $d$ factor in \Cref{eq:k-only-error-main} stems from summing over all iterations.

Consider the single-iteration error at iteration $j$.
Recall that in the algorithm, one iteration corresponds to evolving under the layer $L_jV_j$.
The idea is to consider splitting the circuit at this layer $j$: backwards evolve the butterfly operator $B$ and forward evolve the measurement operator $M$ until this layer (see \Cref{fig:lightcone-overlap}).
Then, the length of the region where their lightcones overlap corresponds to the diameter of $S_A$.
This is because only $V_{j,i}$ with $i$ inside of the overlapping region can affect the $\mathrm{OTOC}^{(1)}$ value, as outside of this interval, at least one of the operators is identity.
This implies that sites carrying non-invariant operators inside of this overlapping region will affect the $\mathrm{OTOC}^{(1)}$ value.

We want to show that if two sites carrying non-invariant operators are separated by a distance $k$, then this term's contribution to the $\mathrm{OTOC}^{(1)}$ value decays exponentially with $k$.
By the above characterization, this is equivalent to showing that the overlap between the evolved butterfly and measurement operators on an interval of length $k$ decays exponentially in $k$.
The point is that, because we have split the circuit at layer $j$, the evolved butterfly and measurement operators rely on independent random gates.
Intuitively, the exponential decay then stems from the fact that two independent random vectors in high-dimensional space have low overlap. 
However, in our two-copy setting, we do not simply have two independent random operators, but four random operators where the same operator across different copies could be correlated but different operators are independent.
Specifically, the two copies of the evolved butterfly operator are correlated with each other, but the evolved butterfly and measurement operators are independent.
Nevertheless, we show that, as long as one copy of the butterfly and measurement operators is sufficiently isotropically random, the overlap still decays exponentially (see \Cref{prop:8} and \Cref{lem:conditions-prop-8}).

\subsection{Error with the window truncation}

In the second step, we show that
\begin{equation}
	\E_{U\sim\mathcal{U}_{n,d}}[|g_k(U) - h_k(U)|^2] \leq \mathcal{O}(d2^{-k}).
\end{equation}
As before, the $d$ factor comes from summing over all iterations.

Consider a term with window width larger than $w=3k$.
We want to show that such a term does not contribute to the $\mathrm{OTOC}^{(1)}$ value significantly, so we can safely discard it.
In this case, we quantify the contribution to the $\mathrm{OTOC}^{(1)}$ value via the average Frobenius norm squared and show that this can be related to the mean-squared error (see \Cref{eq:mse-to-err}).

Here, if a term $A$ has window width larger than $w=3k$, the window is larger than the non-invariant interval we truncated in the previous step.
Consider breaking up our $n$-qubit system into many intervals: a prefix, a left region $L$ of $k$ qubits, a middle region of $k$ qubits, a right region $R$ of $k$ qubits, and a suffix of the remaining qubits.
We can argue that, for all sites $i$ in at least one of the regions $L$ or $R$, $A_i$ must lie in $\mathrm{span}\{\I,\flip\}$.
This is clear because the non-invariant interval has length at most $k$ (by the previous step) while regions $L$ and $R$ are at least distance $k+1$ from each other.
Writing $\mathrm{Proj}_L$ for the projection of sites $i \in L$ onto $\mathrm{span}\{\I_i,\flip_i\}$ (and similarly for $\mathrm{Proj}_R$), this can be described formally as
\begin{equation}
	(\mathrm{id} - \mathrm{Proj}_L)(\mathrm{id} - \mathrm{Proj}_R)A = 0.
\end{equation}
This implies that the average Frobenius norm squared of the term $A$ is bounded by
\begin{equation}
	\E\norm{A}_F^2 \leq \E\norm{\mathrm{Proj}_L A}_F^2 + \E\norm{\mathrm{Proj}_R A}_F^2,
\end{equation}
so it suffices to bound each of these terms, where the average is taken over the randomness in the terms~$A$.

Without loss of generality, consider the projection $\mathrm{Proj}_L$.
For the sake of exposition, suppose that, on $L$, the term $A$ is a two-copy random Pauli $P \otimes P$, where $P$ is a $k$-qubit Pauli.
Then, the average Frobenius norm squared is
\begin{equation}
	\E\norm{\mathrm{Proj}_L(P\otimes P)}_F^2 = \prod_{i\in L}\E\norm{\mathrm{Proj}_i(P_i \otimes P_i)}_F^2
\end{equation}
by independence of the Paulis on each site.
Recall that projecting onto the invariant subspace $\mathrm{span}\{\I,\flip\}$ is equivalent to applying the second Haar twirl~\cite{mele2024introduction} (see also \Cref{lem:second-moment}).
Thus, the projection acting on this term is
\begin{equation}
	\mathrm{Proj}_i(P_i\otimes P_i) = \begin{cases}
		I_i \otimes I_i & \text{if } P_i = I\\
		\frac{1}{3}(X_i\otimes X_i + Y_i \otimes Y_i + Z_i \otimes Z_i) & \text{if } P_i = Q_i \neq I\\
		0 & \text{else}
	\end{cases}.
\end{equation}
The average Frobenius norm squared is then clearly
\begin{equation}
	\E\norm{\mathrm{Proj}_L(P\otimes P)}_F^2 = \left(\frac{1}{4} + \frac{3}{4} \cdot \frac{1}{3}\right)^k = 2^{-k}.
\end{equation}
This example intuitively demonstrates how the exponential decay arises.
However, in the proof, the terms (arising from the Heisenberg evolved butterfly operator) are not simply two-copy independent random Paulis.
Regardless, with some additional bookkeeping, we can show that the same exponential decay holds (see \Cref{lem:single-layer-error-W}).

\section{Numerical experiments}
\label{sec:numerics-main}

\begin{figure}[h]
    \centering
    \includegraphics[width=\linewidth]{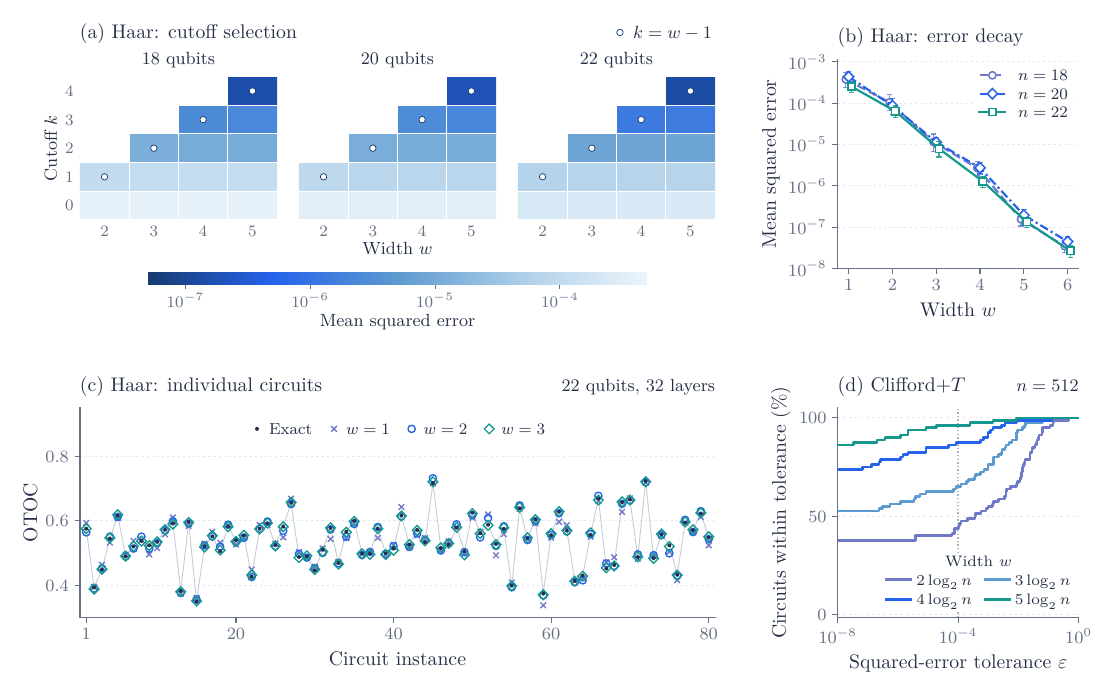}
    \caption{\textbf{Accuracy of our algorithm.}
    For subfigures (a--c), we consider Haar-random circuits on $n$ qubits with depth $d$ for
    $(n,d)\in\{(18,26),(20,30),(22,32)\}$.
    For each pair $(n,d)$, all choices of truncation parameters are evaluated on the same $80$ circuits.
    \textbf{(a)} Mean-squared error (MSE) for each choice of truncation parameters $k,w$.
    The shading of each block indicates the MSE for a given choice of $k,w$, where darker means lower MSE.
    The white circles mark the smallest MSE at each fixed $w$, which is attained at $k=w-1$ in every tested case.
    At this value, the diameter truncation is redundant, so our algorithm uses only the window truncation.
    \textbf{(b)} MSE of the window-only truncation as a function of the window width $w$.
    For each system size, the MSE decays rapidly with $w$.
    \textbf{(c)} Comparison of our algorithm with exact simulation for $80$ instances of Haar-random brickwork circuits on $22$ qubits with depth $32$.
    Black circles indicate exact simulation, and colored symbols show the results of our algorithm for the choices of $w \in \{1,2,3\}$.
    \textbf{(d)} Percentage of Clifford+$T$ circuits whose squared error $|h_w(U)-f^{(1)}(U)|^2$ is at most $\epsilon$.
    All widths use the same $80$ circuit instances, each with $512$ qubits, $850$ Clifford layers and $768$ uniformly placed $T$-gates.}
\label{fig:numerics-main}
\end{figure}

In this section, we present numerical experiments to assess the performance of our algorithm in practice.
The results illustrate the optimal choices of the truncation parameters in practice, the exponential decay of the mean-squared error with the window width, and excellent agreement with exact simulation for Haar-random circuits and Clifford-plus-$T$ circuits.
Further details of our implementation and additional plots are given in \Cref{app:width-numerics}.

Recall that our window propagation algorithm in \Cref{alg:sim-main} uses two truncation parameters: the diameter cutoff $k$ (which bounds the maximum distance between sites that are non-invariant under the action of two-copy single-qubit unitaries), and the window width $w$ (which bounds the length of the retained window, within which operators are allowed to be non-invariant under two-qubit gates).
For the proofs, as described in \Cref{sec:proof-ideas}, we choose $w=3k$ to control the errors introduced by both truncations.
However, in the numerical experiments, to determine how each cutoff affects accuracy in practice, we instead allow $k$ and $w$ to vary independently.
In particular, we denote the estimate obtained with truncation parameters $k,w$ by $h_{k,w}(U)$.

We consider ensembles of unitaries drawn either from Haar-random brickwork circuits or Clifford-plus-$T$ circuits where the $T$-gates are placed uniformly at random.
For these ensembles, we consider system sizes at most $22$ qubits and $512$ qubits, respectively.
At each system size, we choose the depth such that the ensemble mean $\mathrm{OTOC}^{(1)}$ is approximately $0.5$, which is in the regime where we expect appreciable instance-to-instance fluctuations around the ensemble mean.
In both cases, we compare the estimates obtained from our algorithm to exact simulation.
We also compute the mean-squared error (MSE) by averaging $|h_{k,w}(U) - f^{(1)}(U)|^2$ over the sampled circuits.

In \Cref{fig:numerics-main}(a), we explore how best to choose the truncation parameters $k,w$ in practice.
We find that, among the tested choices, applying only the window truncation results in the smallest error at each fixed width.
In particular, at each tested width $w$, the smallest MSE is attained at $k=w-1$.
For this choice, the diameter truncation is redundant: the window already confines the non-invariant sites $S_A$ to an interval of at most $w$ sites, whose diameter is at most $w-1$.
The estimate therefore coincides with the window-only approximation, which we denote by $h_w(U)\triangleq h_{w-1,w}(U)$ (see \Cref{app:haar-exact}).

\Cref{fig:numerics-main}(b) shows that the mean-squared error decreases rapidly with the window width $w$ for each fixed system size of up to $22$ qubits.
Interestingly, this plot also suggests that the MSE is only weakly dependent on the system size for fixed width $w$.
The weak system size dependence observed in \Cref{fig:numerics-main}(b) suggests that $w = \mathcal{O}(\log(1/\epsilon))$ suffices to achieve an MSE of $\epsilon$. In contrast, our rigorous guarantee uses $w = 3k$ with $k = \mathcal{O}(\log(n/\epsilon))$, and proving that the window-truncation-only algorithm with $w = \mathcal{O}(\log(1/\epsilon))$ suggested by our numerics achieves an MSE of $\epsilon$ remains open.

We also compare the outputs of our algorithm to exact simulation.
\Cref{fig:numerics-main}(c) illustrates that our algorithm matches exact simulation well for Haar-random brickwork circuits on $22$ qubits, even for small values of $w$.
To extend the comparison to larger system sizes, we also consider $512$-qubit Clifford circuits with $768$ $T$-gates placed uniformly at random (see \Cref{fig:numerics-main}(d) and \Cref{fig:clifford-t-width}(a--b)).
The additional structure of these circuits allows us to perform exact simulation even at these large system sizes to benchmark our algorithm.

For $22$-qubit Haar-random brickwork circuits of depth $32$, the median runtime of our algorithm is $0.070$ seconds per circuit over $80$ independent circuits, with a range of $0.066$--$0.079$ seconds for $w=3$.
For the $512$-qubit Clifford-plus-$T$ circuits, the median runtime is $0.195$ seconds per circuit over $80$ circuits for $w=5\log_2 n$.
Both benchmarks were performed on an Apple M4 Max, excluding the time taken for circuit generation and the exact simulation reference calculations.

We further assess the computational cost of our algorithm by running it at fixed width $w=3$ on circuit instances with system sizes up to $100$ qubits.
For $100$-qubit Haar-random brickwork circuits of depth $162$, the median runtime is $2.11$ seconds per circuit over $80$ circuits, with a range of $2.06$--$2.20$ seconds for $w = 3$.
For the Haar-random circuits at $w=3$, a power-law fit to the median runtimes over $14$--$100$ qubits gives
\begin{equation}
    T_{3}(n)\simeq 0.0716\,\mathrm{s}\left(\frac{n}{22}\right)^{2.26}.
\end{equation}
The measured runtime therefore grows approximately quadratically over the tested range, and it is consistent with updating local
window contributions across a circuit whose spatial extent and depth both grow with $n$.
This growth is milder than the conservative operation count in \Cref{sec:runtime}, which gives $\mathcal O(n^3d)$ at fixed $k,w$.
This is $\mathcal O(n^4)$ when the depth grows linearly with $n$, as is the case in our setting.
Our implementation exploits local contractions and avoids operations on invariant regions, which are not fully accounted for in this upper bound.
We note that this is only a discussion of the computational cost as the large-system-size regime is beyond the reach of exact simulation (see \Cref{fig:haar-runtime-scaling} in the appendix for more details).

\section{Related work}

In this section, we discuss previous proposals for the classical simulation of OTOCs.
To our knowledge, no prior works contain efficient classical algorithms for simulating OTOCs with rigorous guarantees, so all works discussed in this section are numerical and heuristics.
First, we note that traditional approaches such as matrix product states~\cite{ostlund1995thermodynamic,vidal2003efficient,hemery2019matrix}, neural quantum states~\cite{carleo2017solving,lin2022scaling,westerhout2020generalization}, or Pauli propagation~\cite{beguvsic2025real,rall2019simulation,aharonov2023polynomial,angrisani2025classically,rudolph2023classical,schuster2025polynomial,gonzalez2025pauli,fontana2025classical,shao2024simulating} are known to perform poorly for simulating OTOCs at larger system sizes (see Section III.C of the supplementary information of~\cite{google2025observation}).
Thus, we must turn to more specialized methods for this problem.
We comment on how our diameter truncation rule can be seen as an extension of Pauli propagation to two-copy operators in \Cref{sec:pauli-prop}.

First, we discuss the cached Monte Carlo method of Ref.~\cite{google2025observation}, which was shown to perform well for simulating $\mathrm{OTOC}^{(1)}$ for system sizes of up to $40$ qubits.
At a high level, the algorithm considers Pauli path propagation for each individual copy of the butterfly operator.
Because $\mathrm{OTOC}^{(1)}$ involves two copies of the butterfly operator, this means the algorithm effectively tracks two Pauli paths throughout the evolution.
The structure of $\mathrm{OTOC}^{(1)}$ dictates that the two Pauli paths must begin and end with the same Pauli string.
However, the paths may diverge and rejoin throughout the evolution.
Cached Monte Carlo effectively performs a truncation which removes pairs of Pauli paths which have large spacetime divergences between each other.
This approach is significantly different to our algorithm, as we keep track of the evolution of the two-copy butterfly operator as a whole rather than reconciling the individual evolutions of the two single-copy butterfly operators.
This, in turn, is why we consider the identity/flip basis of two-copy operators as opposed to the Pauli basis for single-copy operators.
Our truncation rules can also be viewed as a solely spatial restriction, while cached Monte Carlo controls spacetime deviations.

The work that appears most similar to ours is Ref.~\cite{liu2026efficient}.
Their algorithm also Heisenberg evolves the two-copy butterfly operator through the circuit and does so by tracking the deviation from the identity/flip structure of the ensemble average, similarly to our window truncation.
The main difference is that, in our algorithm, the deviation from the ensemble average tracked via the window persists throughout the whole simulation.
In contrast, the algorithm in Ref.~\cite{liu2026efficient} additionally restricts the duration of the deviations.
In other words, their analogue of the window size is constrained in spacetime, while ours is only restricted spatially.
Our algorithm also applies the diameter truncation, which has no analogue in Ref.~\cite{liu2026efficient}.
Moreover, we emphasize that Ref.~\cite{liu2026efficient} does not have an end-to-end rigorous proof of correctness or runtime guarantees.

\section{Natural generalizations of our algorithm}
\label{sec:generalization}

In this section, we discuss the prospect of generalizing our algorithm beyond the case of $\mathrm{OTOC}^{(1)}$ in 1D.
In particular, we consider natural approaches to simulate $\mathrm{OTOC}^{(2)}$ or $\mathrm{OTOC}^{(1)}$ in higher spatial dimensions based on our algorithm and explain why they may or may not succeed.

\subsection{Generalization to higher-order OTOCs}
A natural question is whether our approach can be extended to higher-order OTOCs, and in particular to $\mathrm{OTOC}^{(2)}$. 
In this case, we want to propagate four copies of the Heisenberg-evolved butterfly operator, rather than two.
The corresponding Haar-invariant subspace is spanned by the operators $\{\Pi_\sigma:\sigma\in S_4\}$, where $\Pi_\sigma$ permutes the four copies.
Thus, a natural generalization is to replace the identity and flip regions appearing in our algorithm for $\mathrm{OTOC}^{(1)}$ by regions labeled by permutations.
Crucially, two-qubit gates leave homogeneous permutation
regions invariant,
\begin{equation}
(G^\dagger)^{\otimes4}
(\Pi_{\sigma,i}\otimes\Pi_{\sigma,i+1})
G^{\otimes4}
=
\Pi_{\sigma,i}\otimes\Pi_{\sigma,i+1}.
\end{equation}
This suggests extending our window truncation by retaining terms with a bounded number of permutation interfaces (i.e., places where the permutation label changes), and allowing arbitrary four-copy operators within small windows around these interfaces.
For a fixed number $s$ of interfaces and windows of width at most $w$, the dimension of this representation scales as $\mathcal O(n^s256^{sw})$.
Hence, a constant number of interfaces and logarithmic window widths would again yield polynomial complexity.

An important difference is that the Haar-averaged
four-copy evolution no longer preserves terms with a single interface: gates crossing an interface can generate additional permutation regions.
Establishing an efficient simulation algorithm along these lines would require controlling the error from discarding terms with additional permutation regions and extending our diameter and window truncation bounds to four-copy evolution.
This remains an interesting question for future work.

\subsection{Generalization to 2D and higher dimensions}

A second question is whether our approach can be extended to 2D circuits.
In this case, our current truncation scheme does not seem to directly yield a polynomial-time algorithm in higher dimensions.
To explain the difficulty, consider $n$ qubits arranged on a square lattice $\Lambda$.
The local Haar-averaging identities remain unchanged, so the averaged two-copy evolution still belongs to the span of products of identities and flips:
\begin{equation}
    \E_U\left[(U^\dagger B U)^{\otimes 2}\right]
    =
    \sum_{S\subseteq\Lambda}c_S
    \left(\bigotimes_{i\in S}\flip_i\right)
    \otimes
    \left(\bigotimes_{i\notin S}\I_i\right).
\end{equation}
In particular, each term assigns a flip to the sites in $S$ and an identity to the remaining sites.

The main difference from our 1D setting is geometric.
In 1D, each averaged configuration consists of identities on the left and flips on the right.
Meanwhile, in 2D, the flip region can have an arbitrary shape, and the boundary separating it from the identity region can extend across the lattice.

This suggests replacing our 1D window with a band surrounding the boundary between the identity and flip regions.
We could retain an arbitrary two-copy operator within a band of width $w$, with identity and flip regions outside of this band on either side.
As before, gates acting entirely within either identity or flip region leave it unchanged, so only gates intersecting the band require nontrivial updates.

However, the computational cost depends on the number of qubits in the band.
Even for a straight boundary crossing a square lattice of side length $\sqrt{n}$, the length of this boundary is $\Theta(\sqrt{n})$.
Thus, a band of width $w\leq\sqrt{n}$ contains $\Theta(w\sqrt{n})$ qubits.
Representing an arbitrary two-copy operator on these sites in the band requires $16^{\Theta(w\sqrt{n})}$ coefficients.
Thus, restricting the band to logarithmic width is insufficient to obtain a polynomial-time algorithm.
We leave it to future work to investigate whether this structure can be exploited by Monte Carlo methods to estimate first-order OTOCs of 2D circuit instances efficiently. 

\section{Outlook}

In the realm of near-term beyond-classical experimental efforts, OTOCs in random quantum circuits~\cite{google2025observation} are a natural successor to random circuit sampling experiments~\cite{arute2019quantum,wu2021strong} due to having reproducible outputs despite using highly random circuits, while still confounding traditional algorithms such as tensor network contraction.
Our work designing a provably efficient classical simulation algorithm for first-order OTOCs in 1D hints at the possibility that the same structure which allows for reproducibility could also be exploited by classical algorithms, though generalizing to higher orders and spatial dimensions remains challenging.
While we provide the first step towards a rigorous understanding of the classical computational complexity of OTOCs, there are still many open directions for future exploration, which we discuss in \Cref{sec:generalization}.
Are OTOCs classically simulable in higher spatial dimensions, e.g., 2D or 3D?
Are higher-order OTOCs also easy for classical computers?

There is something quite rare about the existence of provably efficient algorithms for non-trivial classes of quantum problems.
Such algorithms often reveal hidden structure about the class of problems, as is the case for Clifford circuits~\cite{gottesman1998heisenberg,aaronson2004improved} or low-entanglement physical systems~\cite{verstraete2008matrix,perez2006matrix}.
The study of OTOCs is motivated by many different aspects of quantum computation and physics, and hence both the algorithms and the structure we discover in our proof may find applications beyond their initial scope.
For example, many algorithms or ansatzes born from an exact solution to a particular problem, e.g., matrix product states, find usefulness well outside that original problem~\cite{schollwock2011density, cirac2021matrix}.
Moreover, our numerical experiments demonstrate that the algorithm our proof relies on is already extremely practical and efficient.
As such, our truncation rules could potentially be used as effective classical simulation techniques for a much wider range of systems than explored here, such as evolutions governed by local Hamiltonians or different families of quantum circuits.

Overall, we believe that understanding this structure is only the beginning for a broad and important class of quantum circuits.
As quantum experiments push forward, we believe our work will help form a reliable and exact reference point for guiding future explorations.

\subsection*{Acknowledgments}
The authors especially thank Ryan Babbush and Tommy Schuster for insightful discussions.
The authors also thank Sergio Boixo, Robin Kothari, Francesco A. Mele, Tom O'Brien, Rolando Somma, and Brayden Ware for valuable discussions surrounding OTOCs and prior efforts on classical simulating them.

L.L. and A.A.M. were student researchers at Google Quantum AI when this work was completed.
L.L. is supported by the U.S. Department of
Energy, Office of Science, Office of Advanced Scientific Computing Research, Department of Energy Computational Science Graduate Fellowship under Award Number DE-SC0026073.
A.A.M. acknowledges support from a 2025 Google PhD Fellowship and Anthropic's AI for Science program, as well as BMFTR through
PhoQuant, QPIC, Hybrid++, QR.N, PASQUOPS, and QSolid; BMWK through EniQmA; DFG through CRC~183; the EU Quantum Flagship through PasQuans2 and Millenion; Munich Quantum Valley; Berlin Quantum; and the European Research Council through the Advanced Grant DebuQC.

This report was prepared as an account of work sponsored by an agency of the United States Government. Neither the United States Government nor any agency thereof, nor any of their employees, makes any warranty, express or implied, or assumes any legal liability or responsibility for the accuracy, completeness, or usefulness of any information, apparatus, product, or process disclosed, or represents that its use would not infringe privately owned rights. Reference herein to any specific commercial product, process, or service by trade name, trademark, manufacturer, or otherwise does not necessarily constitute or imply its endorsement, recommendation, or favoring by the United States Government or any agency thereof. The views and opinions of authors expressed herein do not necessarily state or reflect those of the United States Government or any agency thereof.

\subsection*{AI use statement}
Many of the core algorithm and proof ideas such as a truncated ANOVA expansion related to the diameter truncation rule were developed by the authors without AI in attempts to reach a rigorous proof.  Key sticking points were 
later resolved through interactive sessions with ChatGPT 6 Astra, after earlier models had failed at the same task.
The authors subsequently verified and simplified the full detailed proof presented here. The writing in this paper was done without AI use, apart from minor sentence refinements and checking for errors.
AI assistance from GPT Codex, Claude 5.5, and Fable was also used for the numerical implementations, experiments and figures.
We take full responsibility for the correctness and exposition.

\bibliographystyle{unsrt}
\bibliography{refs}

\newpage
\resumetoc
\appendix
\appendixpage

\tableofcontents

\section{Preliminaries}

\label{sec:prelim}
Throughout, $[m] \triangleq \{1,\dots, m\}$.
The \emph{diameter} of a set $S \subseteq [m]$, denoted $\mathrm{diam}(S)$, is defined as $\mathrm{diam}(S) \triangleq \max S - \min S$.

We let $n$ denote the number of qubits and $N \triangleq 2^n$ the dimension of the corresponding Hilbert space.
When acting on an $N$-dimensional operator, $\ntr \triangleq \tr/N$.
Let $I$ denote the $2\times 2$ identity operator.
We let $\mathcal{P}_n \triangleq \{I, X, Y,Z\}^{\otimes n}$ denote the set of $4^n$ $n$-qubit Pauli matrices.
For a Pauli $P \in \pauli_n$, we define the \emph{support} of $P$ as $\supp(P) \triangleq \{i \in [n] : P_i \neq I\}$.
For a set $S \subseteq [n]$, we also write $\pauli_S$ to denote the set of $|S|$-qubit Paulis supported on $S$.

Let $\mathcal{B}(\mathcal{H})$ denote the set of bounded linear operators on a finite-dimensional Hilbert space $\mathcal{H}$.
This set $\mathcal{B}(\mathcal{H})$ can be viewed as a Hilbert space when endowed with the Hilbert-Schmidt inner product
\begin{equation}
    \langle A, B \rangle \triangleq \tr(A^\dagger B),
\end{equation}
where $A,B \in \mathcal{B}(\mathcal{H})$.
We use $\norm{\cdot}_F$ to denote the norm induced by this inner product, namely the Hilbert-Schmidt or Frobenius norm.
In particular,
\begin{equation}
    \norm{A}_F^2 \triangleq \langle A, A \rangle = \tr(A^\dagger A).
\end{equation}

\begin{lemma}[Parseval's identity; Proposition 10 in~\cite{montanaro2008quantum}]
    \label{lem:parseval}
    Let $A \in \mathcal{B}(\mathbb{C}^{2^n})$.
    Writing $A$ in the Pauli basis, $A = \sum_{P \in \pauli_n} a_P P$.
    Then,
    \begin{equation}
        \frac{1}{2^n}\norm{A}_F^2 = \sum_{P \in \pauli_n} |a_P|^2.
    \end{equation}
\end{lemma}

We prove the following basic Cauchy-Schwarz inequality for random operators.

\begin{lemma}
	\label{lem:cauchy-schwarz}
	Let $m \geq 1$.
	Let $(C,D)$ and $(E,F)$ be independent random pairs of $m$-qubit operators.
	Let $(C',D')$ and $(E',F')$ be independent copies of these operators.
	Then,
	\begin{equation}
		\frac{1}{4^{m}}\E[|\trace(C^\dagger E D^\dagger F)|^2] \leq \frac{1}{4^m}\sqrt{\E[|\trace(C^\dagger C')\trace(D^\dagger D')|^2]} \sqrt{\E[|\trace(E^\dagger E') \trace(F^\dagger F')|^2]}.
	\end{equation}
\end{lemma}

\begin{proof}
	Expanding in terms of the matrix entries of each of these operators, we can write
	\begin{align}
		\trace(C^\dagger E D^\dagger F) = \sum_{\substack{i,j,k,l=1}}^{2^m} \overline{C_{ji}} \overline{D_{lk}} E_{jk} F_{li}.
	\end{align}
	Taking the expectation and the magnitude squared, we have
	\begin{align}
	 	\frac{1}{4^{m}}\E[|\trace(C^\dagger E D^\dagger F)|^2] &= \frac{1}{4^m}\sum_{\substack{i,j,k,l\\i',j',k',l'}} \E[\overline{C_{ji}} \overline{D_{lk}} C_{j'i'} D_{l'k'}] \E[E_{jk} F_{li} \overline{E_{j'k'}} \overline{F_{l'i'}}]\\
	 	&\leq \frac{1}{4^m} \sqrt{\sum_{\substack{i,j,k,l\\i',j',k',l'}} |\E[\overline{C_{ji}} \overline{D_{lk}} C_{j'i'} D_{l'k'}]|^2} \sqrt{\sum_{\substack{i,j,k,l\\i',j',k',l'}} |\E[E_{jk} F_{li} \overline{E_{j'k'}} \overline{F_{l'i'}}]|^2}\\
	 	&= \frac{1}{4^m}\sqrt{\E[|\trace(C^\dagger C')\trace(D^\dagger D')|^2]} \sqrt{\E[|\trace(E^\dagger E') \trace(F^\dagger F')|^2]}.
	 \end{align} 
	 In the first line, we use independence.
	 In the second line, we use the usual form of Cauchy-Schwarz.
\end{proof}

\subsection{Haar measure}
\label{sec:haar}

\begin{definition}[Haar measure]
    Let $U(N)$ denote the group of $N \times N$ complex unitary matrices.
    The \emph{Haar measure} on $U(N)$ is the unique probability measure with the following property: if $U$ is distributed according to the Haar measure, then for any unitary $V \in U(N)$, both $V U $ and $UV$ are distributed according to the Haar measure.
    We denote a unitary drawn from the Haar measure on $U(N)$ as $U \sim \mathrm{Haar}(U(N))$.
\end{definition}

\begin{definition}[Identity and flip operators]
	The \emph{identity operator} $\I = I \otimes I$ on two qubits is a linear operator which leaves any tensor product state unchanged, i.e., for all $\ket{\psi}, \ket{\phi} \in \mathbb{C}^2$,
	\begin{equation}
		\I(\ket{\psi}\otimes \ket{\phi}) = \ket{\psi} \otimes \ket{\phi}.
	\end{equation}
	The \emph{flip} or \emph{swap operator} $\flip$ on two qubits is a linear operator that interchanges the tensor factors of any product state, i.e., for all $\ket{\psi}, \ket{\phi} \in \mathbb{C}^2$,
	\begin{equation}
		\flip(\ket{\psi} \otimes \ket{\phi}) = \ket{\phi} \otimes \ket{\psi}.
	\end{equation}
\end{definition}

On $n$ pairs of qubits, we often write $\flip_{\mathrm{all}} \triangleq \bigotimes_{i=1}^n \flip_i$ for the flip operator on all $2n$ qubits.
A key property of the flip operator is that for any operators $A, B \in \mathcal{B}(\mathbb{C}^N)$,
\begin{equation}
	\tr((A\otimes B)\flip_{\mathrm{all}}) = \tr(AB).
\end{equation}
This is sometimes referred to as the \emph{swap trick}.

\begin{fact}[Identity and flip are invariant under two-copy rotations]
	\label{fact:i-flip-invariant}
	Let $U \in U(2)$.
	Then,
	\begin{equation}
		U^{\dagger \otimes 2} \,\I\, U^{\otimes 2} = \I, \quad U^{\dagger \otimes 2} \,\flip\, U^{\otimes 2} = \flip.
	\end{equation}
\end{fact}

\begin{lemma}[Second Haar moments; Corollary 13 of~\cite{mele2024introduction}]
	\label{lem:second-moment}
	Let $O \in \mathcal{B}(\mathbb{C}^N \otimes \mathbb{C}^N)$.
	Then,
	\begin{equation}
		\E_{U \sim \mathrm{Haar}(U(N))}[U^{\otimes 2} O U^{\dagger \otimes 2}] = c_{\I^{\otimes n}, O} \I^{\otimes n} + c_{\flip_{\mathrm{all}}, O} \flip_{\mathrm{all}},
	\end{equation}
    where 
    \begin{equation}
        c_{\I^{\otimes n},O} = \frac{\tr(O) - N^{-1}\tr(\flip_{\mathrm{all}} O)}{N^2-1},\quad\quad c_{\flip_{\mathrm{all},O}} = \frac{\tr(\flip_{\mathrm{all}} O) - N^{-1}\tr(O)}{N^2-1}.
    \end{equation}
	In other words, the second Haar moment of an operator $O$ projects it onto $\mathrm{span}\{\I^{\otimes n}, \flip_{\mathrm{all}}\}$.
\end{lemma}

\begin{fact}[Second Haar moments of Paulis]
    \label{lem:second-moment-pauli}
    Let $P, Q \in \pauli_1$.
    Then,
    \begin{equation}
        \E_{U\sim \mathrm{Haar}(U(2))}[U^{\dagger\otimes 2} (P\otimes Q) U^{\otimes 2}] = \begin{cases}
            I \otimes I & \text{if } P = Q= I\\
            \frac{1}{3}\sum_{R\in \{X,Y,Z\}} R\otimes R & \text{if } P= Q \neq I\\
            0 & \text{else}
        \end{cases}.
    \end{equation}
\end{fact}

We prove a standard fact about Haar-random gates.
Similar statements have appeared in, e.g., Refs.~\cite{aharonov2023polynomial,harrow2009random}, and we provide a proof for completeness.

\begin{lemma}
	\label{lem:haar-conj}
	Let $G \sim \mathrm{Haar}(U(N))$ be a Haar-random gate from $U(N)$.
	Then, for any two Paulis $P,Q \in \mathcal{P}_n \setminus \{I^{\otimes n}\}$,
	\begin{equation}
		\label{eq:haar-conj-1}
		\E_G[\ntr(PGQG^\dagger)] = 0.
	\end{equation}
	Moreover, for any Paulis $P,P', Q,Q' \in \mathcal{P}_n \setminus \{I^{\otimes n}\}$, then
	\begin{equation}
		\label{eq:haar-conj-2}
        \E_G[\ntr(PGQG^\dagger)\ntr(P'GQ'G^\dagger)] = \frac{\indicator\{P=P'\}\indicator\{Q=Q'\}}{N^2 - 1}.
	\end{equation}
\end{lemma}

\begin{proof}
	For any $P \in \pauli_n \setminus \{I^{\otimes n}\}$, there exists a Pauli $R$ anticommuting with it.
	By invariance of the Haar measure, we have
	\begin{equation}
		\E_G[\ntr(PGQG^\dagger)] = \E_G[\ntr(PRGQ(RG)^\dagger)] = \E_G[\ntr(RPRGQG^\dagger)] = -\E_G[\ntr(PGQG^\dagger)].
	\end{equation}
	This proves \Cref{eq:haar-conj-1}.

	Now, suppose $P \neq P'$.
	Choose a site $i\in[n]$ for which $P_i \neq P_i'$.
	There exists a single-qubit Pauli which anticommutes with one of $P_i, P_i'$ and commutes with the other.
	Let $R$ be the corresponding $n$-qubit Pauli padded with identities on all other sites.
	Then, by the same argument as above,
	\begin{equation}
		\E_G[\ntr(PGQG^\dagger)\ntr(P'GQ'G^\dagger)] = \E_G[\ntr(PRGQ(RG)^\dagger)\ntr(P'RGQ'(RG)^\dagger)] = -\E_G[\ntr(PGQG^\dagger)\ntr(P'GQ'G^\dagger)].
	\end{equation}
	Thus, this expectation is zero.
	An analogous argument works if $Q \neq Q'$ instead.
	Hence, the expectation is only nonzero when $P = P'$ and $Q = Q'$.

	Fix $Q \in \pauli_n \setminus\{I^{\otimes n}\}$.
	For fixed $G$, we have
	\begin{equation}
		\sum_{P \in \pauli_n \setminus \{I^{\otimes n}\}} \ntr(PGQG^\dagger)^2 = \sum_{P \in \pauli_n} \ntr(PGQG^\dagger)^2 = \frac{1}{N}\norm{GQG^\dagger}_F^2 = 1.
	\end{equation}
	In the first equality, we use that $\ntr(GQG^\dagger) = \ntr(Q) = 0$.
	In the second equality, we use Parseval's identity (\Cref{lem:parseval}).
	Taking expectations on both sides over $G$, we have
	\begin{equation}
		\label{eq:sum-equal-1}
		\sum_{P \in \pauli_n \setminus \{I^{\otimes n}\}} \E_G[\ntr(PGQG^\dagger)^2] = 1.
	\end{equation}
	Now, we notice that
	\begin{equation}
		\E_G[\ntr(PGQG^\dagger)^2] = \E_G[\ntr(P'GQG^\dagger)^2]
	\end{equation}
	for any $P,P' \in \pauli_n \setminus \{I^{\otimes n}\}$.
	This is because any two non-identity Paulis can be related by a unitary $W$: $P' = WPW^\dagger$.
	Thus, we can write
	\begin{equation}
		\E_G[\ntr(WPW^\dagger GQG^\dagger)^2] = \E_G[\ntr(PW^\dagger GQ(W^\dagger G)^\dagger)^2] = \E_G[\ntr(PGQG^\dagger)^2],
	\end{equation}
	where in the last equality, we use Haar invariance.
	Using this in \Cref{eq:sum-equal-1}, then we get that
	\begin{equation}
		(N^2 - 1)\E_G[\ntr(PGQG^\dagger)^2] = 1,
	\end{equation}
	which proves \Cref{eq:haar-conj-2}.
\end{proof}

\subsection{Brickwork random quantum circuits}
\label{sec:brickwork}

In the following definition, we consider $n,d \geq 2$ even for simplicity.
Our results also hold for $n,d$ odd, but this makes the notation simpler.

\begin{definition}[Brickwork random quantum circuits in 1D]
    \label{def:brickwork}
    Consider a 1D chain of $n$ qubits.
    A \emph{brickwork random quantum circuit} $U$ of depth $d$ on these $n$ qubits can be written as $U = U_d \cdots U_1$, where $U_j$ are each of the layers of the circuit.
    Each layer consists of independent two-qubit Haar-random gates acting on adjacent qubits.
    Explicitly,
    \begin{equation}
	   U_j = \begin{cases}
		\bigotimes_{i \in \{2,4,\dots, n-2\}} G_j(i,i+1) & \text{if $j$ even}\\
		\bigotimes_{i \in \{1,3,\dots,n-1\}} G_j(i,i+1) & \text{if $j$ odd}
	\end{cases},
\end{equation}
where each $G_j(i,i+1) \sim \mathrm{Haar}(U(4))$ independently.
Denote this distribution over quantum circuits as $\mathcal{U}_{n,d}$.
Sometimes, we also write $U = L_{d/2} \cdots L_1$ when grouping the layers in groups of two, i.e., $L_j = U_{2j} U_{2j-1}$.
\end{definition}

\begin{lemma}
    \label{lem:pauli-second-moment-B}
	Let $\ell \in [n]$.
	Let $U$ be a brickwork random quantum circuit on $n$ qubits in 1D.
	Let $B \neq I$ be a single-qubit Pauli supported on qubit $n$.
	Write $U^\dagger B U = \sum_{P \in \pauli_n} b_P P$.
	Define
	\begin{equation}
		p_\ell \triangleq \E_U\Bigg[\sum_{\substack{P \in \pauli_n\\ P_1 = \cdots = P_{\ell-1} = I, P_\ell \neq I}} |b_P|^2\Bigg].
        \label{eq:pauli-B-leftmost-weight}
	\end{equation}
	Then, $p_\ell$ forms a probability distribution over $[n]$, i.e., $p_\ell \geq 0$ and $\sum_{\ell=1}^n p_\ell = 1$.
	Moreover, for every $P,Q \in \pauli_n \setminus \{I^{\otimes n}\}$, 
	\begin{equation}
		\E_U[b_P b_Q] = \frac{p_\ell}{3\cdot 4^{n-\ell}}\indicator\{P = Q\},
	\end{equation}
	where $\ell = \min\{i \in [n] : P_i \neq I\}$.
\end{lemma}

This lemma intuitively tells us that, conditional on the leftmost non-identity site being $\ell$, the second-moment Pauli distribution of $U^\dagger B U$ is: $P_\ell$ is uniformly distributed over $\{X,Y,Z\}$, $P_i$ for $i > \ell$ is independently uniformly distributed over $\{I,X,Y,Z\}$, and $P_i$ for $i < \ell$ is identity.
This gives the $1/(3\cdot 4^{n-\ell})$ factor.
Then, $p_\ell$ is the probability that the leftmost non-identity site is $\ell$.

\begin{proof}
    For each site $\ell$, we collect all strings whose leftmost non-identity site is~$\ell$ into one equal-weight sum,
    \begin{align}
        D_\ell &\triangleq \frac{1}{3\cdot 4^{n-\ell}}
        \sum_{\substack{P\in\pauli_n\\ P_1=\cdots=P_{\ell-1}=I,\ P_\ell\neq I}} P\otimes P.
    \end{align}
    There are three choices for the Pauli at site $\ell$ and four choices at each site greater than $\ell$, so the coefficients in $D_\ell$ sum to one.
    It suffices to prove that the averaged, evolved two-copy butterfly operator is a convex combination of the $D_\ell$, i.e.,
    \begin{align}
        \label{eq:convex-combo}
        \E_U[(U^\dagger B U)^{\otimes 2}]
        &= \sum_{\ell=1}^n c_\ell D_\ell,
    \end{align}
    for coefficients $c_\ell \geq 0$ such that $\sum_{\ell=1}^n c_\ell = 1$.
    To see why this implies the claim, consider expanding the evolved two-copy butterfly operator in the Pauli basis:
    \begin{align}
        \E_U[(U^\dagger B U)^{\otimes 2}]
        &= \sum_{P,Q\in\pauli_n}\E_U[b_Pb_Q]\,P\otimes Q.
    \end{align}
    Every term in $\sum_\ell c_\ell D_\ell$ has the form $P\otimes P$. Thus, if the average has this form, comparing coefficients gives $\E_U[b_Pb_Q]=0$ for $P\neq Q$.

    To see why $\E[b_P^2] = p_\ell/(3\cdot 4^{n-\ell})$, fix a string $P$ with leftmost non-identity site $\ell$. In $D_\ell$, the coefficient of $P\otimes P$ is $1/(3\cdot 4^{n-\ell})$. Only $D_\ell$ contributes to this term, so in $\sum_j c_jD_j$ its coefficient is $c_\ell/(3\cdot 4^{n-\ell})$. In the Pauli expansion of the average, the coefficient of $P\otimes P$ is $\E_U[b_P^2]$. Equating the two coefficients gives
    \begin{align}
        \E_U[|b_P|^2] &= \E_U[b_P^2] = \frac{c_\ell}{3\cdot 4^{n-\ell}},
    \end{align}
    where the first equality uses that $b_P$ is real because $U^\dagger B U$ is Hermitian.
    By the definition of $p_\ell$ in \Cref{eq:pauli-B-leftmost-weight}, summing over the $3\cdot 4^{n-\ell}$ strings with leftmost site $\ell$ gives $p_\ell=(3\cdot 4^{n-\ell})\,[c_\ell/(3\cdot 4^{n-\ell})]=c_\ell$.
    Thus, the $p_\ell$ are the weights of the convex combination.
    This completes the proof of the claim, assuming \Cref{eq:convex-combo}.

    We now show that the average indeed has the stated convex-combination form in \Cref{eq:convex-combo}.
    Since the gates are independent, we can average them one at a time.
    A pair of Paulis $I \otimes I$ stays fixed.
    For a two-qubit Haar-random gate $G$ and a non-identity two-qubit Pauli $R$, \Cref{lem:haar-conj} with $N=4$ gives
    \begin{align}
        \label{eq:haar-conj-applied}
        \E_G[(G^\dagger R G)^{\otimes 2}]
        &= \frac{1}{15}\sum_{S\in\pauli_2\setminus\{I\otimes I\}} S\otimes S.
    \end{align}
    The $15$ pairs of $S \otimes S$ have equal weight. Of these, $12$ are non-identity on the first site and arbitrary on the second, while $3$ are identity on the first site and non-identity on the second.
    The first two-qubit gate touching $B$ must act on sites $n$ and $n-1$.
    For such a gate $G$, we have
    \begin{equation}
        \E_G[(G^\dagger (I\otimes B) G)^{\otimes 2}] = \frac{4}{5}D_{n-1} + \frac{1}{5}D_n.
    \end{equation}
    Inductively, consider a term $D_{\ell}$ and a two-qubit gate $G$ acting on sites $i$ and $i+1$.
    We show that averaging over this gate preserves the convex-combination form of the average evolved butterfly operator.
    In particular, we consider $\E_G[(G^\dagger D_{\ell}G)^{\otimes 2}]$ and show that this remains a convex combination of terms.
    We break this up into cases:
\begin{itemize}
        \item If both $i$ and $i+1$ are less than $\ell$, then every term in $D_\ell$ is $I \otimes I$ on sites $i$ and $i+1$. Thus, the gate acts on $I \otimes I$ and leaves $D_\ell$ fixed, so $\E_G[(G^\dagger D_\ell G)^{\otimes 2}] = D_\ell$.
        \item If both $i$ and $i+1$ are greater than $\ell$, then $D_\ell$ is simply a uniform sum of two-copy Paulis on sites $i$ and $i+1$, i.e., $\frac{1}{16}\sum_{R \in \pauli_2} R \otimes R$.
        Then, we have
        \begin{align}
           \frac{1}{16}\E_G\left[G^{\dagger\otimes 2} \left(\sum_{R \in \pauli_2} R \otimes R\right)G^{\otimes 2}\right] &= \frac{1}{16}(I\otimes I)^{\otimes 2} + \frac{1}{16}\sum_{R \in \pauli_2 \setminus \{I\otimes I\}} \E_G[(G^\dagger R G)^{\otimes 2}]\\
           &= \frac{1}{16}(I\otimes I)^{\otimes 2} + \frac{1}{16\cdot 15}\sum_{R \in \pauli_2 \setminus \{I\otimes I\}} \sum_{S \in \pauli_2 \setminus \{I \otimes I\}} S \otimes S\\
           &= \frac{1}{16}(I\otimes I)^{\otimes 2} + \frac{1}{16} \sum_{S \in \pauli_2 \setminus \{I \otimes I\}} S \otimes S\\
           &= \frac{1}{16}\sum_{R \in \pauli_2} R \otimes R,
        \end{align}
        where in the second line, we used \Cref{eq:haar-conj-applied}.
        The Paulis at other sites are unaffected by the gate, so this shows that $\E_G[(G^\dagger D_\ell G)^{\otimes 2}] = D_\ell$ in this case as well.
        \item If $\ell = i$, then $D_\ell$ is $\frac{1}{12}\sum_{P \neq I, Q \in \pauli_1} (P\otimes Q)^{\otimes 2}$ on sites $i,i+1$. Then, we have
        \begin{align}
            &\frac{1}{12}\E_G\Bigg[\Bigg(G^\dagger \Bigg(\sum_{\substack{P\neq I\\Q \in \pauli_1}} P\otimes Q\Bigg) G\Bigg)^{\otimes 2}\Bigg]\\
            &= \frac{1}{15}\sum_{S \in \pauli_2 \setminus \{I\otimes I\}} S\otimes S\\
            &= \frac{1}{15}\sum_{\substack{C_1 \in \{X, Y,Z\}\\C_2 \in \pauli_1}} (C_1\otimes C_2)^{\otimes 2} + \frac{1}{15}\sum_{C_2 \in \{X,Y,Z\}}(I\otimes C_2)^{\otimes 2}\\
            &= \frac{4}{5}\Bigg(\frac{1}{12}\sum_{\substack{C_1 \in \{X, Y,Z\}\\C_2 \in \pauli_1}} (C_1\otimes C_2)^{\otimes 2}\Bigg) + \frac{1}{5}\Bigg(\frac{1}{3}\sum_{C_2 \in \{X,Y,Z\}}(I\otimes C_2)^{\otimes 2}\Bigg),
        \end{align}
        where the second line follows by a similar calculation as the previous bullet.
        Note that the two terms in parentheses are $D_i$ and $D_{i+1}$, respectively, restricted to sites $i$ and $i+1$.
        The other sites are unaffected by the gate, so this shows that $\E_G[(G^\dagger D_\ell G)^{\otimes 2}] = 4D_i/5 + D_{i+1}/5$.
        \item If $\ell=i+1$, then by an extremely similar calculation to the previous bullet, one can show that $\E_G[(G^\dagger D_\ell G)^{\otimes 2}] = 4D_i/5 + D_{i+1}/5$ in this case as well.
    \end{itemize}
    Thus, averaging over each gate maps a convex combination of the $D_\ell$ to another such combination.
    By induction, this completes the proof of the convex-combination form in \Cref{eq:convex-combo}.
\end{proof}

We also have an analogous statement when the single-qubit Pauli is supported on qubit $1$ instead.
The proof is completely analogous, so we omit its details.

\begin{lemma}
    \label{lem:pauli-second-moment-M}
	Let $r \in [n]$.
	Let $U$ be a brickwork random quantum circuit on $n$ qubits in 1D.
	Let $M \neq I$ be a single-qubit Pauli supported on qubit $1$.
	Write $U^\dagger M U = \sum_{P \in \pauli_n} m_P P$.
	Define
	\begin{equation}
		q_r \triangleq \E_U\Bigg[\sum_{\substack{P \in \pauli_n\\P_r \neq I, P_{r+1} = \cdots = P_n = I}} |m_P|^2\Bigg].
	\end{equation}
	Then, $q_r$ forms a probability distribution over $[n]$, i.e., $q_r \geq 0$ and $\sum_{r=1}^n q_r = 1$.
	Moreover, for every $P,Q \in \pauli_n \setminus \{I^{\otimes n}\}$,
	\begin{equation}
		\E_U[m_P m_Q] = \frac{q_r}{3 \cdot 4^{r-1}}\indicator\{P = Q\},
	\end{equation}
	where $r = \max\{i \in [n] : P_i \neq I\}$.
\end{lemma}

\subsection{Two-copy operators}
\label{sec:two-copy}

In this section, we refer to operators in $\mathcal{B}(\mathcal{H}\otimes \mathcal{H})$ as \emph{two-copy operators} because they act on two copies of the Hilbert space $\mathcal{H}$.

\begin{definition}[Two-copy operator basis]
    \label{def:basis}
    Let $\{A_1,\dots, A_{14}\}$ be an orthonormal basis (with respect to the Hilbert-Schmidt inner product) of $\mathrm{span}\{\I, \flip\}^\perp$.
    Then, we can write a basis of $\mathcal{B}(\mathbb{C}^2 \otimes \mathbb{C}^2)$ as $\mathcal{A}_1 \triangleq \{\I, \flip, A_1, \dots, A_{14}\}$.
    Under the identification $\mathbb{C}^N \otimes \mathbb{C}^N \cong (\mathbb{C}^2 \otimes \mathbb{C}^2)^{\otimes n}$, we can define a basis of $\mathcal{B}(\mathbb{C}^N \otimes \mathbb{C}^N)$ of two-copy operators on $n$ qubits as
    \begin{equation}
        \mathcal{A} \triangleq \mathcal{A}_1^{\otimes n} = \{\I, \flip,A_1,\dots, A_{14}\}^{\otimes n}.
    \end{equation}
\end{definition}

Here, we single out the two basis elements $\I, \flip$ due to their invariant properties discussed in \Cref{fact:i-flip-invariant}.
We note that this basis $\mathcal{A}$ is not an orthonormal basis.

\begin{definition}[Non-invariant sites]
    \label{def:sa}
    For $A \in \mathcal{A}$, we write the set of sites for which $A$ is non-invariant under two-copy rotations as
    \begin{equation}
        S_A \triangleq \{i \in [n] : A_i \notin \mathrm{span}\{\I, \flip\}\}.
    \end{equation}
\end{definition}

In the analysis of our algorithm, it will be helpful to define the following orthogonal projector, which is inspired by the ANOVA expansion (see, e.g., Section 8.3 of~\cite{o2021analysis}).
Let $k \geq 1$, and let $V = V_1 \otimes \cdots \otimes V_n$ be a layer of single-qubit Haar-random gates, where $V_i \sim \mathrm{Haar}(U(2))$.
Define the orthogonal projector
\begin{equation}
	\label{eq:pi}
    \Pi \triangleq \sum_{\substack{S \subseteq [n]\\\mathrm{diam}(S) \leq k}} \prod_{i \in S} (\mathrm{id} - \E_{V_i}) \prod_{i \notin S} \E_{V_i},
\end{equation}
where $\E_{V_i}$ denotes averaging over the Haar-random gate $V_i$ and leaving all other gates fixed.
In the following lemma, we show that this is indeeed an orthogonal projector.

\begin{lemma}[Orthogonal projector]
    The operator $\Pi$ defined in \Cref{eq:pi} is an orthogonal projector.
\end{lemma}

\begin{proof}
    First, notice that each of the $\E_{V_i}$'s is an orthogonal projector:
    \begin{equation}
        \langle f, \E_{V_i}[g] \rangle = \E_{V_{\neq i}}[\overline{\E_{V_i}[f(V)]} \E_{V_i}[g(V)]] = \langle \E_{V_i}[f], g\rangle.
    \end{equation}
    The fact that $\E_{V_i}$ is its own square is clear.
    Moreover, these projectors commute by independence of the Haar-random gates $V_i$, i.e., $\E_{V_i}\E_{V_{i'}} = \E_{V_{i'}}\E_{V_i}$ for all $i,i'$.
    Thus, because the product of commuting orthogonal projectors is also an orthogonal projectors, then each term in the sum $\Pi_S \triangleq \prod_{i \in S} (\mathrm{id} - \E_{V_i}) \Pi_{i\notin S} \E_{V_i}$ is an orthogonal projector.
    
    We also observe that $\Pi_S, \Pi_{S'}$ are orthogonal for $S \neq S'$.
    Since $S \neq S'$, there exists some $i$ in the symmetric difference $S \triangle S'$.
    Without loss of generality, suppose $i \in S$ and $i \notin S'$.
    Then, $\Pi_S$ contains a factor of $\mathrm{id} - \E_{V_i}$ while $\Pi_{S'}$ contains a factor of $\E_{V_i}$.
    Because $(\mathrm{id} - \E_{V_i})\E_{V_i} = 0$, then $\Pi_S \Pi_{S'} = 0$ if $S \neq S'$.

    With this, we can show that $\Pi$ is an orthogonal projector:
    \begin{equation}
        \Pi^2 = \sum_{\substack{S,S' \subseteq [n]\\\mathrm{diam}(S), \mathrm{diam}(S') \leq k}} \Pi_S \Pi_{S'} = \sum_{\substack{S \subseteq [n]\\\mathrm{diam}(S) \leq k}} \Pi_S = \Pi.
    \end{equation}
    The property that $\Pi^\dagger = \Pi$ is clear because $\Pi_S$ also has this property.
\end{proof}

The following lemma shows that, when applying the projector to a two-copy operator that is Heisenberg-evolved by $V^{\otimes 2}$, it applies a specific truncation rule to its expansion in the basis $\mathcal{A}$.

\begin{lemma}
	\label{lem:pi}
    Let $V = V_1 \otimes \cdots \otimes V_n$ be a layer of single-qubit Haar-random gates, where $V_i \sim \mathrm{Haar}(U(2))$.
    Let $O \in \mathcal{B}(\mathbb{C}^N \otimes \mathbb{C}^N)$.
    Expand $V^{\dagger \otimes 2} O V^{\otimes 2}$ in the basis $\mathcal{A}$:
    \begin{equation}
        V^{\dagger \otimes 2} O V^{\otimes 2} = \sum_{A \in \mathcal{A}} c_A A.
    \end{equation}
    Then, 
    \begin{equation}
        \Pi (V^{\dagger \otimes 2} O V^{\otimes 2}) = \sum_{\substack{A \in \mathcal{A}\\\mathrm{diam}(S_A) \leq k}} c_A A,
    \end{equation}
    where $S_A$ is defined in \Cref{def:sa}.
\end{lemma}

\begin{proof}
    Averaging over $V_i$ is simply a second Haar moment, which we know projects onto $\mathrm{span}\{\I, \flip\}$ by \Cref{lem:second-moment}.
    Thus, it is clear that
    \begin{equation}
    	\E_{V_i}[V^{\dagger \otimes 2} O V^{\otimes 2}] = \sum_{\substack{A \in \mathcal{A}\\i \notin S_A}} c_A A.
    \end{equation}
    This implies that, for a set $S \subseteq [n]$,
    \begin{equation}
    	\left[\prod_{i \in S} (\mathrm{id} - \E_{V_i})\prod_{i \notin S} \E_{V_i}\right](V^{\dagger \otimes 2} O V^{\otimes 2}) = \sum_{\substack{A \in \mathcal{A}\\ S_A = S}} c_A A.
    \end{equation}
    Summing over the sets in the definition of $\Pi$ completes the claim.
\end{proof}

\section{Classical simulation algorithm}
\label{sec:algo}

Our main result is an efficient classical algorithm for simulating $\mathrm{OTOC}^{(1)}$.
More precisely, we prove the following theorem, which is a restatement of \Cref{thm:sim-main} from the main text.

\setcounter{theorem}{0}
\begin{theorem}[Classical simulation with mean-squared error]
	\label{thm:sim}
	Let $0 < \epsilon < 1$.
	There exists a deterministic classical algorithm (\Cref{alg:sim}) that, when given a quantum circuit $U \sim \mathcal{U}_{n,d}$, outputs a real number $h(U)$ such that
	\begin{equation}
		\E_{U \sim \mathcal{U}_{n,d}}[|h(U) - f^{(1)}(U)|^2] \leq \epsilon.
	\end{equation}
	The runtime of this algorithm is $\mathrm{poly}(n,d,1/\epsilon)$.
\end{theorem}
\setcounter{theorem}{1}

As a simple corollary, via Markov's inequality, we can transform this mean-squared-error statement into an additive error guarantee with high probability.

\begin{corollary}[Classical simulation with high probability]
    \label{coro:sim-whp}
    Let $0 < \epsilon,\ \delta < 1$.
    Given a quantum circuit $U \sim \mathcal{U}_{n,d}$, then \Cref{alg:sim} with truncation parameter $k = \mathcal{O}(\log(d/(\epsilon^2\delta)))$ outputs a real number $h(U)$ such that
    \begin{equation}
        \Pr_{U\sim \mathcal{U}_{n,d}}[|h(U) - f^{(1)}(U)| \leq \epsilon] \geq 1 -\delta.
    \end{equation}
    The runtime of this algorithm is $\mathrm{poly}(n,d,1/\epsilon,1/\delta)$.
\end{corollary}

\begin{proof}
    This follows from a simple application of Markov's inequality:
    \begin{align}
        \Pr_{U \sim \mathcal{U}_{n,d}}[|h(U) - f^{(1)}(U)| > \epsilon] &= \Pr_{U \sim \mathcal{U}_{n,d}}[|h(U) - f^{(1)}(U)|^2 > \epsilon^2]
        \leq \frac{\E_{U \sim \mathcal{U}_{n,d}}[|h(U) - f^{(1)}(U)|^2]}{\epsilon^2}
        \leq \delta.
    \end{align}
    The last inequality and the runtime follow by applying \Cref{thm:sim} with $\epsilon \leftarrow \epsilon^2 \delta$.
\end{proof}

We now describe the algorithm that achieves \Cref{thm:sim}.
Notice that we can rewrite the $\mathrm{OTOC}^{(1)}$ function via the swap trick:
\begin{equation}
	f^{(1)}(U) = \frac{1}{N}\tr((U^\dagger B UM)^2) = \frac{1}{N}\tr((U^\dagger BUM)^{\otimes 2} \flip_{\mathrm{all}}),
\end{equation}
where $\flip_{\mathrm{all}} = \bigotimes_{i=1}^n \flip_i$.
At a high level, our algorithm operates similarly to Pauli propagation but for the two-copy observable $B \otimes B$ and with different truncation rules.
Namely, the algorithm Heisenberg evolves the two-copy butterfly operator through each layer of the circuit.
At each layer, it decomposes the evolved observable in the basis $\mathcal{A}$ from \Cref{def:basis} and truncates to only include certain basis elements\footnote{For Pauli propagation, one decomposes in the Pauli basis and only keeps low-weight Paulis, but our two truncation rules are more subtle.}.
The algorithm is detailed in \Cref{alg:sim}.

\begin{algorithm}
    \caption{Window propagation algorithm}
    \label{alg:sim}
    \KwData{A 1D brickwork, depth-$d$, $n$-qubit quantum circuit $U = L_{d/2} \cdots L_1$; butterfly operator $B$; measurement operator $M$; error parameter $\epsilon > 0$.}
    \KwResult{An estimate $h(U)$ such that
    \begin{equation}
    	\E_{U \sim \mathcal{U}_{n,d}}[|h(U) - f^{(1)}(U)|^2] \leq \epsilon.
    \end{equation}}
    Set truncation parameters $k \triangleq \mathcal{O}(\log (d/\epsilon)), w \triangleq 3k$.\\
    For $A \in \mathcal{A}$, define $S_A = \{i \in [n] : A_i \notin \mathrm{span}\{\I, \flip\}\}$.\\
    Define
    \begin{equation}
        \mathcal{A}_{\mathrm{front}}
        \triangleq
        \left\{
          \I^{\otimes(\ell-1)} \otimes W_\ell \otimes \cdots \otimes W_r \otimes \flip^{\otimes(n-r)}
          \;\middle|\;
          \begin{gathered}
            0 \leq r \leq n,\quad 1 \leq \ell \leq r+1,\\
            r-\ell+1 \leq w,\\
            W_i \in \mathcal{A}_1 \quad (\ell \leq i \leq r)
          \end{gathered}
        \right\}
        \subseteq \mathcal{A}.
    \end{equation}
    \\
    Define the orthogonal projection $\mathcal{F}_{w} \triangleq \mathrm{Proj}_{\mathrm{span}\,\mathcal{A}_{\mathrm{front}}}$.\\
    Initialize $O_{d/2} = B \otimes B$.\\
    \For{$j = d/2,d/2 - 1\dots, 1$}{
    	Compute the Heisenberg evolved observable $L_j^{\dagger \otimes 2} O_j L_j^{\otimes 2}$ in the basis $\mathcal{A}$:
    	\begin{equation}
    		L_j^{\dagger \otimes 2} O_j L_j^{\otimes 2} = \sum_{A \in \mathcal{A}} c_A A.
    	\end{equation}\\
    	Apply the diameter truncation rule:
    	\begin{equation}
    		O_{j-1}' \triangleq \sum_{\substack{A \in \mathcal{A}\\\mathrm{diam}(S_A) \leq k}} c_A A.
    	\end{equation}\\
    	Apply the window truncation rule:
    	\begin{equation}
    		O_{j-1} \triangleq \mathcal{F}_{w}(O_{j-1}').
    	\end{equation}\\
    }
    Output $h(U) \triangleq h_k(U) \triangleq \frac{1}{N}\tr(O_0 (M\otimes M)\flip_{\mathrm{all}})$. 
\end{algorithm}

We note here that because $\mathcal{A}$ (\Cref{def:basis}) is not an orthonormal basis, then we cannot write the window truncation by simply restricting the terms included in the expansion of $O_{j-1}'$.
In contrast, for the diameter truncation, we know that restricting the terms performs an orthogonal projection (\Cref{lem:pi}).

The diameter truncation rule is somewhat similar to the low-weight truncation in Pauli propagation and the heuristic cached Monte Carlo algorithm of~\cite{google2025observation}.
However, crucially, our algorithm differs in that it does not truncate based on the \emph{size} of $S_A$ but its \emph{diameter}.
This ensures that we keep the basis elements $A$ such that the sites for which $A$ is non-invariant under two-copy rotations (i.e., $A_i \notin \mathrm{span}\{\I, \flip\}$) are constrained to a spatially local region.
In contrast, if we truncated based on the size of $S_A$, this would only enforce that, for kept basis elements $A$, there are few total sites for which $A$ is non-invariant under two-copy rotations, without any spatial closeness for such sites.

While we show that the algorithm with only the diameter truncation approximates the first-order OTOC function well, we highlight that it is not enough for an \emph{efficient} algorithm.
This is because the diameter truncation rule only constrains the non-invariant subspace.
In particular, the algorithm could potentially be keeping track of the coefficients of all elements of the exponentially large set $\{\I, \flip\}^{\otimes n}$ at some iteration, which is clearly not efficient.

Thus, we introduce the window truncation rule, which additionally constrains elements in $\mathrm{span}\{\I, \flip\}^{\otimes n}$.
Here, we project onto the span of elements of the particular form given in $\mathcal{A}_{\mathrm{front}}$, consisting of identities on the left, an arbitrary window of length at most $3k$, and flips on the right.
Because the dimension of this space is roughly at most $n \cdot 16^{3k}$, this is roughly the number of terms one must keep track of, which is polynomial because we choose $k = \mathcal{O}(\log (d/\epsilon))$.
Moreover, after applying a layer of the circuit, this number does not blow up significantly precisely because $\I$ and $\flip$ are invariant under two-copy rotations (\Cref{fact:i-flip-invariant}).
Thus, any gate that is applied solely to sites within the identity/flip tails has no effect.
Only gates that are applied to sites at the boundary of the identity/flip tails can enlarge the window, and even then can only enlarge it by at most one site at each end.
Thus, applying one layer $L_j$ enlarges the window size by at most four sites. 

\subsection{Relationship between window propagation and Pauli propagation}
\label{sec:pauli-prop}

For the sake of exposition, we describe how design principles similar to those used in window propagation presented in~\Cref{alg:sim}, can also give rise to a simple modification of the well-studied Pauli propagation algorithm~\cite{rall2019simulation,aharonov2023polynomial,angrisani2025classically,rudolph2023classical,schuster2025polynomial,gonzalez2025pauli,fontana2025classical,shao2024simulating}.
In Pauli propagation, one application is to classically estimate \emph{time-ordered correlators} of the form $\ntr(U^\dagger B U M)$.
The algorithm does so by Heisenberg evolving $B$ through each layer of the circuit, and at each layer, it removes high-weight Paulis from the Pauli decomposition of the evolved observable.
In the following, we detail the key choices made to design \Cref{alg:sim} and apply them to the task of predicting expectation values of this form $\ntr(U^\dagger BUM)$.

The steps in designing \Cref{alg:sim} are roughly as follows:
\begin{enumerate}
	\item\label{step:1} Consider a two-copy observable on a single pair of qubits. Choose a basis with respect to which we decompose the observable at each layer of the circuit. In the case of \Cref{alg:sim}, because we Heisenberg evolve the \emph{two-copy} butterfly operator $B \otimes B$, then it was important that the basis distinguished elements that are invariant under two-copy rotations, i.e., $\I$ and $\flip$.
	\item\label{step:2} Extend the basis to $n$ pairs of qubits. In the case of \Cref{alg:sim}, this resulted in the basis \Cref{def:basis}.
	\item\label{step:3} Define the truncation rules. In the case of \Cref{alg:sim}, these were as follows:
	\begin{enumerate}
		\item\label{step:3a} The diameter truncation rule restricts the size of the non-invariant subspace $\mathrm{span}\{\I, \flip\}^\perp$.
		\item\label{step:3b} The window truncation rule restricts elements of the invariant subspace $\mathrm{span}\{\I,\flip\}$ to be of a specific form, ensuring computational efficiency.
	\end{enumerate}
\end{enumerate}

Let us now follow these steps to define an algorithm for Heisenberg evolving a \emph{single-copy} observable $B$, as in the setting for Pauli propagation.
In Step~\ref{step:1}, we consider an observable on one qubit.
The only restriction on our basis is that it contains $I$, which is the only element invariant under single-copy rotations.
Then, a basis of the $2\times 2$ operators can be written as $\{I,A_1,A_2,A_3\}$, where $A_1,A_2,A_3$ is an orthonormal basis of $\mathrm{span}\{I\}^\perp$.
Suggestively, a typical choice of basis is $A_1 = X, A_2 = Y, A_3 = Z$.

In Step~\ref{step:2}, we can easily extend this to a basis on $n$ qubits by taking the tensor product: $\pauli_n = \{I, X,Y,Z\}^{\otimes n}$.
In Step~\ref{step:3}, consider $U = L_{d/2} \cdots L_1$.
Then, Heisenberg evolving the observable $O_j$ at step $j$ and writing it in this basis gives
\begin{equation}
	L_j^\dagger O_j L_j = \sum_{P \in \pauli_n} c_P P.
\end{equation}
In this case, the diameter truncation gives
\begin{equation}
	O_{j-1}' = \sum_{\substack{P \in \pauli_n\\\mathrm{diam}(S_P) \leq k}} c_P P,
\end{equation}
where $S_P = \{i \in [n] : P_i \neq I\}$.
This restricts the size of the non-invariant subspace $\mathrm{span}\{I\}^{\perp}$, as in the case of \Cref{alg:sim}.
At this point, notice the similarity to Pauli propagation: the only difference is that Pauli propagation typically uses the truncation rule $|S_P| \leq k$ rather than $\mathrm{diam}(S_P) \leq k$.

Moreover, Pauli propagation does not include a second truncation as in Step~\ref{step:3b}.
Recall that the main purpose of Step~\ref{step:3b} is to improve the computational efficiency.
However, for single-copy operators, the diameter truncation alone already reduces the number of coefficients to keep track of to $\mathcal{O}(n^k)$ in each layer.
Standard analyses of Pauli propagation show that $k = \log(1/\epsilon)$ is sufficient for a low mean-squared error, so this is already efficient enough.
In contrast, in the two-copy setting, after the diameter truncation, one may still have to keep track of $2^n$ terms because the algorithm retains everything in $\mathrm{span}\{\I,\flip\}$, necessitating the window truncation.

\section{Proof of {\Cref{thm:sim-main}}}
\label{sec:proof}

\subsection{Overview of the proof}
\label{sec:overview}
Recall that our goal is to prove that
\begin{equation}
	\E_{U \sim \mathcal{U}_{n,d}}[|h(U) - f^{(1)}(U)|^2] \leq \epsilon
\end{equation}
for $h(U)$ defined in \Cref{alg:sim}.
The key intermediate step is to consider the estimate obtained when only the diameter truncation rule is applied in \Cref{alg:sim}.
Call the resulting estimate $g_k(U)$.
Explicitly, we define this in \Cref{alg:gk}.
We emphasize that this is not the algorithm that we execute.
It is only used in the proof.

\begin{algorithm}
    \caption{Diameter-truncation-only algorithm}
    \label{alg:gk}
    \KwData{A 1D brickwork, depth-$d$, $n$-qubit quantum circuit $U = L_{d/2} \cdots L_1$; butterfly operator $B$; measurement operator $M$; truncation parameter $k \geq 1$; error parameter $\epsilon > 0$.}
    \KwResult{An intermediate estimate $g_k(U)$}
    Initialize $\widehat{O}_{d/2} = B \otimes B$.\\
    \For{$j = d/2,d/2 - 1\dots, 1$}{
    	Compute the Heisenberg evolved observable $L_j^{\dagger \otimes 2} \widehat{O}_j L_j^{\otimes 2}$ in the basis $\mathcal{A}$:
    	\begin{equation}
    		L_j^{\dagger \otimes 2} \widehat{O}_j L_j^{\otimes 2} = \sum_{A \in \mathcal{A}} c_A A.
    	\end{equation}\\
    	Let $S_A = \{i \in [n] : A_i \notin \mathrm{span}\{\I, \flip\}\}$.\\
    	Apply the diameter truncation rule:
    	\begin{equation}
    		\widehat{O}_{j-1} \triangleq \sum_{\substack{A \in \mathcal{A}\\\mathrm{diam}(S_A) \leq k}} c_A A.
    	\end{equation}
    }
    Output $g_k(U) \triangleq \frac{1}{N}\tr(\widehat{O}_0 (M\otimes M)\flip_{\mathrm{all}})$.
\end{algorithm}

The proof of \Cref{thm:sim} then consists of two parts.
First, we bound the error between this intermediate estimate $g_k(U)$ and the true $\mathrm{OTOC}^{(1)}$ value $f^{(1)}(U)$ in \Cref{sec:k-trunc-proof}, giving
\begin{equation}
    \label{eq:k-trunc-overview}
	\E_{U \sim \mathcal{U}_{n,d}}[|g_k(U) - f^{(1)}(U)|^2] \leq \mathcal{O}(d \alpha^k),
\end{equation}
for some constant $\alpha < 1$.
Second, we bound the error between the intermediate estimate $g_k(U)$ and the actual output of our algorithm $h(U) = h_k(U)$ in \Cref{sec:w-trunc-proof}, resulting in
\begin{equation}
    \label{eq:w-trunc-overview}
	\E_{U \sim \mathcal{U}_{n,d}}[|g_k(U) - h_k(U)|^2] \leq \mathcal{O}(d 2^{-k}).
\end{equation}
We choose the truncation parameter $k = \mathcal{O}(\log(d/\epsilon))$ so that both of these errors can be bounded by $\epsilon/4$.
We can then combine these two pieces via Cauchy-Schwarz:
\begin{equation}
	\E_{U \sim \mathcal{U}_{n,d}}[|f^{(1)}(U) - h(U)|^2] \leq 2 \E_{U \sim \mathcal{U}_{n,d}}[|f^{(1)}(U) - g_k(U)|^2] + 2 \E_{U \sim \mathcal{U}_{n,d}}[|g_k(U) - h_k(U)|^2] \leq \epsilon,
\end{equation}
as required.
We also analyze the runtime of our algorithm in \Cref{sec:runtime}.

In both parts of the proof, a key observation is that we can draw out layers of single-qubit independent Haar-random gates on all qubits via invariance of the Haar measure.
Namely, for every layer $j \in [d/2]$ and every site $i \in [n]$, consider independent single-qubit Haar-random gates $V_{j,i} \sim \mathrm{Haar}(U(2))$.
Define $V_j \triangleq \bigotimes_{i=1}^n V_{j,i}$.
Then, we can define a unitary $\tilde{U} \triangleq \tilde{L}_{d/2} \cdots \tilde{L}_1$, where $\tilde{L}_j \triangleq L_j V_j$.
Here, in each layer $L_j$, every qubit is acted upon by at least one Haar-random two-qubit gate.
Thus, each introduced single-qubit Haar-random gate is adjacent to a Haar-random two-qubit gate, so by invariance of the Haar measure, $\tilde{U}$ and $U$ have the same distribution.
Thus, we can write the errors as
\begin{equation}
    \E_{U \sim \mathcal{U}_{n,d}}[|f^{(1)}(U) - g_k(U)|^2] = \E_{\tilde{U} \sim \mathcal{U}_{n,d}}[|f^{(1)}(\tilde{U}) - g_k(\tilde{U})|^2]
\end{equation}
\begin{equation}
    \E_{U \sim \mathcal{U}_{n,d}}[|g_k(U) - h_k(U)|^2] = \E_{\tilde{U} \sim \mathcal{U}_{n,d}}[|g_k(\tilde{U}) - h_k(\tilde{U})|^2].
\end{equation}
We use this throughout the proofs.

We prove the bound between $g_k$ and $f^{(1)}$ in \Cref{eq:k-trunc-overview} in several steps.
First, in \Cref{sec:layer-by-layer-k}, we show that it suffices to bound the mean-squared error one layer at a time.
Focusing on one layer, we formalize the intuition given in \Cref{fig:lightcone-overlap} of splitting the circuit at this layer.
We forward evolve the measurement operator $M$ and backward evolve the butterfly operator until this layer and decompose these evolved operators in a new basis based on the position of its leftmost/rightmost non-identity Pauli, respectively.
For this reason, we refer to this as a \emph{spatial decomposition}.
Then, we bound the single-layer error in terms of expectation values of components of the spatial decomposition in \Cref{sec:layer-bound-T}.
In \Cref{sec:simplify-spatial}, we further simplify the form of these expectation values, which is an important step to complete the single-layer error bound.
With this, assuming a crucial proposition which states that two pairs of random operators have exponentially decaying overlap, as long as at least one operator in each pair is isotropically random (\Cref{prop:8}), we prove \Cref{eq:k-trunc-overview} in \Cref{sec:diameter-proof}.
Finally, we prove \Cref{prop:8} in \Cref{sec:proof-technical-prop}.

The proof of the error bound between $g_k$ and $h_k$ in \Cref{eq:w-trunc-overview} can also be organized into multiple steps.
In this case, one also hopes to show that the mean-squared error can be bounded one layer at a time.
To this end, we define a new error function which upper bounds the mean-squared error but is easier to bound.
We then show that it satisfies a nice monotonicity property in \Cref{sec:layer-by-layer-W}.
As a simple corollary, we obtain the desired layer-by-layer error bound in \Cref{sec:layer-by-layer-W}.
Finally, we bound the error for a single layer in \Cref{sec:single-layer-W}, following the intuition from \Cref{sec:proof-ideas}.

\subsection{Diameter truncation}
\label{sec:k-trunc-proof}

In this section, we bound the error between the intermediate estimate $g_k(U)$ and the true OTOC value $f^{(1)}(U)$.

\begin{theorem}[Diameter truncation error]
	\label{prop:k-trunc}
	There exists a constant $0 < \alpha < 1$ such that the following holds. Let $g_k(U)$ be the estimate obtained when only the diameter truncation rule is applied in \Cref{alg:sim}.
	Then,
    \begin{equation}
		\E_{U \sim \mathcal{U}_{n,d}}[|g_k(U) - f^{(1)}(U)|^2] \leq \frac{512d}{3(1-\sqrt{\alpha})^2}\alpha^{k+2}.
	\end{equation}
\end{theorem}

To prove this, we first find it useful to draw out layers $V_j$ of single-qubit independent Haar-random gates on all qubits via invariance of the Haar-measure as discussed in \Cref{sec:overview}.
Call the new circuit $\tilde{U}$.
\begin{equation}
	\E_{U \sim \mathcal{U}_{n,d}} [|f^{(1)}(U) - g_k(U)|^2] = \E_{\tilde{U} \sim \mathcal{U}_{n,d}}[|f^{(1)}(\tilde{U}) - g_k(\tilde{U})|^2].
\end{equation}
Thus, it suffices to prove the error bound for $\tilde{U}$.

\subsubsection{Layer-by-layer error}
\label{sec:layer-by-layer-k}

In this section, we show that it suffices to bound the mean-squared error one layer at a time.
In the next lemma, we show that one can write $g_k$ as a truncated ANOVA expansion (see, e.g., Section 8.3 of~\cite{o2021analysis}) with respect to the single-qubit gates, which implies this claim.

\begin{lemma}[Layer-by-layer error for diameter truncation]
    \label{lem:layer-by-layer-k}
	Let $\Pi_j$ be an orthogonal projector defined as in \Cref{eq:pi} for the layer $V_j$.
	Then,
	\begin{equation}
        \label{eq:gk-pi-form}
		g_k = \Big[\prod_{j=1}^{d/2} \Pi_j\Big] f^{(1)}.
	\end{equation}
	Moreover,
	\begin{equation}
		\label{eq:layer-by-layer-k}
		\E_{\tilde{U} \sim \mathcal{U}_{n,d}}[|f^{(1)}(\tilde{U}) - g_k(\tilde{U})|^2] \leq \sum_{j=1}^{d/2} \E_{\tilde{U}\sim \mathcal{U}_{n,d}}[|(\mathrm{id} - \Pi_j)f^{(1)}(\tilde{U})|^2].
	\end{equation}
\end{lemma}

\begin{proof}
	Consider the algorithm at the $j^\mathrm{th}$-iteration with $\tilde{U}(G,V)$.
    \Cref{lem:pi} applied to the layer $V_j$ and observable $L_j^{\dagger \otimes 2} \widehat{O}_j L_j^{\otimes 2}$ already tells us that
    \begin{equation}
        \Pi_j(\tilde{L}_j^{\dagger \otimes 2} \widehat{O}_j \tilde{L}_j^{\otimes 2}) = \widehat{O}_{j-1}.
    \end{equation}
	Applying these projections for each iteration of the algorithm completes the proof of \Cref{eq:gk-pi-form}.
    
	Finally, to show \Cref{eq:layer-by-layer-k}, we note that
	\begin{equation}
		(\mathrm{id} - \Pi_j)\prod_{p=1}^{j-1} \Pi_p = \prod_{p=1}^{j-1} \Pi_p - \prod_{p=1}^j \Pi_p.
	\end{equation}
	Thus, by a telescoping sum, then
	\begin{equation}
		\label{eq:telescope}
		\sum_{j=1}^{d/2} (\mathrm{id} - \Pi_j) \prod_{p=1}^{j-1} \Pi_p = \mathrm{id} - \prod_{j=1}^{d/2} \Pi_j.
	\end{equation}
	Note that distinct elements of the sum on the left-hand side are orthogonal.
	Thus, we have
	\begin{align}
		\E_{\tilde{U}\sim \mathcal{U}_{n,d}}[|f^{(1)}(\tilde{U}) - g_k(\tilde{U})|^2] &= \E_{\tilde{U} \sim \mathcal{U}_{n,d}}\Big[\Big|\Big(\mathrm{id} - \prod_{j=1}^{d/2} \Pi_j\Big)f^{(1)}(\tilde{U})\Big|^2\Big]\\
		&= \E_{\tilde{U}\sim \mathcal{U}_{n,d}}\left[\left|\left(\sum_{j=1}^{d/2}(\mathrm{id} - \Pi_j)\prod_{p=1}^{j-1} \Pi_p\right)f^{(1)}(\tilde{U})\right|^2\right]\\
		&= \sum_{j=1}^{d/2} \E_{\tilde{U} \sim \mathcal{U}_{n,d}}\left[\left|\left((\mathrm{id} - \Pi_j)\prod_{p=1}^{j-1}\Pi_p\right)f^{(1)}(\tilde{U})\right|^2\right]\\
		&\leq \sum_{j=1}^{d/2} \E_{\tilde{U} \sim \mathcal{U}_{n,d}}\left[\left|(\mathrm{id} - \Pi_j)f^{(1)}(\tilde{U})\right|^2\right].
	\end{align}
	In the first line, we use that $g_k = \prod_{j=1}^{d/2} \Pi_j f$.
	In the second line, we use \Cref{eq:telescope}.
	In the third line, we use orthogonality.
	In the last line, we use that orthogonal projectors are contractive.
\end{proof}

This tell us that the projector $\Pi_j$ is the same as applying the diameter truncation rule in \Cref{alg:sim} at the $j^\mathrm{th}$-layer of the circuit.
Thus, by \Cref{eq:layer-by-layer-k}, it suffices to bound the error between the true $\mathrm{OTOC}^{(1)}$ function $f^{(1)}$ and the diameter truncation rule applied \emph{only} at the $j^\mathrm{th}$-layer $\Pi_j f^{(1)}$.
In other words, we can consider the error accumulated one layer at a time.
In particular, we will show that
\begin{equation}
	\E_{\tilde{U}\sim \mathcal{U}_{n,d}}\left[\left|(\mathrm{id} - \Pi_j)f^{(1)}(\tilde{U})\right|^2\right] \leq \frac{1024}{3(1-\sqrt{\alpha})^2}\alpha^{k+2}.
\end{equation}

\subsubsection{Bound on single-layer error in terms of spatial components}
\label{sec:layer-bound-T}

From here on, we fix a layer $j \in [d/2]$.
In this section, we formalize the intuition in \Cref{sec:proof-ideas} about splitting the evolution $\tilde{U}^\dagger B \tilde{U}M$ at the $j^\mathrm{th}$-layer of single-qubit gates.
In particular, we forward evolve the measurement operator $M$ until the $j^\mathrm{th}$-layer and backward evolve the butterfly operator until this layer (see \Cref{fig:lightcone-overlap}).
Explicitly, we can rewrite $\tilde{U}$ as
\begin{equation}
	\tilde{U} = \tilde{L}_{d/2} \cdots \tilde{L}_1 = (\tilde{L}_{d/2} \cdots \tilde{L}_{j+1} L_j) V_j (\tilde{L}_{j-1} \cdots \tilde{L}_1).
\end{equation}
Then, we can define
\begin{align}
    \label{eq:bj-notation}
	B_j &\triangleq (L_j^\dagger \tilde{L}_{j+1}^\dagger \cdots \tilde{L}_{d/2}^\dagger) B (\tilde{L}_{d/2} \cdots \tilde{L}_{j+1}L_j)\\
	M_j &\triangleq (\tilde{L}_{j-1} \cdots \tilde{L}_1) M ( \tilde{L}_1^\dagger \cdots \tilde{L}_{j-1}^\dagger).
\end{align}
By cyclicity of the trace, we can write
\begin{equation}
	\label{eq:f-cut}
	f^{(1)}(\tilde{U}) = \ntr((\tilde{U}^\dagger B\tilde{U}M)^2) = \ntr((V_j^\dagger B_j V_j M_j)^2) = \ntr(V_j^{\dagger \otimes 2} B_j^{\otimes 2} V_j^{\otimes 2} M_j^{\otimes 2} \flip_{\mathrm{all}}).
\end{equation}
The idea behind bounding the error between $f^{(1)}$ and $\Pi_j f^{(1)}$ is to consider the sites where the backwards lightcone of $B$ intersects the forward lightcone of $M$ (see \Cref{fig:lightcone-overlap}).
The length of the overlapping region corresponds to the diameter of $S_A$.
We then show that long overlapping regions do not contribute much to the $\mathrm{OTOC}^{(1)}$ value.

To formalize this idea, we decompose $B_j^{\otimes 2}$ and $M_j^{\otimes 2}$ into components that only act on qubits in a specific region.
Writing $B_j = \sum_{P \in \mathcal{P}_n} b_{j,P} P$, $M_j = \sum_{P \in \mathcal{P}_n} m_{j,P} P$, we define
\begin{equation}
	B_j^{\geq \ell} \triangleq \sum_{\substack{P \in \pauli_n\\P_1 = \cdots = P_{\ell-1} = I}} b_{j,P} P, \quad\quad M_j^{\leq r} \triangleq \sum_{\substack{P \in \pauli_n\\P_{r+1} = \cdots = P_n = I}} m_{j,P} P
\end{equation}
for $\ell, r \in [n]$.
Here, $B_j^{\geq \ell}$ only acts non-trivially on qubits in $[\ell, n]$.
Similarly, $M_j^{\leq r}$ only acts non-trivially on qubits in $[1,r]$.
Then, $B_j^{=\ell} \triangleq B_j^{\geq \ell} - B_j^{\geq \ell+1}$ and $M_j^{=r} \triangleq M_j^{\leq r} - M_j^{\leq r-1}$.
Here, we define $B_j^{\geq n+1} = 0$ and $M_j^{\leq 0} = 0$.
$B_j^{=\ell}$ collects the terms whose first (leftmost) non-trivial Pauli acts on site $\ell$, and $M_j^{=r}$ collects the terms whose last (rightmost) non-trivial Pauli acts on site $r$.
The following lemma shows that we can decompose $B_j^{\otimes 2}$ and $M_j^{\otimes 2}$ in terms of these components.

\begin{lemma}[Spatial decomposition]
	For the definitions given above, then for $j \in [d/2]$,
	\begin{align}
		B_j^{\otimes 2} &= \sum_{\ell =1}^n (B_j^{=\ell} \otimes B_j^{\geq \ell} + B_j^{\geq \ell + 1} \otimes B_j^{=\ell})\\
		M_j^{\otimes 2} &= \sum_{r=1}^n (M_j^{=r} \otimes M_j^{\leq r} + M_j^{\leq r-1} \otimes M_j^{=r}).
	\end{align}
\end{lemma}

\begin{proof}
	First, note that $B_j, M_j$ are traceless because they are evolutions of traceless Paulis.
	Then, we can also notice that $B_j^{\geq 1} = B_j$, $B_j^{\geq n + 1} = 0$, $M_j^{\leq n} = M_j$, and $M_j^{\leq 0} = 0$ by definition.
	Now, to prove the expression for $B_j^{\otimes 2}$, we expand the term in the sum for fixed $\ell \in [n]$:
	\begin{align}
		B_j^{=\ell} \otimes B_j^{\geq \ell} + B_j^{\geq \ell + 1} \otimes B_j^{=\ell} &= (B_j^{\geq \ell} - B_j^{\geq \ell+1}) \otimes B_j^{\geq \ell} + B_j^{\geq \ell+1} \otimes (B_j^{\geq \ell} - B_j^{\geq \ell+1})\\
		&= (B_j^{\geq \ell})^{\otimes 2} - B_j^{\geq \ell+1} \otimes B_j^{\geq \ell} + B_j^{\geq \ell+1} \otimes B_j^{\geq \ell} - (B_j^{\geq \ell+1})^{\otimes 2}\\
		&= (B_j^{\geq \ell})^{\otimes 2} - (B_j^{\geq \ell+1})^{\otimes 2}.
	\end{align}
	In the first line, we use $B_j^{=\ell} = B_j^{\geq \ell} - B_j^{\geq \ell +1}$ by definition.
	Summing over $\ell$, we have a telescoping sum
	\begin{equation}
		\sum_{\ell=1}^n (B_j^{=\ell} \otimes B_j^{\geq \ell} + B_j^{\geq \ell+1} \otimes B_j^{=\ell}) = \sum_{\ell=1}^n (B_j^{\geq \ell})^{\otimes 2} - (B_j^{\geq \ell+1})^{\otimes 2} = (B_j^{\geq 1})^{\otimes 2} - (B_j^{\geq n+1})^{\otimes 2} = B_j^{\otimes 2},
	\end{equation}
	where in the last equality, we use that $B_j^{\geq 1} = B_j$ and $B_j^{\geq n+1} = 0$.

	The expression for $M_j^{\otimes 2}$ follows similarly.
	For fixed $r \in [n]$, then
	\begin{align}
		M_j^{=r} \otimes M_j^{\leq r} + M_j^{\leq r-1} \otimes M_j^{=r} &= (M_j^{\leq r} - M_j^{\leq r-1}) \otimes M_j^{\leq r} + M_j^{\leq r -1}\otimes (M_j^{\leq r} - M_j^{\leq r-1})\\
		&= (M_j^{\leq r})^{\otimes 2} - M_j^{\leq r -1}\otimes M_j^{\leq r} + M_j^{\leq r-1} \otimes M_j^{\leq r} - (M_j^{\leq r-1})^{\otimes 2}\\
		&= (M_j^{\leq r})^{\otimes 2} - (M_j^{\leq r-1})^{\otimes 2}.
	\end{align}	
	In the first line, we use that $M_j^{= r} = M_j^{\leq r} - M_j^{\leq r-1}$ by definition.
	Again, this is a telescoping sum when summing over $r$:
	\begin{equation}
		\sum_{r=1}^n (M_j^{=r} \otimes M_j^{\leq r} + M_j^{\leq r-1} \otimes M_j^{=r}) = \sum_{r=1}^n(M_j^{\leq r})^{\otimes 2} - (M_j^{\leq r-1})^{\otimes 2} = (M_j^{\leq n})^{\otimes 2} - (M_j^{\leq 0})^{\otimes 2} = M_j^{\otimes 2}.
	\end{equation}
\end{proof}

Plugging this decomposition into \Cref{eq:f-cut}, we have
\begin{equation}
	f^{(1)} = \sum_{\ell,r=1}^n \sum_{a,b=0}^1 T_j^{\ell, r}(a,b),
\end{equation}
where
\begin{equation}
	\label{eq:tlr}
\begin{split}
	T_j^{\ell,r}(0,0) &\triangleq \ntr(V_j^\dagger B_j^{=\ell} V_j M_j^{=r} V_j^\dagger B_j^{\geq \ell} V_j M_j^{\leq r})\\
	T_j^{\ell,r}(0,1) &\triangleq \ntr(V_j^\dagger B_j^{=\ell} V_j M_j^{\leq r-1} V_j^\dagger B_j^{\geq \ell} V_j M_j^{= r})\\
	T_j^{\ell,r}(1,0) &\triangleq \ntr(V_j^\dagger B_j^{\geq \ell +1} V_j M_j^{=r} V_j^\dagger B_j^{= \ell} V_j M_j^{\leq r})\\
	T_j^{\ell,r}(1,1) &\triangleq \ntr(V_j^\dagger B_j^{\geq \ell+1} V_j M_j^{\leq r-1} V_j^\dagger B_j^{=\ell} V_j M_j^{= r}).
\end{split}
\end{equation}
Notably, in each term $T_j^{\ell,r}$, the action of the butterfly and measurement operators is restricted to a spatially local region, the qubits $[\ell, r]$.
We make this intuition formal in the following lemma.

\begin{lemma}[Single-layer error in terms of spatial decomposition components]
    \label{lem:layer-bound-T}
	Let $j \in [d/2]$.
	Write $f^{(1)} = \sum_{\ell,r=1}^n \sum_{a,b=0}^1 T_j^{\ell,r}(a,b)$ as above.
	Then, for all $a, b\in \{0,1\}$, $T_j^{\ell,r}(a,b)$ only depends on the single-qubit gates $V_{j,\ell},\dots, V_{j,r}$ for $\ell < r$.
    For completeness, note that terms with $\ell>r$ do not depend on $V_j$ at all (disjoint supports), so they are fixed by $\Pi_j$.
	Moreover, we can bound
	\begin{equation}
		\sqrt{\E_{\tilde{U} \sim \mathcal{U}_{n,d}}[|(\mathrm{id} - \Pi_j)f^{(1)}(\tilde{U})|^2]} \leq \sum_{\substack{\ell,r=1\\r - \ell > k}}^n \sum_{a,b=0}^1 \sqrt{\E_{\tilde{U} \sim \mathcal{U}_{n,d}}[|T_j^{\ell, r}(a,b)|^2]},
	\end{equation}
	where $T_j^{\ell,r}(a,b)$ depends on $\tilde{U}$ implicitly.
\end{lemma}

\begin{proof}
	First, we prove that each $T_j^{\ell,r}(a,b)$ only depends on the single qubit gates $V_{j,\ell},\dots, V_{j,r}$ for $\ell < r$.
	We prove this for $T_j^{\ell,r}(0,0)$.
	The other cases are essentially the same because the only properties we will use are that $B_j^{=\ell}, B_j^{\geq \ell}$ are identity on sites $[1,\ell-1]$ and $M_j^{=r}, M_j^{\leq r}$ are identity on sites $[r+1,n]$.
	These properties hold for the terms in other $T_j^{\ell,r}(a,b)$ as well.

	We focus on
	\begin{equation}
		T_j^{\ell,r}(0,0) = \ntr(V_j^\dagger B_j^{=\ell} V_j M_j^{=r} V_j^\dagger B_j^{\geq \ell} V_j M_j^{\leq r}).
	\end{equation}
	Here, write $V_j = V_{j,[1,\ell-1]} V_{j, [\ell, n]}$, where $V_{j,[1,\ell-1]} = V_{j,1}\otimes \cdots \otimes V_{j,\ell-1} \otimes I^{\otimes (n-\ell + 1)}$.
	Then, because $B_j^{=\ell}$ and $B_j^{\geq \ell}$ are identity on $[1,\ell-1]$, $V_{j,[1,\ell-1]}$ commutes with these terms.
	Thus, we get
	\begin{equation}
		(V_{j,[1,\ell-1]}^\dagger V_{j,[\ell,n]}^\dagger) B_j^{=\ell} (V_{j,[1,\ell-1]} V_{j,[\ell,n]}) = (V_{j,[1,\ell-1]}^\dagger V_{j,[1,\ell-1]}) (V_{j,[\ell,n]}^\dagger B_j^{=\ell} V_{j,[\ell,n]}) = V_{j,[\ell,n]}^\dagger B_j^{=\ell} V_{j,[\ell,n]}
	\end{equation}
	and similarly for $B_j^{\geq \ell}$.
	Hence,
	\begin{equation}
		T_j^{\ell,r}(0,0) = \ntr(V_{j,[\ell,n]}^\dagger B_j^{=\ell} V_{j,[\ell,n]} M_j^{=r} V_{j,[\ell,n]}^\dagger B_j^{\geq \ell} V_{j,[\ell,n]} M_j^{\leq r}).
	\end{equation}
	We can proceed with a similar argument for the $M_j$ terms to remove the dependence on sites $[r+1,n]$.
	Further decompose $V_{j,[\ell,n]} = V_{j,[\ell,r]} V_{j,[r+1,n]}$.
	Then, because $M_j^{=r}$ and $M_j^{\leq r}$ are identity on $[r+1,n]$, $V_{j,[r+1,n]}$ commutes with these terms.
	Thus, those rotations also cancel with each other, and we can write
	\begin{equation}
		T_j^{\ell,r}(0,0) = \ntr(V_{j,[\ell,r]}^\dagger B_j^{=\ell} V_{j,[\ell,r]} M_j^{=r} V_{j,[\ell,r]}^\dagger B_j^{\geq \ell} V_{j,[\ell,r]} M_j^{\leq r}).
	\end{equation}
	This completes the proof that $T_j^{\ell,r}$ only depends on the rotations $V_{j,\ell},\dots, V_{j,r}$.

	Now, we prove the inequality.
	Recall the definition of the projector $\Pi_j$ (see \Cref{eq:pi}, \Cref{lem:layer-by-layer-k}):
	\begin{equation}
		\Pi_j \triangleq \sum_{\substack{S \subseteq [n]\\\mathrm{diam}(S) \leq k}} \prod_{i\in S} (\mathrm{id} - \E_{V_{j,i}}) \prod_{i\notin S} \E_{V_{j,i}}.
	\end{equation}
	Here, $\E_{V_{j,i}}[T_j^{\ell,r}(a,b)] = T_j^{\ell,r}(a,b)$ for $i \notin [\ell, r]$ because $T_j^{\ell,r}$ does not even depend on these single-qubit gates.
	Thus, terms in the sum corresponding to $S \not\subseteq [\ell,r]$ are zero.
    Note that the case when $\ell = r$ has diameter zero.
    Also, note that when $\ell > r$, the butterfly and measurement components ($B_j^{=\ell}, B_j^{\geq \ell}$ and $M_j^{=r}, M_j^{\leq r}$) have disjoint supports.
    Thus, in a term $T_j^{\ell,r}$, all on-site rotations $V_{j,i}$ cancel so that the term does not depend on $V_{j,i}$.
    
	This means that, in this setting, the sum over $S$ in the definition of $\Pi_j$ must be over sets $S \subseteq [\ell, r]$ with $\ell \leq r$.
	Such sets have diameter at most $r - \ell$, so if we require that $r - \ell \leq k$, then no terms are truncated from this sum.
	In other words, when $r- \ell \leq k$, then $\Pi_j T_j^{\ell,r}(a,b) = T_j^{\ell,r}(a,b)$.
	This implies that
	\begin{equation}
		(\mathrm{id} - \Pi_j)f^{(1)} = (\mathrm{id} - \Pi_j)\sum_{\substack{\ell,r \in [n]\\r - \ell > k}} \sum_{a,b=0}^1 T_j^{\ell,r}(a,b).
	\end{equation}
	Then, we can bound this as follows:
	\begin{align}
		\sqrt{\E_{\tilde{U} \sim \mathcal{U}_{n,d}}[|(\mathrm{id} - \Pi_j)f^{(1)}(\tilde{U})|^2]} &= \sqrt{\E_{\tilde{U} \sim \mathcal{U}_{n,d}}\Big[\Big|(\mathrm{id} - \Pi_j)\sum_{\substack{\ell,r \in [n]\\r - \ell > k}} \sum_{a,b=0}^1 T_j^{\ell,r}(a,b)\Big|^2\Big]}\\
		&\leq \sqrt{\E_{\tilde{U} \sim \mathcal{U}_{n,d}}\Big[\Big|\sum_{\substack{\ell,r \in [n]\\r - \ell > k}} \sum_{a,b=0}^1 T_j^{\ell,r}(a,b)\Big|^2\Big]}\\
		&\leq \sum_{\substack{\ell,r=1\\r - \ell > k}}^n \sum_{a,b=0}^1 \sqrt{\E_{\tilde{U} \sim \mathcal{U}_{n,d}}[|T_j^{\ell, r}(a,b)|^2]}.
	\end{align}
	In the second line, we use contractivity since $(\mathrm{id} -\Pi_j)$ is an orthogonal projector.
	In the last line, we use triangle inequality.
\end{proof}

This proof shows precisely the intuition conveyed above.
Namely, each term $T_j^{\ell,r}$ of the $\mathrm{OTOC}^{(1)}$ function only depends on the on-site gates where two components of the backward-evolved butterfly and the forward-evolved measurement operators overlap (on qubits $[\ell, r]$).
Moreover, the length of this region indeed corresponds to the diameter of $S_A$ in the algorithm description, which is why only terms with $r - \ell > k$ are counted in the error.

\subsubsection{Simplifying the spatial components}
\label{sec:simplify-spatial}
It remains to bound $\E[|T_j^{\ell,r}(a,b)|^2]$ for $\ell,r\in [n]$ satisfying $r - \ell > k$ and for all $a,b \in \{0,1\}$.
Towards this end, in this section, we simplify the expression for the terms $T_j^{\ell,r}(a,b)$ which will be crucial in proving such a bound.

Notice that one can write the expressions for $T_j^{\ell,r}(a,b)$ in a more unified way as
\begin{equation}
    T_j^{\ell, r}(a,b) = \begin{cases}
        \ntr(V_j^\dagger B_j^{=\ell} V_j M_j^{=r} V_j^\dagger B_j^{\geq \ell + a} V_j M_j^{\leq r - b}) & a + b\;\text{ even}\\\\
        \overline{\ntr(V_j^\dagger B_j^{=\ell} V_j M_j^{=r} V_j^\dagger B_j^{\geq \ell + a} V_j M_j^{\leq r - b})} & a + b\;\text{ odd}
    \end{cases}
\end{equation}
This is true because each of the terms $V_j^\dagger B_j^{=\ell} V_j$, $M_j^{=r}$, $V_j^\dagger B_j^{\geq \ell + a} V_j$, and $M_j^{\leq r- b}$ are Hermitian.
Thus, swapping the order of, e.g., $B_j^{=\ell}$ and $B_j^{\geq \ell + a}$, conjugates the trace.
Thus, we have that
\begin{equation}
    |T_j^{\ell,r}(a,b)| = |\ntr(V_j^\dagger B_j^{=\ell} V_j M_j^{=r} V_j^\dagger B_j^{\geq \ell + a} V_j M_j^{\leq r - b})|.
\end{equation}
In what follows, the only properties we use about the specific terms are those that hold for all $a,b$, i.e., that $B_j^{\geq \ell + a}$ is identity on sites $[1,\ell-1]$.
Hence, for concreteness going forward, we prove the bound for the case of $a=b=0$.
The bounds on the other terms are the same.

First, we can simplify the form of these terms further.
Since we are distinguishing the overlap region $[\ell, r]$, we can rewrite $B_j^{=\ell}$ as follows:
\begin{equation}
	B_j^{=\ell} = \sum_{P \in \pauli_{[r+1,n]}} I^{\otimes (\ell -1)} \otimes B_P \otimes P,
\end{equation} 
where $B_P$ is an operator supported on $[\ell, r]$.
Similarly, we have
\begin{align}
	B_j^{\geq \ell} &= \sum_{P \in \pauli_{[r+1,n]}} I^{\otimes (\ell-1)} \otimes B_P' \otimes P\\
	M_j^{= r} &= \sum_{Q \in \pauli_{[1,\ell-1]}} Q \otimes M_Q \otimes I^{\otimes (n-r)}\\
	M_j^{\leq r} &= \sum_{Q \in \pauli_{[1,\ell-1]}} Q \otimes M_Q' \otimes I^{\otimes (n-r)},
\end{align}
where $B_P', M_Q, M_Q'$ are supported on $[\ell, r]$.
The following lemma shows that we can trace out all sites outside of $[\ell, r]$.

\begin{lemma}[Restricting to the overlap region]
	\label{lem:T-terms}
	Let $j \in [d/2]$ and $\ell,r \in [n]$ such that $r - \ell > k$.
	Then,
	\begin{equation}
		T_j^{\ell,r}(0,0) = \frac{1}{2^{r - \ell + 1}}\sum_{\substack{P \in \pauli_{[r+1,n]}\\Q \in \pauli_{[1,\ell-1]}}} \tr(V_{j,[\ell,r]}^\dagger B_P V_{j,[\ell,r]} M_Q V_{j,[\ell,r]}^\dagger B_P' V_{j,[\ell,r]} M_Q'),
	\end{equation}
	where $V_{j,[\ell,r]}$ denotes $V_{j,\ell} \otimes \cdots \otimes V_{j,r}$.
\end{lemma}

\begin{proof}
	We know from \Cref{lem:layer-bound-T} that
	\begin{equation}
		\label{eq:T-depend-lr}
		T_j^{\ell,r}(0,0) = \frac{1}{N}\tr(V_{j,[\ell,r]}^\dagger B_j^{=\ell} V_{j,[\ell,r]} M_j^{=r} V_{j,[\ell,r]}^\dagger B_j^{\geq \ell} V_{j,[\ell,r]} M_j^{\leq r}).
	\end{equation}
	We abuse notation here, using $V_{j,[\ell,r]}$ to also denote the version padded with identities.
	The normalization factor makes it clear when we do/do not pad with identities.
	Using the decomposition of $B_j^{=\ell}$ and $B_j^{\geq \ell}$ above, we can write
	\begin{equation}
		V_{j,[\ell,r]}^\dagger B_j^{=\ell} V_{j,[\ell,r]} = \sum_{P \in \pauli_{[r+1,n]}} I^{\otimes (\ell-1)} \otimes V_{j,[\ell,r]}^\dagger B_P V_{j,[\ell,r]} \otimes P
	\end{equation}
	\begin{equation}
		V_{j,[\ell,r]}^\dagger  B_j^{\geq \ell} V_{j,[\ell,r]} = \sum_{P \in \pauli_{[r+1,n]}} I^{\otimes (\ell-1)} \otimes V_{j,[\ell,r]}^\dagger B_P' V_{j,[\ell,r]} \otimes P.
	\end{equation}
	Then, plugging this into \Cref{eq:T-depend-lr} along with the decompositions of $M_j^{=r}, M_j^{\leq r}$ and using that $\tr(A\otimes B) = \tr(A) \tr(B)$, we have
	\begin{align}
		T_j^{\ell,r}(0,0) &= \frac{1}{N}\sum_{\substack{P,P' \in \pauli_{[r+1,n]}\\Q,Q' \in \pauli_{[1,\ell-1]}}}\tr(QQ') \tr(V_{j,[\ell,r]}^\dagger B_P V_{j,[\ell,r]} M_Q V_{j,[\ell,r]}^\dagger B_{P'}' V_{j,[\ell,r]} M_{Q'}') \tr(PP')\\
		&= \frac{1}{2^{r-\ell+1}}\sum_{\substack{P \in \pauli_{[r+1,n]}\\Q \in \pauli_{[1,\ell-1]}}}\tr(V_{j,[\ell,r]}^\dagger B_P V_{j,[\ell,r]} M_Q V_{j,[\ell,r]}^\dagger B_{P}' V_{j,[\ell,r]} M_{Q}'),
	\end{align}
	as claimed.
	In the second line, we use that $\tr(PP') = 2^{n-r} \indicator\{P = P'\}$ and $\tr(QQ') = 2^{\ell-1} \indicator\{Q = Q'\}$.
\end{proof}

\subsubsection{Proof of diameter truncation correctness}
\label{sec:diameter-proof}

With this simplified form of $T_j^{\ell,r}$, we can bound $\E[|T_j^{\ell,r}(a,b)|^2]$, which is the focus of this section.
The sketch of the idea is as follows.
Consider adding more single-qubit Haar-random gates via Haar invariance: let $R, R' \sim \mathrm{Haar}(U(2))$ be independent Haar-random gates for write $\tilde{V} = R' R^\dagger$.
$\tilde{V}$ has the same distribution as $V$, our layer of single-qubit Haar-random gates.
Then, schematically, if we had quantities such as $\tr(V^\dagger B_P V M_Q)$, we could rewrite them as
\begin{equation}
	\tr(R^{\prime \dagger} B_P R' R^\dagger M_Q R).
\end{equation}
Here, this is simply the overlap of two independently random operators $R^{\prime \dagger} B_P R'$ and $R^\dagger M_Q R$.
By independence, we expect the overlap to decay with the support of the operators.
However, the terms in the sum in \Cref{lem:T-terms} are more complicated: we can write them schematically as
\begin{equation}
	\tr(R^{\prime \dagger} B_P R' R^\dagger M_Q R R^{\prime \dagger} B_P' R' R^\dagger M_Q' R).
\end{equation}
The point is that $(R^{\prime \dagger} B_P R', R^{\prime \dagger} B_P' R')$ and $(R^\dagger M_Q R, R^\dagger M_Q' R)$ are independent random pairs of operators.
However, $R^{\prime \dagger} B_P R'$ and $R^{\prime \dagger} B_P' R'$ (and similarly for $M_Q$) could be correlated with each other.
Nevertheless, the following proposition says that as long as one of them is sufficiently isotropically random, then the overlap is still small.

\begin{prop}[Isotropically random operators imply exponentially decaying overlap]
	\label{prop:8}
    There exists a universal constant $0 < \alpha < 1$ such that the following holds.
	Let $m \geq 1$.
	Let $(C,D)$ and $(E,F)$ be independent random pairs of $m$-qubit operators satisfying $\norm{C}_F^2 = \norm{D}_F^2 = \norm{E}_F^2 = \norm{F}_F^2 = 2^m$.
	Suppose that, for any $R = R_1 \otimes \cdots R_m$, where $R_i \in U(2)$, the joint distribution of $(C,D)$ is the same as that of $(R CR^\dagger, RDR^\dagger)$, and similarly for $(E,F)$.
	Further suppose that for any fixed $m$-qubit operator $G$,
	\begin{equation}
		\label{eq:low-average}
		\E[|\ntr(C^\dagger G)|^2] \leq \frac{c}{8^m} \norm{G}_F^2, \quad \E[|\ntr(E^\dagger G)|^2] \leq \frac{c}{8^m} \norm{G}_F^2
	\end{equation}
	for some constant $c > 0$.
	Then,
	\begin{equation}
		\E[|\ntr(C^\dagger E D^\dagger F)|^2] \leq 16c\alpha^m.
	\end{equation}
\end{prop}

This proposition provides exactly the exponential decay we are aiming to prove.
We highlight that similar conditions to \Cref{eq:low-average} have appeared in the literature for designing classical algorithms for simulating noisy quantum circuits (see, e.g., Definition 1 in~\cite{schuster2025polynomial}).
There, the low-average condition plays a similar role, where combined with noise, it allows one to prove that the overlap of scrambled observables exponentially decays with their locality.
Instead, we combine this related condition with the randomness of our operators to prove the desired exponential decay.
We defer the proof of this proposition to \Cref{sec:proof-technical-prop} and show that it indeed implies \Cref{prop:k-trunc}.

First, we show that the conditions of \Cref{prop:8} are satisfied for our case, using the terms from \Cref{lem:T-terms} as a starting point.

\begin{lemma}[Conditions of {\Cref{prop:8}} are satisfied]
	\label{lem:conditions-prop-8}
	Let $j \in [d/2]$.
	For $\ell,r\in[n]$, define $p_{j,\ell} \triangleq \E[\norm{B_j^{=\ell}}_F^2]/2^n$ and $q_{j,r} \triangleq \E[\norm{M_j^{=r}}_F^2]/2^n$.
	Fix $\ell < r$ such that $p_{j,\ell} q_{j,r} > 0$.
	Let $m = r-\ell + 1$.
	Consider the operators $B_P,B_P', M_Q, M_Q'$ from \Cref{lem:T-terms}.
	Define the following weighted expectation over $(B_j, P)$ and $(M_j,Q)$ respectively by
	\begin{equation}
		{\E\limits_B}^*[\cdot] \triangleq \frac{1}{2^{m}p_{j,\ell}}\E_{B_j}\left[\sum_{P\in \pauli_{[r+1,n]}}\norm{B_P}_F^2 (\cdot)\right],
	\end{equation}
	\begin{equation}
		{\E\limits_M}^*[\cdot] \triangleq \frac{1}{2^{m}q_{j,r}}\E_{M_j}\left[\sum_{Q \in \pauli_{[1, \ell-1]}}\norm{M_Q}_F^2 (\cdot)\right].
	\end{equation}

	Let $R = R_{\ell} \otimes \cdots \otimes R_r,R' = R_\ell' \otimes \cdots \otimes R_r'$, where $R_i,R_i' \sim\mathrm{Haar}(U(2))$.
	Consider the notation established throughout this section.
	Define 
	\begin{equation}
		C \triangleq \frac{2^{m/2}}{\norm{B_P}_F} R^\dagger B_P R,\quad D \triangleq \frac{2^{m/2}}{\norm{B_P'}_F} R^\dagger B_P' R
	\end{equation}
	\begin{equation}
		E \triangleq \frac{2^{m/2}}{\norm{M_Q}_F} R^{\prime \dagger} M_Q R',\quad F \triangleq \frac{2^{m/2}}{\|M_Q'\|_F} R^{\prime \dagger} M_Q' R'.
	\end{equation}
	Note that zero values of $B_P$ or $M_Q$ have zero weighted probability.
	Then, $(C,D)$ and $(E,F)$ satisfy the conditions of \Cref{prop:8} with $m = r-\ell + 1$ and $c = 4/3$.
	In particular, $\norm{C}_F^2 = \norm{D}_F^2 = \norm{E}_F^2 = \norm{F}_F^2 = 2^m$, and the joint distribution of $(C,D)$ is invariant under conjugation by a layer of single-qubit gates (and similarly for $(E,F)$).
	Moreover, for any fixed $m$-qubit operator $G$,
	\begin{equation}
		\label{eq:low-average-satisfied}
		{\E\limits_B}^*\E_R[|\ntr(C^\dagger G)|^2] \leq \frac{4}{3 \cdot 8^{m}} \norm{G}_F^2, \quad {\E\limits_M}^*\E_{R'}[|\ntr(E^\dagger G)|^2] \leq \frac{4}{3 \cdot 8^{m}} \norm{G}_F^2.
	\end{equation}
\end{lemma}

While the statement of this lemma may look unwieldy at first, the reason for defining these weighted expectations becomes clear in the proof.

\begin{proof}
	First, we justify that the distributions under which the reweighted expectations are taken, are valid probability distributions.
	For $P,P' \in \pauli_{[r+1,n]}$, then $\tr(PP') = 2^{n-r}\indicator\{P = P'\}$.
	Thus, we can write
	\begin{equation}
        \label{eq:b-norm}
		\norm{B_j^{=\ell}}_F^2 = 2^{\ell - 1}2^{n-r} \sum_{P \in \pauli_{[r+1,n]}}\norm{B_P}_F^2.
	\end{equation}
	Dividing by $2^n$ and taking the (unweighted) expectation, we have
	\begin{equation}
		\E_{B_j}\left[\sum_{P \in \pauli_{[r+1,n]}} \norm{B_P}_F^2\right] = 2^m p_{j,\ell},
	\end{equation}
	where we use the definition of $p_{j,\ell} = \E[\norm{B_j^{=\ell}}_F^2]/2^n$.
	This shows that ${\E\limits_B}^*[1] = 1$, as required.
	An analogous proof holds for the weighted expectation ${\E\limits_M}^*[\cdot]$.

	Now, we begin verifying the conditions of \Cref{prop:8}.
	$(C,D)$ and $(E,F)$ are clearly independent because $B_P, B_P'$ and $M_Q,M_Q'$ have independent sources of randomness and $R,R'$ are independent.
	Moreover, we can easily check the norm condition:
	\begin{equation}
		\norm{C}_F^2 = \frac{2^m}{\norm{B_P}_F^2}\norm{R^\dagger B_P R}_F^2 = 2^m,
	\end{equation}
	where the last equality follows by unitary invariance of the Frobenius norm.
	The same calculation holds for $D,E,$ and $F$.
	The invariant-under-conjugation condition also clearly holds by Haar invariance.

	Thus, it remains to show \Cref{eq:low-average-satisfied}.
	We show this for $C$, as the proof for $E$ is analogous.
	To demonstrate why we need the weighted expectations, we first attempt to prove \Cref{eq:low-average-satisfied} with the unweighted expectations.
	Recall that we wrote
	\begin{equation}
		B_j^{=\ell} = \sum_{\substack{K \in \pauli_n\\K_1 = \cdots = K_{\ell-1} = I, K_\ell \neq I}} b_{j,K} K = \sum_{P \in \pauli_{[r+1,n]}} I^{\otimes (\ell - 1)} \otimes B_P \otimes P.
	\end{equation}
	Thus, we can write
	\begin{equation}
		B_P = \sum_{\substack{T \in \pauli_{[\ell,r]}\\T_\ell \neq I}} b_{j,I^{\otimes (\ell -1)}\otimes T \otimes P} T.
	\end{equation}
	Then, for $T_\ell,T'_\ell\neq I$, applying \Cref{lem:pauli-second-moment-B} shows that
	\begin{equation}
		\label{eq:lem-3-applied}
		\E_{B_j}[b_{j,I^{\otimes (\ell-1)} \otimes T \otimes P}b_{j,I^{\otimes (\ell-1)} \otimes T' \otimes P}] = \frac{p_{j,\ell}}{3 \cdot 4^{n-\ell}}\indicator\{T = T'\}.
	\end{equation}
	Then, for a fixed $m$-qubit operator $G$, we can attempt to bound $2^{-2m}\E_{B_j}[|\trace(C^\dagger G)|^2]$.
	\begin{align}
		\frac{1}{2^{2m}}\E_{B_j}[|\trace(C^\dagger G)|^2] &= \frac{1}{2^{m}\norm{B_P}_F^2}\sum_{\substack{T,T' \in \pauli_{[\ell,r]}\\T_\ell,T_\ell' \neq I}}\E_{B_j}[b_{j,I^{\otimes (\ell-1)} \otimes T\otimes P} b_{j,I^{\otimes (\ell-1)} \otimes T' \otimes P} \overline{\trace(R^\dagger T^\dagger R G)}\trace(R^\dagger T^{\prime \dagger}R G)]\\
		&= \frac{1}{2^{m}\norm{B_P}_F^2} \frac{p_{j,\ell}}{3\cdot 4^{n-\ell}} \sum_{\substack{T \in \pauli_{[\ell,r]}\\T_\ell \neq I}} |\tr(R^\dagger T^\dagger R G)|^2\\
		&\leq \frac{1}{2^{m} \norm{B_P}_F^2} \frac{p_{j,\ell}}{3\cdot 4^{n-\ell}} \sum_{\substack{T \in \pauli_{[\ell,r]}}} |\tr(R^\dagger T^\dagger R G)|^2\\
		&= \frac{1}{\norm{B_P}_F^2} \frac{p_{j,\ell}}{3\cdot 4^{n-\ell}} \norm{RGR^\dagger}_F^2\\
		&= \frac{1}{\norm{B_P}_F^2} \frac{p_{j,\ell}}{3\cdot 4^{n-\ell}} \norm{G}_F^2.\label{eq:attempt-fail}
	\end{align}
	In the second line, we use \Cref{eq:lem-3-applied}.
	In the second to last line, we use Parseval's identity (\Cref{lem:parseval}).
	In the last line, we use the unitary invariance of the Frobenius norm.
	This is not quite the right bound we want: there is an extra factor of $p_{j,\ell}$ and the normalization is not right.
	We introduce the weighted expectations to fix these issues.
	Note that to fix the normalization, simply multiplying by a correction factor of $\norm{B_P}_F^2 4^{n-r-1}$ would not result in a valid probability distribution.
	This is why we sum over $\sum_{P \in \pauli_{[r+1,n]}} \norm{B_P}_F^2$ in the weighted expectation definition instead.

	Thus, following a similar calculation for the weighted expectation, we have
	\begin{align}
		\frac{1}{2^{2m}}{\E\limits_B}^*[|\trace(C^\dagger G)|^2] &= \frac{1}{2^{3m} p_{j,\ell}}\E_{B_j}\left[\sum_{P \in \pauli_{[r+1,n]}} \norm{B_P}_F^2 \frac{2^m}{\norm{B_P}_F^2} |\trace(R^\dagger B_P^\dagger R G)|^2\right]\\
		&= \frac{1}{2^{2m} p_{j,\ell}}\E_{B_j}\left[\sum_{P \in \pauli_{[r+1,n]}} |\trace(R^\dagger B_P^\dagger R G)|^2\right]\\
		&= \frac{1}{2^{2m}} \frac{1}{3 \cdot 4^{n-\ell}} \sum_{P\in \pauli_{[r+1,n]}}\sum_{\substack{T \in \pauli_{[\ell,r]}\\T_\ell \neq I}} |\trace(R^\dagger T^\dagger R G)|^2\\
		&\leq  \frac{1}{2^{2m}} \frac{4}{3\cdot 4^m} \sum_{T \in \pauli_{[\ell,r]}} |\trace(R^\dagger T^\dagger RG)|^2\\
		&= \frac{4}{3\cdot 8^m}\norm{RGR^\dagger}_F^2\\
		&= \frac{4}{3\cdot 8^m}\norm{G}_F^2.
	\end{align}
	In the first line, we use the definition of the weighted expectation.
	In the third line, we use \Cref{eq:lem-3-applied}.
	In the fifth line, we use Parseval's identity (\Cref{lem:parseval}).
	In the last line, we use unitary invariance of the Frobenius norm.
	This completes the proof.

	With this proof, our design choices are (hopefully) now clear.
	In the definition of $C$ (and $D,E,F$), we conjugate $B_P$ by Haar-random rotations to ensure the invariance condition.
	We also add appropriate normalization factors to ensure that $\norm{C}_F^2 = 2^m$.
	Then, we define these weighted expectations in order to ensure the conditions of \Cref{prop:8} to correct for the bound achieved by the unweighted expectations as in \Cref{eq:attempt-fail}.
\end{proof}

Now, we show how to prove \Cref{prop:k-trunc} by applying \Cref{prop:8}.

\begin{proof}[Proof of \Cref{prop:k-trunc}]
	Putting everything together, we showed in \Cref{lem:layer-by-layer-k} that
	\begin{equation}
		\label{eq:to-plug-layer}
		\E_{\tilde{U}\sim \mathcal{U}_{n,d}}[|f^{(1)}(\tilde{U})-g_k(\tilde{U})|^2] \leq \sum_{j=1}^{d/2} \E_{\tilde{U} \sim \mathcal{U}_{n,d}}[|(\mathrm{id} - \Pi_j)f^{(1)}(\tilde{U})|^2].
	\end{equation}
	For fixed $j \in [d/2]$, then we showed in \Cref{lem:layer-bound-T} that
	\begin{equation}
		\label{eq:to-plug-T}
		\sqrt{\E_{\tilde{U} \sim \mathcal{U}_{n,d}}[|(\mathrm{id} - \Pi_j)f^{(1)}(\tilde{U})|^2]} \leq \sum_{\substack{\ell,r=1\\r - \ell > k}}^n \sum_{a,b=0}^1 \sqrt{\E_{\tilde{U} \sim \mathcal{U}_{n,d}}[|T_j^{\ell, r}(a,b)|^2]},
	\end{equation}
	for terms $T_j^{\ell,r}(a,b)$ defined in \Cref{eq:tlr}.
	Now, we use \Cref{prop:8} to bound $\E[|T_j^{\ell,r}(a,b)|^2]$.
	As we discussed previously, it suffices to consider the case of $a = b = 0$.
	The same bound applies for the other terms.
	By \Cref{lem:T-terms}, we can write
	\begin{equation}
		T_j^{\ell,r}(0,0) = \frac{1}{2^{r - \ell + 1}}\sum_{\substack{P \in \pauli_{[r+1,n]}\\Q \in \pauli_{[1,\ell-1]}}} \tr(V_{j,[\ell,r]}^\dagger B_P V_{j,[\ell,r]} M_Q V_{j,[\ell,r]}^\dagger B_P' V_{j,[\ell,r]} M_Q'),
	\end{equation}
	where $V_{j,[\ell,r]} = V_{j,\ell} \otimes \cdots \otimes V_{j,r}$ for $V_{j,i} \sim \mathrm{Haar}(U(2))$.
	We need to massage this into the form that can be bounded by applying \Cref{prop:8}.
	Defining $R = R_{\ell} \otimes \cdots \otimes R_r$ and $R' = R_\ell' \otimes \cdots \otimes R_r'$, where $R_i,R_i' \sim \mathrm{Haar}(U(2))$, notice that $RR^{\prime \dagger}$ has the same distribution as $V_{j,[\ell,r]}$.
	Thus, letting $m = r-\ell + 1$, we can write
	\begin{align}
		\E_{\tilde{U}}[|T_j^{\ell,r}(0,0)|^2] &= \frac{1}{2^{2m}}\E_{B_j,M_j,V}\Bigg[\Bigg|\sum_{\substack{P \in \pauli_{[r+1,n]}\\Q \in \pauli_{[1,\ell-1]}}} \tr(V_{j,[\ell,r]}^\dagger B_P V_{j,[\ell,r]} M_Q V_{j,[\ell,r]}^\dagger B_P' V_{j,[\ell,r]} M_Q')\Bigg|^2\Bigg]\\
		&= \frac{1}{2^{2m}}\E_{B_j, M_j}\E_{R,R'}\Bigg[\Bigg|\sum_{\substack{P \in \pauli_{[r+1,n]}\\Q \in \pauli_{[1,\ell-1]}}} \tr(R' R^\dagger B_P R R^{\prime \dagger} M_Q R' R^\dagger B_P' R R^{\prime \dagger}M_Q')\Bigg|^2\Bigg]\\
		&= \frac{1}{2^{6m}}\E_{B_j, M_j}\E_{R,R'}\Bigg[\Bigg|\sum_{\substack{P \in \pauli_{[r+1,n]}\\Q \in \pauli_{[1,\ell-1]}}} \norm{B_P'}_F \norm{M_Q'}_F \norm{B_P}_F \norm{M_Q}_F \tr(C^\dagger ED^\dagger F) \Bigg|^2\Bigg],
	\end{align}
	where in the last line, we use the definitions of $C,D,E,F$ in \Cref{lem:conditions-prop-8}.
	By Cauchy-Schwarz, we have
	\begin{align}
		&\E_{\tilde{U}}[|T_j^{\ell,r}(0,0)|^2]\\
		&\leq \frac{1}{2^{6m}} \E_{B_j, M_j}\E_{R,R'}\Bigg[\Bigg(\sum_{P \in \pauli_{[r+1,n]}} \norm{B_P'}_F^2\Bigg)\Bigg(\sum_{Q \in \pauli_{[1,\ell-1]}} \norm{M_Q'}_F^2\Bigg) \sum_{\substack{P \in \pauli_{[r+1,n]}\\Q \in \pauli_{[1,\ell-1]}}} \norm{B_P}_F^2 \norm{M_Q}_F^2 | \tr(C^\dagger E D^\dagger F)|^2\Bigg]\\
		&\leq \frac{1}{2^{4m}}\E_{B_j, M_j}\E_{R,R'}\Bigg[\sum_{\substack{P \in \pauli_{[r+1,n]}\\Q \in \pauli_{[1,\ell-1]}}} \norm{B_P}_F^2 \norm{M_Q}_F^2 | \tr(C^\dagger E D^\dagger F)|^2\Bigg].
	\end{align}
	In the first line, we use Cauchy-Schwarz.
	In the second line, we use that
	\begin{equation}
		\frac{1}{2^m}\sum_{P\in \pauli_{[r+1,n]}} \norm{B_P'}_F^2 = \frac{1}{2^n}\norm{B_j^{\geq \ell}}_F^2 \leq 1,
	\end{equation}
	similarly to \Cref{eq:b-norm}.
	The same also holds for $M_Q'$.
	Now, we can rewrite this expectation in terms of our weighted expectations in \Cref{lem:conditions-prop-8}:
	\begin{align}
		\E_{\tilde{U}}[|T_j^{\ell,r}(0,0)|^2] \leq \frac{p_{j,\ell} q_{j,r}}{2^{2m}} {\E\limits_B}^* {\E\limits_M}^* \E_{R,R'}[|\trace(C^\dagger E D^\dagger F)|^2] \leq \frac{64}{3} p_{j,\ell} q_{j,r} \alpha^m.
	\end{align}
	In the second inequality, we apply \Cref{prop:8}, where we know the conditions are satisfied by \Cref{lem:conditions-prop-8}.
    Note that the starred expectations involve dividing by $p_{j,\ell} q_{j,r}$.
    If this quantity is zero, then $B_P = 0$ or $M_P = 0$, so the terms in $T_j^{\ell,r}(0,0)$ are all zero.
    Thus, we can simply consider the case when $p_{j,\ell} q_{j,r} \neq 0$.

	Now, we need to plug this bound back into \Cref{eq:to-plug-T,eq:to-plug-layer}.
	Plugging into \Cref{eq:to-plug-T}, we have
	\begin{align}
		\sqrt{\E_{\tilde{U} \sim \mathcal{U}_{n,d}}[|(\mathrm{id} - \Pi_j)f^{(1)}(\tilde{U})|^2]} &\leq \sum_{\substack{\ell,r=1\\r - \ell > k}}^n \sum_{a,b=0}^1 \frac{8}{\sqrt{3}} \sqrt{p_{j,\ell}}\sqrt{q_{j,r}}\alpha^{m/2}\\
		&= \sum_{r=k+1}^{n-1} \sum_{\ell=1}^{n-r} \sum_{a,b=0}^1 \frac{8}{\sqrt{3}} \sqrt{p_{j,\ell}} \sqrt{q_{j,\ell+r}} \alpha^{(r+1)/2}\\
		&\leq \frac{32}{\sqrt{3}} \sum_{r=k+1}^{n-1} \alpha^{(r+1)/2}\\
		&\leq \frac{32}{\sqrt{3}} \sum_{r=k+1}^\infty \alpha^{(r+1)/2}\\
		&= \frac{32}{\sqrt{3}(1-\sqrt{\alpha})} \alpha^{(k+2)/2}.
	\end{align}
	In the second line, we shift the index and use $m = r - \ell +1$.
	In the third line, we use that, by Cauchy-Schwarz, for $r \in [n-1]$,
	\begin{equation}
		\sum_{\ell=1}^{n-r} \sqrt{p_{j,\ell}} \sqrt{q_{j,\ell+r}} \leq \sqrt{\sum_{\ell=1}^{n-r} p_{j,\ell}} \sqrt{\sum_{\ell=1}^{n-r} q_{j,\ell+r}} \leq 1.
	\end{equation}
	This also uses that $\sum_{\ell=1}^n p_{j,\ell} = 1$ and $\sum_{r=1}^n q_{j,r} = 1$ because these are probabilities by \Cref{lem:pauli-second-moment-B,lem:pauli-second-moment-M}.
	In the fifth line, we use the formula for a geometric series.

	Finally, we plug this into \Cref{eq:to-plug-layer}.
	\begin{align}
		\E_{\tilde{U}\sim \mathcal{U}_{n,d}}[|f^{(1)}(\tilde{U}) - g_k(\tilde{U})|^2] \leq \frac{d}{2} \frac{1024}{3(1-\sqrt{\alpha})^2}\alpha^{k+2} = \frac{512d}{3(1-\sqrt{\alpha})^2} \alpha^{k+2},
	\end{align}
	which proves the claim.
\end{proof}

\subsubsection{Proof of \Cref{prop:8}}
\label{sec:proof-technical-prop}
Finally, we turn our attention to proving \Cref{prop:8}.
First, we prove the following key technical lemma.

\begin{lemma}
	\label{lem:7}
    There exists a constant $0 < \beta < 1$ such that the following holds.
	Let $m \geq 1$.
	Let $R = R_1 \otimes \cdots \otimes R_m$, where $R_i \sim \mathrm{Haar}(U(2))$.
	Let $\phi(R)$ be a function that is a linear combination of $\prod_{i=1}^m \ntr(P_i R_i Q_i R_i^\dagger)$, where $P_i, Q_i$ are single-qubit Paulis.
	Let $G,H$ be $m$-qubit operators whose Pauli expansions contain no identity factor at any site and satisfy $\norm{G}_F^2 = \norm{H}_F^2 = 1$.
	Then,
	\begin{equation}
		\E_R[|\phi(R)|^2 |\tr(G^\dagger R H R^\dagger)|^2] \leq \beta^m \E_R[|\phi(R)|^2].
	\end{equation}
\end{lemma}

\begin{proof}
    We first bound the weighted overlap defined in the lemma statement for a single qubit, i.e., when $m=1$.
    For $m=1$, let $V,V' \sim \mathrm{Haar}(U(2))$ be independent Haar-random unitaries and write $R=V D_t V^{\prime\dagger}$, where $D_t=\operatorname{diag}(z,1)$ and $z=e^{it}$ is independently uniform over the fifth roots of unity. For every fixed $t$ and $V'$, the unitary $R$ is Haar-random by right invariance.
    We can therefore compute the Haar average over $R$ by averaging over $t$ at fixed $V,V'$ and then average over $V,V'$.

    Conjugation by $D_t$ fixes diagonal entries and multiplies the upper-right and lower-left entries by $z$ and $z^{-1}$, respectively, i.e.,
    \begin{equation}
        D_t \begin{pmatrix}
            a &b\\c & d
        \end{pmatrix} D_t^\dagger = \begin{pmatrix}
            a & zb\\z^{-1}c & d
        \end{pmatrix}.
    \end{equation}
    Thus, each trace in $\phi(R)$ contains only the powers of $z$ in $I \triangleq \{-1,0,1\}$ because
    \begin{equation}
        \ntr(PRQR^\dagger) = \ntr(V^\dagger P VD_t V^{\prime \dagger} Q V' D_t^\dagger) = \ntr(\tilde{P}D_t\tilde{Q}D_t^\dagger),
    \end{equation}
    where we define $\tilde{P} \triangleq V^\dagger PV$, $\tilde{Q} \triangleq V^{\prime \dagger} Q V'$.
    Hence, because $D_t\tilde{Q}D_t^\dagger$ only depends on $z^{-1}, 1, z$ then so does the trace, as claimed.
    This means that we can write $\phi(R)$ as
    \begin{equation}
        \label{eq:phi-decomp}
        \phi(R) = \sum_{b \in I} \phi_b z^b.
    \end{equation}
    To expand $\tr(G^\dagger RHR^\dagger)$ in the same form, we can similarly write $\widetilde G\triangleq V^\dagger G V$ and $\widetilde H\triangleq V^{\prime\dagger}H V'$.
    Consider the orthonormal basis of traceless matrices given by
    \begin{align}
        \label{eq:onb-traceless}
        (E_{-1},E_0,E_1) = \left(\frac{X - iY}{2}, \frac{Z}{\sqrt{2}}, \frac{X+iY}{2}\right).
    \end{align}
    Then, because $G,H$ are non-identity, they are traceless.
    Conjugation preserves the trace, so $\tilde{G}, \tilde{H}$ are also traceless.
    Decomposing them in this basis, we have $\tilde{G} = \sum_{a \in I} g_a E_a$ and $\tilde{H} = \sum_{b \in I} h_b E_b$.
    It is easy to check that each basis element satisfies $D_t E_a D_t^\dagger=z^a E_a$, so we have
    \begin{align}
        \tr(G^\dagger RHR^\dagger) = \tr(\widetilde G^\dagger D_t \widetilde H D_t^\dagger) = \sum_{a,b\in\Lambda} \overline{g_b}h_a z^a\,\tr(E_b^\dagger E_a) = \sum_{a\in\Lambda}\overline{g_a}h_a z^a.
    \end{align}
    In the first equality, we substitute $R=V D_t V^{\prime\dagger}$ and use cyclicity of the trace.
    In the second equality, we expand the two matrices in the $E_a$ basis.
    In the last equality, we use $\trace(E_b^\dagger E_a)=\indicator\{a=b\}$ by orthonormality.

    Multiplying this expression by \Cref{eq:phi-decomp}, we have
    \begin{equation}
        \phi(R) \tr(G^\dagger RHR^\dagger) = \sum_{a,b\in I} \phi_b \overline{g_a} h_a z^{a + b} = \sum_{k=-2}^2 c_k z^k,
    \end{equation}
    where we defined $c_k \triangleq \sum_{a \in I_k} \phi_{k-a} \overline{g_a} h_a$ and $I_k \triangleq \{a \in I : k - a \in I\}$.
    Taking the expectation over $t$, we have
    \begin{align}
        \E_t[|\phi(R)\tr(G^\dagger RHR^\dagger)|^2] = \E_t\left[\left|\sum_{k=-2}^2 c_k z^k\right|^2\right] &= \E_t\left[\sum_{k,k'=-2}^2 c_k \overline{c_{k'}} z^{k-k'}\right]\\
        &= \sum_{k=-2}^2 |c_k|^2\\
        &= \sum_{k=-2}^2 \left|\sum_{a \in I_k} \phi_{k-a} \overline{g_a} h_a\right|^2\\
        &\leq \sum_{k=-2}^2 \left(\sum_{b\in I_k} |g_b|^2\right)\left(\sum_{a \in I_k} |\phi_{k-a}|^2 |h_a|^2\right)\\
        &= \sum_{\substack{a,b,e\in I\\e+ a - b \in I}} |\phi_e|^2 |h_a|^2 |g_b|^2.\label{eq:weighted-rotation-restricted-sum}
    \end{align}
    The second line follows because
    \begin{equation}
        \E_t[z^q] = \frac{1}{5}\sum_{j=0}^4 e^{2\pi ijq/5} \begin{cases}
            1 & 5 \mid q\\
            0 & 5 \nmid q
        \end{cases}
    \end{equation}
    for an integer $q$.
    We apply this to $q = k-k'$, which lies in $-4$ to $4$, so the only multiple of $5$ is $0$.
    In the third line, we use the definition of $c_k$.
    In the fourth line, we use Cauchy-Schwarz.
    In the fifth line, we set $e = k-a$ so that $a \in I_k$ implies $e \in I$ and $b \in I_k$ implies $k-b = e + a -b \in I$.

    Notice that if we consider the last line without the condition that $e + a -b \in I$, then we have
    \begin{equation}
        \sum_{a,b,e \in I} |\phi_e|^2 |h_a|^2 |g_b|^2 = \E_t[|\phi(R)|^2].\label{eq:weighted-rotation-unrestricted-sum},
    \end{equation}
    where we use that $\norm{H}_F^2 = \norm{G}_F^2 = 1$.
    This already shows that claim for $\beta = 1$, but we want to prove it with $\beta < 1$.
    To this end, we argue that the sum with the condition that $e + a - b \in I$ contains strictly fewer terms than the unrestricted sum.

    Notice that for every $e \in I$, at least one of $e + 2$ and $e - 2$ lies outside of $I = \{-1, 0,1\}$ since they differ by four and so cannot lie in an interval of three points.
    In particular, since the sum requires that $e + a - b \in I$, this means that the terms for $a -b \in \{2,-2\}$ are excluded, which correspond to $(a,b) = (-1,1)$ or $(-1,1)$.
    Provided that $h_a, g_b$ are non-zero for these values of $a,b$, this implies that the sum with the condition $e + a - b \in I$ is strictly smaller than the unrestricted sum.

    We now argue that $h_{\pm 1}, g_{\pm 1}$ are non-zero almost surely.
    For example, consider $g_1$.
    Suppose for the sake of contradiction that $g_1 = 0$.
    Then, because $E_1 = \begin{pmatrix}
        0 & 1 \\ 0 & 0 
    \end{pmatrix}$, this implies that the top right entry of $\tilde{G} = V^\dagger GV$ is $0$.
    Then, this implies that the second standard basis vector $v_2 = (0,1)$ is an eigenvector of $V^\dagger G V$, i.e., $V^\dagger G V v_2 = \lambda e_2$.
    Multiplying by $V$, we see that $GVv_2 = \lambda Vv_2$, i.e., the second column of $V$, given by $Vv_2$ is an eigenvector of $G$.
    However, the probability of this occurring is zero because $V$ is Haar-random, and a non-zero traceless $2\times 2$ matrix has at most two eigenlines.
    Hence, we can conclude that $g_1 \neq 0$, as required.
    A similar argument holds for $h_{\pm 1}$ and $g_{-1}$.

    Thus, almost surely over $V,V'$, an excluded pair has positive weight whenever $\phi_e$ is nonzero.
    For nonzero $\phi$, this occurs with positive probability because $\E_{V,V'}[\sum_e|\phi_e|^2]=\E_R[|\phi(R)|^2]>0$.
    The restricted sum is therefore strictly smaller than the full sum after averaging over $V,V'$, i.e.,
    \begin{equation}
        \label{eq:averaged-ineq}
        \E_{V,V'}\Bigg[\sum_{\substack{e,a,b \in I\\e + a - b \in I}}|\phi_e|^2 |h_a|^2 |g_b|^2\Bigg] < \E_{V,V'}\left[\sum_{\substack{e,a,b \in I}}|\phi_e|^2 |h_a|^2 |g_b|^2\right] = \E_R[|\phi(R)|^2].
    \end{equation}

    To obtain a constant $\beta$ that works for all inputs, normalize $\phi(R)$ such that $\E_R[|\phi(R)|^2] = 1$.
    By \Cref{eq:weighted-rotation-unrestricted-sum}, the average of the full sum is then one.
    Then, $\phi$ belongs to the finite-dimensional function space $\mathcal{L} = \mathrm{span}\{1, R \mapsto \ntr(PRQR^\dagger) : P, Q \in \{X,Y,Z\}\}$, which is a compact space.
    Because the lefthand side of \Cref{eq:averaged-ineq} is a continuous function over this compact space, by the Extreme Value Theorem, then it achieves its maximum.
    Call this maximum $\beta^*$, and note that $\beta^* < 1$ by \Cref{eq:averaged-ineq} since $\E_R[|\phi(R)|^2] = 1$ by normalization.
    Choose $0 < \beta < 1$ such that $\beta^* \leq \beta < 1$.
    Undoing the normalization, then we have
    \begin{equation}
        \E_{V,V'}\Bigg[\sum_{\substack{e,a,b\in I\\e + a-b \in I}} |\phi_e|^2 |h_a|^2 |g_b|^2\Bigg] \leq \beta \E_R[|\phi(R)|^2].\label{eq:weighted-rotation-one-site}
    \end{equation}
    Applying \Cref{eq:weighted-rotation-restricted-sum} completes the claim for $m = 1$.

    For $m > 1$, we do the same as the above for each qubit.
    Write $R_i=V_i D_{t_i}V_i^{\prime\dagger}$ independently for each $i \in [m]$, and let $V=\bigotimes_{i=1}^m V_i$, $V'=\bigotimes_{i=1}^m V_i'$, $D_{t_i} = \mathrm{diag}(z_i, 1)$, where $z_i = e^{i t_i}$.
    We also consider vector indices $e,a,b \in I^m$.
    Similarly to before, we can write
    \begin{equation}
        \phi(R) = \sum_{b\in I^m}\phi_b \prod_{i=1}^m z_i^{b_i},\quad
        \tr(G^\dagger RHR^\dagger) = \sum_{a\in I^m}\overline{g_a}h_a \prod_{i=1}^m z_i^{a_i},
    \end{equation}
    where $g_a,h_a$ are the coefficients of $V^\dagger G V,V^{\prime\dagger}H V'$ in the tensor-product basis of \Cref{eq:onb-traceless}.
    The same argument as in \Cref{eq:weighted-rotation-restricted-sum} holds with vector indices now:
    \begin{equation}
        \E_t[|\phi(R)\tr(G^\dagger RHR^\dagger)|^2] \leq \sum_{\substack{e,a,b\in I^m\\e + a-b\in I^m}} |\phi_e|^2 |h_a|^2 |g_b|^2,
    \end{equation}
    where the condition $e + a - b \in I^m$ means that $e_i + a_i - b_i \in I$ for all $i \in [m]$.

    To apply \Cref{eq:weighted-rotation-one-site} one site at a time, let $A_p$ be the average with the condition imposed on the first $p$ sites,
    \begin{align}
        A_p \triangleq \E_{V,V'}\Bigg[\sum_{\substack{e,a,b\in I^m\\ e_i+a_i-b_i\in I\ \text{for } i \in [p]}} |\phi_e|^2 |h_a|^2 |g_b|^2\Bigg],
    \end{align}
    for $p \in \{0,\dots, m\}$.
    By the above, $\E_R[|\phi(R)\tr(G^\dagger RHR^\dagger)|^2]\leq A_m$. Meanwhile, the same argument as in \eqref{eq:weighted-rotation-unrestricted-sum} gives $A_0=\E_R[|\phi(R)|^2]$.

    We show that $A_p\leq\beta A_{p-1}$.
    Fix the unitaries $V_i,V_i'$ and the labels $e_i,a_i,b_i$ for every $i\neq j$.
    Fixing these leave us with coefficients $\phi_e, h_a, g_b$ which are now effectively functions of $e_p, a_p, b_p$.
    Then, applying the $m=1$ bound from \Cref{eq:weighted-rotation-one-site}, we have
    \begin{align}
        \E_{V_j,V_j'}\Bigg[\sum_{\substack{e_p,a_p,b_p\in I\\ n_p+a_p-b_p\in I}}|\phi_e|^2 |h_a|^2 |g_b|^2\Bigg] \leq\beta\E_{V_j,V_j'}\Bigg[\sum_{e_p,a_p,b_p\in I} |\phi_e|^2 |h_a|^2 |g_b|^2\Bigg].
    \end{align}
    Summing over the fixed labels that satisfy the first $p-1$ conditions and averaging over the fixed unitaries gives $A_p\leq\beta A_{p-1}$. Iterating this bound for $p=m,\ldots,1$, we obtain
    \begin{align}
        \E_R[|\phi(R)\tr(G^\dagger RHR^\dagger)|^2]
        &\leq A_m \leq \beta A_{m-1} \leq \cdots \leq \beta^m A_0 = \beta^m \E_R[|\phi(R)|^2],
    \end{align}
    which proves the claim.
\end{proof}

With this, we can prove \Cref{prop:8}.

\begin{proof}[Proof of \Cref{prop:8}]
	Let $(C,D)$ and $(E,F)$ satisfy the hypotheses of \Cref{prop:8}.
	Recall that we want to bound $\E[|\ntr(C^\dagger ED^\dagger F)|^2]$.
	Consider the Pauli decomposition of $D$ and $F$: $D = \sum_{P \in \pauli_m} d_P P$ and $F = \sum_{P \in \pauli_m} f_P P$.
	The key idea is to break up the Pauli decomposition into high-weight and low-weight parts.
	Denote $\pauli_m^{\mathrm{low}} \triangleq \{P \in \pauli_m : |\supp(P)| \leq \lfloor m/100 \rfloor\}$ and $\pauli_m^{\mathrm{high}} \triangleq \{P \in \pauli_m : |\supp(P)| > \lfloor m/100 \rfloor\}$.
	We write
	\begin{equation}
		D^{\mathrm{low}} \triangleq \sum_{P \in \pauli_m^{\mathrm{low}}} d_P P,\quad\quad D^{\mathrm{high}} \triangleq \sum_{P \in \pauli_m^{\mathrm{high}}} d_P P,
	\end{equation}
	and similarly for $F^{\mathrm{low}}, F^{\mathrm{high}}$.
	Then, by definition $D = \dlow + \dhigh$ and $F = \flow + \fhigh$.
	Thus, we can write
	\begin{align}
		\ntr(C^\dagger E D^\dagger F) &= \ntr(C^\dagger E (\dlow + \dhigh)^\dagger (\flow + \fhigh))\\
		&= \ntr(C^\dagger E \dlowd \flow) + \ntr(C^\dagger E \dlowd \fhigh) + \ntr(C^\dagger E \dhighd F).
	\end{align}
	Then, we have
	\begin{equation}
		\label{eq:low-high-decomp}
		\E[|\ntr(C^\dagger E D^\dagger F)|^2] \leq 3\E[|\ntr(C^\dagger E \dlowd \flow)|^2] + 3\E[|\ntr(C^\dagger E \dlowd \fhigh)|^2] + 3 \E[|\ntr(C^\dagger E \dhighd F)|^2].
	\end{equation}

	We bound each of these terms separately.
	The point is that in the low-low term, there are few total low-weight Paulis.
	Meanwhile, for terms involving a high-weight component, we can apply \Cref{lem:7}.
	Both cases give an exponential decay in $m$, as required.

	First, we bound the low-low term.
	\begin{align}
		\E[|\ntr(C^\dagger E \dlowd \flow)|^2] &= \E\left[\left|\sum_{P,Q \in \pauli_m^{\mathrm{low}}} \overline{d_P} f_Q \ntr(C^\dagger E PQ)\right|^2\right]\\
		&\leq \E\left[\left(\sum_{P \in \pauli_m} |d_P|^2\right)\left(\sum_{Q \in \pauli_m} |f_Q|^2\right) \sum_{P,Q \in \pauli_m^{\mathrm{low}}} |\ntr(C^\dagger E PQ)|^2\right]\\
		&= \sum_{P,Q \in \pauli_m^{\mathrm{low}}}\E[|\ntr(C^\dagger E PQ)|^2]\\
		&\leq \sum_{P,Q \in \pauli_m^{\mathrm{low}}}\frac{c}{8^m}\norm{EPQ}_F^2\\
		&= \sum_{P,Q \in \pauli_m^{\mathrm{low}}}\frac{c}{8^m}\norm{E}_F^2\\
		&= \frac{c}{4^m} \left(\sum_{s=0}^{\lfloor m/100\rfloor} \binom{m}{s} 3^s\right)^2\\
		&\leq \frac{c}{4^m} \left(4^{\lfloor m/100 \rfloor} \sum_{s=0}^{\lfloor m/100\rfloor} \binom{m}{s} \left(\frac{3}{4}\right)^s\right)^2\\
		&\leq \frac{c}{4^m} \left(4^{m/100} \sum_{s=0}^m \binom{m}{s}\left(\frac{3}{4}\right)^s \right)^2\\
		&= \frac{c}{4^m} \cdot 4^{m/50} (1 + 3/4)^{2m}\\
		&\leq c \left(\frac{81}{100}\right)^m\label{eq:low-low}
	\end{align}
	In the first line, we use the definition of $\dlow$ and $\flow$.
	In the second line, we use Cauchy-Schwarz.
	In the third line, we use Parseval's identity (\Cref{lem:parseval}) and the assumption that $\norm{D}_F^2 = \norm{F}_F^2 = 2^m$.
	In the fourth line, we condition on $E$ (via independence), and apply the hypothesis in \Cref{eq:low-average}.
	In the fifth line, we use unitary invariance of the Frobenius norm.
	In the sixth line, we use that $\norm{E}_F^2 = 2^m$ and count the number of low-weight terms.
	In the eighth line, we use the binomial theorem.
	In the last line, we use that $4^{1/100} \cdot 7/4 \leq 9/5$.

	Now, we bound the low-high term.
	By \Cref{lem:cauchy-schwarz}, then
	\begin{align}
		\E[|\ntr(C^\dagger E \dlowd \fhigh)|^2] &\leq \frac{1}{4^m}\sqrt{\E[|\trace(C^\dagger C') \trace(\dlowd D^{\mathrm{low \prime}})|^2]}\sqrt{\E[|\trace(E^\dagger E')\trace(F^{\mathrm{high}\dagger} F^{\mathrm{high} \prime})|^2]}\\
		&\leq \frac{\sqrt{c}}{2^m}\sqrt{\E[|\trace(E^\dagger E')\trace(F^{\mathrm{high}\dagger} F^{\mathrm{high} \prime})|^2]},\label{eq:low-high-inter}
	\end{align}
	where $(C', D')$ and $(E', F')$ are indepedent copies of $(C, D)$ and $(E,F)$.
	In the second line, we notice that
	\begin{equation}
		\label{eq:triv-bound}
		\E[|\trace(C^\dagger C')\tr(\dlowd D^{\mathrm{low} \prime})|^2] \leq 4^m \E[|\trace(C^\dagger C')|^2] \leq c 2^m \E[\norm{C'}_F^2] = c4^m,
	\end{equation}
	where in the first inequality we use that $|\tr(\dlowd D^{\mathrm{low} \prime})|^2 \leq \norm{\dlow}_F^2 \norm{D^{\mathrm{low}\prime}}_F^2 \leq 4^m$.
	In the second inequality, we use the hypothesis (\Cref{eq:low-average}).
	In the third inequality, we use that $\norm{C'}_F^2 = 2^m$.
	Thus, we need to bound terms like \Cref{eq:low-high-inter}.
	Let $R = R_1 \otimes \cdots\otimes R_m$, where $R_i \sim \mathrm{Haar}(U(2))$.
	By the invariance hypothesis of $(E,F)$, then
	\begin{equation}
		\label{eq:invariance-applied}
		\E[|\trace(E^\dagger E')\trace(F^{\mathrm{high}\dagger} F^{\mathrm{high} \prime})|^2] = \E[|\trace(E^\dagger R E' R^\dagger)|^2|\trace(F^{\mathrm{high}\dagger} R F^{\mathrm{high}\prime} R^\dagger)|^2].
	\end{equation}
	Let $\phi(R) = \trace(E^\dagger R E' R^\dagger)$.
	We can't quite apply \Cref{lem:7} directly yet because it requires that $F^{\mathrm{high} \dagger}$ and $F^{\mathrm{high}\prime}$ contain no identity factors at any site.
	Thus, we group the parts of $\fhigh$ with exactly the same support: for a set $S \subseteq [m]$, define
	\begin{equation}
		F_S \triangleq \sum_{\substack{P \in \pauli_m\\\supp(P) = S}} f_PP
	\end{equation}
	and analogously for $F'_S$.
	Then,
	\begin{equation}
		\fhigh = \sum_{\substack{S \subseteq [m]\\|S| > \lfloor m/100\rfloor}} F_S
	\end{equation}
	and analogously for $F^{\mathrm{high}\prime}$.
	Then, considering first the expectation only over $R$, we have
	\begin{align}
		&\sqrt{\E_R[|\tr(E^\dagger E') \tr(F^{\mathrm{high}\dagger} F^{\mathrm{high}\prime})|^2]}\\
		&\leq \sum_{\substack{S \subseteq [m]\\|S| > \lfloor m/100 \rfloor}}\sqrt{\E_R[|\phi(R)|^2|\trace(F_S^\dagger R F_S' R^\dagger)|^2]}\\
		&\leq \sum_{\substack{S \subseteq [m]\\|S| > \lfloor m/100 \rfloor}} \beta^{|S|/2} \norm{F_S}_F \norm{F_S'}_F \sqrt{\E_R[|\phi(R)|^2]}\\
		&\leq \beta^{(\lfloor m/100\rfloor + 1)/2} \sqrt{\E_R[|\phi(R)|^2]} \sum_{\substack{S \subseteq [m]\\|S| > \lfloor m/100 \rfloor}} \norm{F_S}_F \norm{F_S'}_F\\
		&\leq \beta^{(\lfloor m/100\rfloor + 1)/2} \sqrt{\E_R[|\phi(R)|^2]} \sqrt{\sum_{\substack{S \subseteq [m]\\|S| > \lfloor m/100 \rfloor}}\norm{F_S}_F^2} \sqrt{\sum_{\substack{S \subseteq [m]\\|S| > \lfloor m/100 \rfloor}} \norm{F_S'}_F^2}\\
		&= \beta^{(\lfloor m/100\rfloor + 1)/2} \sqrt{\E_R[|\phi(R)|^2]} \norm{\fhigh}_F \norm{F^{\mathrm{high}\prime}}_F\\
		&\leq 2^m \beta^{(\lfloor m/100\rfloor + 1)/2} \sqrt{\E_R[|\phi(R)|^2]}\\
		&= 2^m \beta^{(\lfloor m/100\rfloor + 1)/2} \sqrt{\E_R[|\trace(E^\dagger R E' R^\dagger)|^2]}
	\end{align}
	In the second line, we use the new form of $\fhigh$ in \Cref{eq:invariance-applied} and triangle inequality.
	In the third line, we use \Cref{lem:7} applied to
    \begin{equation}
        \frac{\tr(F_S^\dagger R F_S' R^\dagger)}{\norm{F_S}_F \norm{F_S'}_F} = \tr\left(\left(\frac{F_S}{\norm{F_S}_F}\right)^\dagger R \left(\frac{F_S'}{\norm{F_S'}_F}\right) R^\dagger\right).
    \end{equation}
	In the fourth line, we use that $|S| > \lfloor m/100 \rfloor$.
	In the fifth line, we use Cauchy-Schwarz.
	In the sixth line, we use orthogonality of different supports.
	In the seventh line, we use $\norm{\fhigh}_F^2 \leq \norm{F}_F^2 = 2^m$.
	In the last line, we use the definition of $\phi(R)$.
	Now, squaring and taking the expectation over all other variables, we have
    \begin{align}
        \E[|\tr(E^\dagger E')\trace(F^{\mathrm{high} \dagger} F^{\mathrm{high}'})|^2] &\leq 4^m \beta^{\lfloor m/100\rfloor + 1} \E[|\tr(E^\dagger R E' R^\dagger)|^2]\\
        &\leq 4^m \beta^{\lfloor m/100\rfloor + 1} c2^{-m} \norm{RE'R^\dagger}_F^2\\
		&= c 4^m \beta^{\lfloor m/100\rfloor + 1}.
    \end{align}
    In the second line, we use the hypothesis in \Cref{eq:low-average}.
	In the last line, we use $\norm{E'}_F^2 = 2^m$.
	Plugging into \Cref{eq:low-high-inter}, we have
	\begin{equation}
		\E[|\ntr(C^\dagger E \dlowd \fhigh)|^2] \leq c \beta^{(\lfloor m/100\rfloor + 1)/2} \leq c \beta^{m/200}.
	\end{equation}
	The calculation is essentially the same for bounding the high term in \Cref{eq:low-high-decomp}, where we can obtain
	\begin{equation}
		\E[|\ntr(C^\dagger E \dhighd F)|^2] \leq c \beta^{m/200}.
	\end{equation}
	Thus, plugging these two and \Cref{eq:low-low} back into \Cref{eq:low-high-decomp}, we have
	\begin{equation}
		\E[|\ntr(C^\dagger E D^\dagger F)|^2] \leq 3c\left(\frac{81}{100}\right)^m + 6c\beta^{m/200}.
	\end{equation}
	Defining $\alpha \triangleq \max\{81/100, \beta^{1/200}\} < 1$, then we have
	\begin{equation}
		\E[|\ntr(C^\dagger E D^\dagger F)|^2] \leq 9c \alpha^m,
	\end{equation}
	which proves the claim.
\end{proof}

\subsection{Window truncation}
\label{sec:w-trunc-proof}

In this section, we bound the error between the intermediate estimate $g_k(U)$ and the actual output of our algorithm $h_k(U)$.
Recall that $g_k(U)$ is defined as the output of \Cref{alg:gk}.

\begin{theorem}[Window truncation error]
	\label{prop:w-trunc}
	Let $g_k(U)$ be the estimate obtained when only the diameter truncation rule is applied, i.e., the output of \Cref{alg:gk}.
    Let $h_k(U)$ be the estimated outputted by \Cref{alg:sim}.
	Then,
	\begin{equation}
		\E_{U \sim \mathcal{U}_{n,d}}[|g_k(U) - h_k(U)|^2] \leq 8d2^{-k}.
	\end{equation}
\end{theorem}

As discussed in \Cref{sec:overview} and as done in the previous section, we draw out layers $V_j$ of single-qubit independent Haar random gates.
Call this new circuit $\tilde{U}$.
Then, we have
\begin{equation}
	\E_{U\sim \mathcal{U}_{n,d}}[|g_k(U) - h_k(U)|^2] = \E_{\tilde{U} \sim \mathcal{U}_{n,d}}[|g_k(\tilde{U}) - h_k(\tilde{U})|^2].
\end{equation}
Thus, it suffices to prove the error bound for $\tilde{U}$.

\subsubsection{Monotonicity of an error function}
\label{sec:monotone}

In this section, we introduce a new error function which upper bounds the mean-squared error and appears easier to bound in our analysis.
We then show that it satisfies a monotonicity property, which is crucial to proving that it suffices to bound the error one layer at a time in \Cref{sec:layer-by-layer-W}, similar to \Cref{sec:layer-by-layer-k}.

At each iteration $j \in \{0,\dots, d/2\}$, the difference between the observables in the actual algorithm (\Cref{alg:sim}) and the algorithm outputting $g_k$ (\Cref{alg:gk}) is $E_j \triangleq \widehat{O}_j - O_j$.
Then, because $g_k = \ntr(\widehat{O}_0 (M \otimes M) \flip_{\mathrm{all}})$ and $h_k = \ntr(O_0 (M\otimes M) \flip_{\mathrm{all}})$, we can write the error between $h_k$ and $g_k$ as
\begin{equation}
	\E_{\tilde{U} \sim \mathcal{U}_{n,d}}[|g_k(\tilde{U}) - h_k(\tilde{U})|^2] = \E_{\tilde{U}\sim \mathcal{U}_{n,d}}[|\ntr(E_0(M\otimes M)\flip_{\mathrm{all}})|^2].
\end{equation}
We bound this error by defining a new error metric.
To do so, let $\Lambda_p$ be a projector that acts on two-copy operators which leaves the first $p$ sites untouched and projects the last $n - p$ sites onto the span of $\flip$ on these sites, for a boundary $p \in \{0,\dots, n\}$.
Then, consider the error function
\begin{equation}
	\mathcal{E}_p(O) \triangleq 4^{-p}\E_{\tilde{U}\sim \mathcal{U}_{n,d}}[\norm{\Lambda_p O}_F^2],
\end{equation}
where the observable $O$ implicitly depends on the circuit $\tilde{U}$.
Intuitively, this measures the average squared size of the component of $O$ which has flips on sites $p+1$ through $n$ (with some normalization).
Notice that
\begin{equation}
	|\ntr(E_0 (M \otimes M)\flip_{\mathrm{all}})|^2 = \frac{1}{N^2}|\langle \Lambda_1(M\otimes M)\flip_{\mathrm{all}}, E_0\rangle|^2 = \frac{1}{N^2}|\langle (M\otimes M)\flip_{\mathrm{all}}, \Lambda_1E_0\rangle|^2 \leq \norm{\Lambda_1 E_0}_F^2,
\end{equation}
where the first equality uses that $(M\otimes M) \flip_{\mathrm{all}}$ is Hermitian and $\Lambda_1 ((M\otimes M) \flip_{\mathrm{all}}) = (M \otimes M) \flip_{\mathrm{all}}$.
The second equality uses that $\Lambda_1$ is an orthogonal projector.
The last inequality uses that $\norm{(M\otimes M)\flip_{\mathrm{all}}}_F = N$ and Cauchy-Schwarz.
Thus, we can see that
\begin{equation}
	\E_{\tilde{U}\sim \mathcal{U}_{n,d}}[|g_k(\tilde{U}) - h_k(\tilde{U})|^2] \leq 4\mathcal{E}_1(E_0).
\end{equation}
In fact, we can further upper bound this by yet another error metric, defined as
\begin{equation}
	\err(O) \triangleq \max_{0 \leq p \leq n} \mathcal{E}_p(O).
\end{equation}
Then, trivially,
\begin{equation}
    \label{eq:mse-to-err}
	\E_{\tilde{U}\sim \mathcal{U}_{n,d}}[|g_k(\tilde{U}) - h_k(\tilde{U})|^2] \leq 4\, \err(E_0).
\end{equation}
The reason we introduce $\err$ is because of the following lemma.
This is only true when varying $p$.
Roughly, the reason is because applying a Haar-random layer can change how much of the observable lies in the subspace with a trailing tail of flips at each possible boundary $p$.
In particular, the proof shows that, after applying a Haar-random layer, the error at boundary $p$ is bounded by a convex combination of the errors at neighboring boundaries.

\begin{lemma}[Monotonicity of $\err$]
    \label{lem:haar-evolve-err}
	Let $O$ be a two-copy operator.
	Let $L$ be a layer of independent Haar-random two-qubit gates independent from $O$.
	Then,
	\begin{equation}
		\err(L^{\dagger \otimes 2} O L^{\otimes 2}) \leq \err(O).
	\end{equation}
	Moreover, if $\Gamma$ is an orthogonal projector commuting with $\Lambda_p$ for every $p \in \{0,\dots, n\}$, then $\err(\Gamma O) \leq \err(O)$.
	Also, $\err(\Pi_j O) \leq \err(O)$, where $\Pi_j$ is the projector from \Cref{lem:layer-by-layer-k}.
\end{lemma}

To prove~\Cref{lem:haar-evolve-err}, it is useful to first bound $\mathcal{E}_p(L^{\dagger \otimes 2}O L^{\otimes 2})$ in the two-qubit setting.

\begin{lemma}
	\label{lem:haar-evolve-two-q}
	Consider a two-qubit system with qubits labeled by integers $i$ and $i+1$.
	Let $G \sim \mathrm{Haar}(U(4))$ be a two-qubit Haar-random gate acting on qubits $i$ and $i+1$.
	Then, for a two-copy operator $O$ on these two qubits,
	\begin{equation}
		\E_G[\norm{\Lambda_i(G^{\dagger\otimes 2} O G^{\otimes 2})}_F^2] \leq \frac{1}{5}\norm{O}_F^2 + \frac{4}{5}\norm{\mathrm{Proj}_{\mathrm{span}\{\flip_i\flip_{i+1}\}}O}_F^2,
	\end{equation}
	where $\mathrm{Proj}_{\mathrm{span}\{\flip_i \flip_{i+1}\}}$ is an orthogonal projector onto $\mathrm{span}\{\flip_i \flip_{i+1}\}$.
	Here, $\Lambda_i$ only acts on the two qubits, i.e., it leaves qubit $i$ untouched and projects qubit $i+1$ onto $\flip$.
	In particular,
	\begin{equation}
		\E_G[G^{\otimes 2}\Lambda_i (G^{\dagger\otimes 2}(\cdot) G^{\otimes 2})G^{\dagger\otimes 2}] \preceq \frac{1}{5}\mathrm{id} + \frac{4}{5}\mathrm{Proj}_{\mathrm{span}\{\flip_i\flip_{i+1}\}}.
	\end{equation}
\end{lemma}

\begin{proof}
	First, we prove the inequality for projector $\tilde{\Lambda}_i$ instead of $\Lambda_i$.
    Here, we define $\tilde{\Lambda}_i$ as the projector which leaves qubit $i$ untouched and projects qubit $i+1$ onto identity.
	Then, we show how to transform this into the desired inequality with $\Lambda_i$.

	Before that, we consider a version of $\tilde{\Lambda}_i$ for just single-copy operators.
	For single-copy operators on two-qubits, then this projects onto
	\begin{align}
		\mathrm{span}\{I\otimes I, X \otimes I, Y \otimes I, Z \otimes I\} &= \mathrm{span}\{I\otimes I\} \oplus \mathrm{span}\{X\otimes I, Y\otimes I, Z \otimes I\}\\
		&\triangleq \mathrm{span}\{I\otimes I\} \oplus S_0.
	\end{align}
	Here, define orthogonal projectors onto each of these components by $\mathrm{Proj}_{\mathrm{span}\{I\otimes I\}}$ and $\mathrm{Proj}_{\mathrm{span}\{I\otimes I\}^\perp}$, respectively.
	Now, consider applying the Haar-random gate $G$.
	This leaves $I\otimes I$ invariant.
	Moreover, because conjugation by $G$ preserves orthogonality and tracelessness, then $S(G) \triangleq \mathrm{span}\{G(P \otimes I)G^\dagger : P \in \{X,Y,Z\}\}$ is a subspace of $\mathrm{span}\{I\otimes I\}^\perp$.
	Hence, we can also write
	\begin{equation}
		\label{eq:proj-inequality}
		0 \preceq \mathrm{Proj}_{S(G)} \preceq \mathrm{Proj}_{\mathrm{span}\{I\otimes I\}^\perp}.
	\end{equation}
	We also find it useful to compute the average of $\mathrm{Proj}_{S(G)}$ over $G$.
	Note that an orthonormal basis of $S(G)$ is $\{G(X\otimes I)G^\dagger/2, G(Y\otimes I)G^\dagger/2, G(Z\otimes I)G^\dagger/2\}$.
	Then, we can write the expectation value of the matrix elements of this projection as follows for $P,Q \in \pauli_2 \setminus \{I^{\otimes 2}\}$:
	\begin{align}
		\frac{1}{4}\E_G\left[\tr(P\mathrm{Proj}_{S(G)}(Q))\right] &= \frac{1}{4}\E_G\left[\sum_{R \in \{X,Y,Z\}} \frac{1}{2}\tr(PG(R\otimes I)G^\dagger) \cdot \frac{1}{2}\tr(QG(R\otimes I)G^\dagger)\right]\\
		&= \sum_{R \in \{X,Y,Z\}} \frac{\indicator\{P=Q\}}{15}\\
		&= \frac{\indicator\{P=Q\}}{5}.
	\end{align}
	In the second line, we use \Cref{lem:haar-conj}.
	For $P$ or $Q$ identity, then the matrix element is zero because $S(G)$ is a subspace of $\mathrm{span}\{I\otimes I\}^\perp$.
	Thus, we see that
	\begin{equation}
		\label{eq:expected-proj}
		\E_G[\mathrm{Proj}_{S(G)}] = \frac{1}{5}\mathrm{Proj}_{\mathrm{span}\{I\otimes I\}^\perp}.
	\end{equation}
    
	So far, we have only been considering single-copy operators.
	Now, we consider two-copy operators using these properties.
	We still only consider $\tilde{\Lambda}_i$ for now.
	In terms of the single-copy projectors, we can write
	\begin{equation}
		\tilde{\Lambda}_i = (\mathrm{Proj}_{\mathrm{span}\{I\otimes I\}} + \mathrm{Proj}_{S_0})^{\otimes 2}.
	\end{equation}
	After applying the gate $G$ on the two-copies, we have the projector
	\begin{equation}
		G^{\otimes 2}\tilde{\Lambda}_i(G^{\dagger\otimes 2} O G^{\otimes 2})G^{\dagger\otimes 2} = (\mathrm{Proj}_{\mathrm{span}\{I\otimes I\}} + \mathrm{Proj}_{S(G)})^{\otimes 2}(O).
	\end{equation}
	We bound this using the properties we showed above for each of the single-copy components.
	\begin{align}
		&\E_G[G^{\otimes 2}\tilde{\Lambda}_i(G^{\dagger\otimes 2} (\cdot) G^{\otimes 2})G^{\dagger\otimes 2}]\\
		&=\E_G[(\mathrm{Proj}_{\mathrm{span}\{I\otimes I\}} + \mathrm{Proj}_{S(G)})^{\otimes 2}]\\
		&= \mathrm{Proj}_{\mathrm{span}\{I\otimes I\}}^{\otimes 2} + \mathrm{Proj}_{\mathrm{span}\{I\otimes I\}} \otimes \E_G[\mathrm{Proj}_{S(G)}] + \E_G[\mathrm{Proj}_{S(G)}] \otimes \mathrm{Proj}_{\mathrm{span}\{I\otimes I\}} + \E_G[\mathrm{Proj}_{S(G)}^{\otimes 2}]\\
		\begin{split}
		&\preceq \mathrm{Proj}_{\mathrm{span}\{I\otimes I\}}^{\otimes 2} + \mathrm{Proj}_{\mathrm{span}\{I\otimes I\}} \otimes \E_G[\mathrm{Proj}_{S(G)}] + \E_G[\mathrm{Proj}_{S(G)}] \otimes \mathrm{Proj}_{\mathrm{span}\{I\otimes I\}}\\
		&+ \E_G[\mathrm{Proj}_{S(G)}] \otimes \mathrm{Proj}_{\mathrm{span}\{I\otimes I\}^{\perp}}
		\end{split}\\
		\begin{split}
		&= \mathrm{Proj}_{\mathrm{span}\{I\otimes I\}}^{\otimes 2} + \frac{1}{5}\mathrm{Proj}_{\mathrm{span}\{I\otimes I\}} \otimes \mathrm{Proj}_{\mathrm{span}\{I\otimes I\}^\perp} + \frac{1}{5}\mathrm{Proj}_{\mathrm{span}\{I\otimes I\}^\perp} \otimes \mathrm{Proj}_{\mathrm{span}\{I\otimes I\}}\\
		&+ \frac{1}{5}\mathrm{Proj}_{\mathrm{span}\{I\otimes I\}^\perp}\otimes \mathrm{Proj}_{\mathrm{span}\{I\otimes I\}^\perp}
		\end{split}\\
		&= \frac{1}{5}\mathrm{id} + \frac{4}{5}\mathrm{Proj}_{\mathrm{span}\{I\otimes I\}}\otimes \mathrm{Proj}_{\mathrm{span}\{I\otimes I\}}\\
		&= \frac{1}{5}\mathrm{id} + \frac{4}{5}\mathrm{Proj}_{\mathrm{span}\{\I_i\I_{i+1}\}}.\label{eq:proj-inequality-done}
	\end{align}
	In the third line, we expand out the tensor product.
	In the fourth line, we use \Cref{eq:proj-inequality}.
	In the fifth line, we use \Cref{eq:expected-proj}.
	In the second to last line, we use that $\mathrm{Proj}_{\mathrm{span}\{I\otimes I\}} + \mathrm{Proj}_{\mathrm{span}\{I\otimes I\}^\perp} = \mathrm{id}$.
	Applying this inequality to a two-copy operator $O$ on two-qubits, we have
	\begin{align}
		\E_G\left[\norm{\tilde{\Lambda}_i G^{\dagger \otimes 2} O G^{\otimes 2}}_F^2\right] = \langle O, \E_G[G^{\otimes 2}\tilde{\Lambda}_i(G^{\dagger\otimes 2}O G^{\otimes 2})G^{\dagger\otimes 2}]\rangle \leq \frac{1}{5}\norm{O}_F^2 + \frac{4}{5}\norm{\mathrm{Proj}_{\mathrm{span}\{\I_i\I_{i+1}\}} O}_F^2,
	\end{align}
	which is the desired inequality but for $\tilde{\Lambda}_i$ instead of $\Lambda_i$.

	To obtain the inequality for $\Lambda_i$ instead, we can consider the transformation $\mathcal{T}(O) \triangleq O \flip_i \flip_{i+1}$.
    Then, $\mathcal{T}$ commutes with $G^{\dagger \otimes 2} (\cdot)G^{\otimes 2}$ and $\Lambda_i = \mathcal{T} \circ \tilde{\Lambda}_i \circ \mathcal{T}^{-1}$.
    Moreover, conjugating by $\mathcal{T}$ sends $\mathrm{span}\{\I_i \I_{i+1}\}$ to $\mathrm{span}\{\flip_i \flip_{i+1}\}$.
    Hence, the claim follows by conjugating \Cref{eq:proj-inequality-done} by $\flip_i \flip_{i+1}$.
\end{proof}

With this, we can prove \Cref{lem:haar-evolve-err}.

\begin{proof}[Proof of \Cref{lem:haar-evolve-err}]
	First, we want to prove $\err(L^{\dagger\otimes 2} O L^{\otimes 2}) \leq \err(O)$.
	It suffices to prove that $\mathcal{E}_p(L^{\dagger\otimes 2} O L^{\otimes 2}) \leq \err(O)$ for all $p \in \{0,\dots,n\}$ by definition of $\err$.
	We notice that only a gate in $L$ acting on sites $p$ and $p+1$ can affect $\mathcal{E}_p$.
	This is because if a gate acts on two sites $\leq p$, then $\Lambda_p$ acts as identity, so $\mathcal{E}_p$ is unchanged due to unitary invariance of the Frobenius norm.
	Meanwhile, if a gate acts on two sites $> p$, then $\Lambda_p$ projects onto $\flip$.
	However, $\flip$ is invariant under two-copy gates, so we can commute the gate's action with $\Lambda_p$, and $\err$ stays the same again due to unitary invariance of the Frobenius norm.
	Thus, it suffices to consider the gate $G$ acting on sites $p$ and $p+1$.
    If no such gate exists, then $\mathcal{E}_p$ is unchanged, so we consider the case when such a gate exists.

	We can write $\Lambda_p$ as
	\begin{equation}
		\Lambda_p = \mathrm{id}_{[1,p]} \otimes \mathrm{Proj}_{\mathrm{span}\{\flip^{\otimes (n-p)}\}} = \mathrm{id}_{[1,p-1]} \otimes (\mathrm{id}_p \otimes \mathrm{Proj}_{\mathrm{span}\{\flip_{p+1}\}})\otimes \mathrm{Proj}_{\mathrm{span}\{\flip^{\otimes (n-p-1)}\}},
	\end{equation}
	where $\mathrm{id}_{[1,p-1]}$ denotes the identity operator on qubits $[1,p-1]$.
	Since the gate $G$ only acts on sites $p$ and $p+1$, applying the operator inequality version of \Cref{lem:haar-evolve-two-q} on those two sites, we have
	\begin{align}
		\E_G[G^{\otimes 2} \Lambda_p(G^{\dagger\otimes 2}(\cdot) G^{\otimes 2})G^{\dagger\otimes 2}] &\preceq \frac{1}{5}\left(\mathrm{id}_{[1,p+1]} \otimes \mathrm{Proj}_{\mathrm{span}\{\flip^{\otimes(n-p-1)}\}}\right)\\
		\begin{split}
		&+ \frac{4}{5}\left(\mathrm{id}_{[1,p-1]} \otimes \mathrm{Proj}_{\mathrm{span}\{\flip_p \flip_{p+1}\}} \otimes \mathrm{Proj}_{\mathrm{span}\{\flip^{\otimes(n-p-1)}\}}\right)
		\end{split}\\
		&= \frac{1}{5}\Lambda_{p+1} + \frac{4}{5}\Lambda_{p-1}.
	\end{align}
	Applying this inequality to a two-copy operator $O$ on $n$ qubits, we have
	\begin{equation}
		\E_G\left[\norm{\Lambda_p G^{\dagger\otimes 2} O G^{\otimes 2}}_F^2\right] = \langle O, \E_G[G^{\otimes 2}\Lambda_p(G^{\dagger\otimes 2}O G^{\otimes 2})G^{\dagger\otimes 2}]\rangle \leq \frac{1}{5}\norm{\Lambda_{p+1}O}_F^2 + \frac{4}{5}\norm{\Lambda_{p-1} O}_F^2.
	\end{equation}
	Because all gates in $L$ act on disjoint pairs of qubits (it is only one layer rather than two layers combined), then by the argument at the beginning of the proof, the same inequality holds for $L$ instead of just a single gate $G$:
	\begin{equation}
		\E_G\left[\norm{\Lambda_p L^{\dagger\otimes 2} O L^{\otimes 2}}_F^2\right] \leq \frac{1}{5}\norm{\Lambda_{p+1}O}_F^2 + \frac{4}{5}\norm{\Lambda_{p-1} O}_F^2.
	\end{equation}
	Finally, multiplying by $4^{-p}$ and averaging over the implicit dependence of $O$ on $\tilde{U}$, then by definition of $\mathcal{E}_p$, we have
	\begin{equation}
		\mathcal{E}_p(L^{\dagger\otimes 2} O L^{\otimes 2}) \leq \frac{4}{5}\mathcal{E}_{p+1}(O) + \frac{1}{5}\mathcal{E}_{p-1}(O) \leq \err(O).
	\end{equation}
	Note that in the last inequality, this is why we must define $\err$ as a maximum over all choices of $p$.
	This completes the proof of the first statement.
	The other statements follow easily from commutativity and contractivity of orthogonal projectors.
\end{proof}

\subsubsection{Layer-by-layer error}
\label{sec:layer-by-layer-W}

In this section, we use \Cref{lem:haar-evolve-err} to show that it suffices to bound the error one layer at a time, similarly to \Cref{lem:layer-by-layer-k}.

\begin{corollary}[Layer-by-layer error for window truncation]
    \label{coro:layer-by-layer-w}
	For $j \in \{0,\dots, d/2\}$, let $E_j = \widehat{O}_j - O_j$, where $O_j, \hat{O}_j$ are defined in \Cref{alg:sim} and \Cref{alg:gk}, respectively.
	Then,
	\begin{equation}
		\E_{\tilde{U}\sim \mathcal{U}_{n,d}}[|g_k(\tilde{U}) - h_k(\tilde{U})|^2] \leq 4\,\err(E_0) \leq 4\sum_{j=0}^{d/2-1} \err((\mathrm{id} - \mathcal{F}_{3k})\hat{O}_{j}).
	\end{equation}
\end{corollary}

\begin{proof}
	The first inequality is already proven above.
	To obtain the second inequality, we want to express the difference between the observables $E_j = \hat{O}_j - O_j$ recursively, i.e., we want to write $E_{j-1}$ in terms of $E_j$.
	We can do so by expanding the definitions of $\hat{O}_j$ and $O_j$ in the algorithms (\Cref{alg:gk,alg:sim}).

	First, consider $\hat{O}_j$.
	From \Cref{lem:layer-by-layer-k}, we know that the operation performed in each iteration of \Cref{alg:gk} is
	\begin{equation}
		\hat{O}_{j-1} = \Pi_j\left(\tilde{L}_j^{\dagger \otimes 2} \hat{O}_j \tilde{L}_j^{\otimes 2}\right),
	\end{equation}
	where $\Pi_j$ is an orthogonal projection defined in \Cref{eq:pi}.
	Similarly, for $O_j$, the operation performed in each iteration of \Cref{alg:sim} is
	\begin{equation}
		O_{j-1} = \mathcal{F}_{3k}\Pi_j\left(\tilde{L}_j^{\dagger \otimes 2} O_j \tilde{L}_j^{\otimes 2}\right).
	\end{equation}
	Then, the difference between these two operators is
	\begin{align}
		E_{j-1} &= \Pi_j\left(\tilde{L}_j^{\dagger \otimes 2} \hat{O}_j \tilde{L}_j^{\otimes 2}\right) - \mathcal{F}_{3k}\Pi_j\left(\tilde{L}_j^{\dagger \otimes 2} O_j \tilde{L}_j^{\otimes 2}\right)\\
		&= \mathcal{F}_{3k}\Pi_j\left(\tilde{L}_j^{\dagger \otimes 2} \hat{O}_j \tilde{L}_j^{\otimes 2}\right) - \mathcal{F}_{3k}\Pi_j\left(\tilde{L}_j^{\dagger \otimes 2} O_j \tilde{L}_j^{\otimes 2}\right) + \Pi_j\left(\tilde{L}_j^{\dagger \otimes 2} \hat{O}_j \tilde{L}_j^{\otimes 2}\right) - \mathcal{F}_{3k}\Pi_j\left(\tilde{L}_j^{\dagger \otimes 2} \hat{O}_j \tilde{L}_j^{\otimes 2}\right)\\
		&= \mathcal{F}_{3k}\Pi_j\left(\tilde{L}_j^{\dagger \otimes 2} E_j \tilde{L}_j^{\otimes 2}\right) + (\mathrm{id} - \mathcal{F}_{3k})\hat{O}_{j-1}.
	\end{align}
	Note that because these two terms are orthogonal (because $\mathcal{F}_{3k}$ is an orthogonal projection), then
	\begin{equation}
		\norm{E_{j-1}}_F^2 = \norm{\mathcal{F}_{3k}\Pi_j\left(\tilde{L}_j^{\dagger \otimes 2} E_j \tilde{L}_j^{\otimes 2}\right)}_F^2 + \norm{(\mathrm{id} - \mathcal{F}_{3k})\hat{O}_{j-1}}_F^2.
	\end{equation}
	The same holds when applying $\Lambda_p$ for any $p \in \{0,\dots,n\}$ because $\Lambda_p$ and $\mathcal{F}_{3k}$ commute.
	Thus, for any $p \in \{0,\dots,n\}$,
	\begin{equation}
		\mathcal{E}_p(E_{j-1}) = \mathcal{E}_p\left(\mathcal{F}_{3k}\Pi_j\left(\tilde{L}_j^{\dagger \otimes 2} E_j \tilde{L}_j^{\otimes 2}\right)\right) + \mathcal{E}_p\left((\mathrm{id} - \mathcal{F}_{3k})\hat{O}_{j-1}\right).
	\end{equation}
	Maximizing over $p$, we have
	\begin{align}
		\err(E_{j-1}) &= \max_{0\leq p \leq n} \left(\mathcal{E}_p\left(\mathcal{F}_{3k}\Pi_j\left(\tilde{L}_j^{\dagger \otimes 2} E_j \tilde{L}_j^{\otimes 2}\right)\right) + \mathcal{E}_p\left((\mathrm{id} - \mathcal{F}_{3k})\hat{O}_{j-1}\right)\right)\\
		&\leq \err\left(\mathcal{F}_{3k}\Pi_j\left(\tilde{L}_j^{\dagger \otimes 2} E_j \tilde{L}_j^{\otimes 2}\right)\right) + \err\left((\mathrm{id} - \mathcal{F}_{3k})\hat{O}_{j-1}\right)\\
		&\leq \err(E_j) + \err\left((\mathrm{id} - \mathcal{F}_{3k})\hat{O}_{j-1}\right),
	\end{align}
	where in the last line, we apply \Cref{lem:haar-evolve-err} twice.
	Iterating this bound for all $j \in \{0,\dots, d/2\}$ and using that $E_{d/2} = 0$ gives the result.
\end{proof}

\subsubsection{Single-layer error}
\label{sec:single-layer-W}

Finally, in this section, we prove a bound on the error arising from applying the window truncation at only a single layer.
In particular, we bound the terms in the sum of \Cref{coro:layer-by-layer-w} for each layer $j$.
We do so in the following lemma, completing the proof of \Cref{prop:w-trunc}.

\begin{lemma}[Single-layer error for window truncation]
    \label{lem:single-layer-error-W}
	For $j \in \{0,\dots, d/2-1\}$, let $\widehat{O}_j$ be defined as in \Cref{alg:gk}.
	Then,
	\begin{equation}
		\err((\mathrm{id} - \mathcal{F}_{3k})\widehat{O}_j) \leq 4 \cdot 2^{-k}.
	\end{equation}
\end{lemma}

\begin{proof}
	By definition of $\err$, it suffices to prove that $\mathcal{E}_p((\mathrm{id} - \mathcal{F}_{3k})\hat{O}_j) \leq 4 \cdot 2^{-k}$ for all $p \in \{0,\dots, n\}$.
	Recall that the range of $\Lambda_p$ consists of operators of the form $W_1\otimes \cdots \otimes W_p \otimes \flip^{\otimes (n-p)}$.
	Thus, if $p \leq 3k$, then the first $p$ arbitrary factors lie within the allowed window and have a tail consisting of all flips.
	Hence, for $p \leq 3k$, $\Lambda_p (\mathrm{id} - \mathcal{F}_{3k}) = 0$, so the bound trivially holds.

	Going forward, we assume that $p > 3k$.
	Then, we can break up our $n$-qubit system as follows: the first $[1,p-3k]$ sites, a left buffer $L$ of qubits $[p-3k+1, p-2k]$, a middle region of qubits $[p-2k+1, p-k]$, a right buffer $R$ of qubits $[p-k+1, p]$, and a tail which will be projected onto all flips on sites $[p+1, n]$ 
	First, notice that we can discard the case when the first $[1,p-3k]$ sites are identity.
	This is because, after applying $\Lambda_p$, we would have all identities, a window of width $3k$, and a tail of all flips, which lies in the range of $\mathcal{F}_{3k}$.
	Thus, we can write
	\begin{align}
		\norm{\Lambda_p(\mathrm{id} - \mathcal{F}_{3k})\hat{O}_j}_F^2 &= \norm{(\mathrm{id} - \mathcal{F}_{3k})(\mathrm{id} - \mathrm{Proj}_{\mathrm{span}\{\I^{\otimes (p-3k)}\}})\Lambda_p\hat{O}_j}_F^2\\
		&\leq \norm{(\mathrm{id} - \mathrm{Proj}_{\mathrm{span}\{\I^{\otimes (p-3k)}\}})\Lambda_p\hat{O}_j}_F^2,\label{eq:identities-removed}
	\end{align}
	where we use that $\Lambda_p$ and $\mathcal{F}_{3k}$ commute.
	In the inequality, we use the contractivity of orthogonal projectors.

	Now, the key observation is that for at least one of $L$ or $R$, a term in $\hat{O}_j$ on sites in $L$ or $R$ must lie entirely in $\mathrm{span}\{\I, \flip\}$.
	To see this, suppose a term in $\hat{O}_j$ contains a non-invariant site in both $L$ and $R$.
	Because of the middle region of $k$ qubits, these two non-invariant sites must be distance at least $k+1$ from each other.
	However, this is not allowed by the diameter truncation rule which $\hat{O}_j$ has undergone.
	Thus, any term of $\hat{O}_j$ must be entirely invariant in one of $L$ or $R$.
	Formally, define a projector
	\begin{equation}
		\mathrm{Proj}_L \triangleq \bigotimes_{i\in L} \mathrm{Proj}_{\mathrm{span}\{\I_i, \flip_i\}}
	\end{equation}
	and similarly for $\mathrm{Proj}_R$.
	Then, the observation that a term in $\hat{O}_j$ on sites in $L$ or $R$ must lie entirely in the invariant subspace can be described formally as
	\begin{equation}
		\label{eq:lr-one-invariant}
		(\mathrm{id} - \mathrm{Proj}_L)(\mathrm{id} - \mathrm{Proj}_R)\hat{O}_j = 0.
	\end{equation}
	This also holds for $\hat{O}_j$ with the additional projectors applied in \Cref{eq:identities-removed}, as these projectors cannot add any non-invariant sites.
	Thus, letting $Z \triangleq (\mathrm{id} - \mathrm{Proj}_{\mathrm{span}\{\I^{\otimes (p-3k)}\}})\Lambda_p\hat{O}_j$ for brevity, then
	\begin{align}
		\norm{\Lambda_p(\mathrm{id} - \mathcal{F}_{3k})\hat{O}_j}_F^2 &\leq \norm{Z}_F^2\\
		&= \langle Z, (\mathrm{Proj}_L + \mathrm{Proj}_R - \mathrm{Proj}_L \mathrm{Proj}_R)Z\rangle \\
		&\leq \norm{\mathrm{Proj}_R Z}_F^2 + \norm{\mathrm{Proj}_L Z}_F^2\\
		&= \sum_{S \in \{L,R\}}\norm{\mathrm{Proj}_S (\mathrm{id} - \mathrm{Proj}_{\mathrm{span}\{\I^{\otimes (p-3k)}\}})\Lambda_p\hat{O}_j}_F^2.
	\end{align}
	In the first line, we use \Cref{eq:identities-removed}.
	In the second line, we use \Cref{eq:lr-one-invariant}.
	In the last line, we use the definition of $Z$.
	Using the notation from \Cref{eq:bj-notation}, let
	\begin{equation}
		V_{j+1}^\dagger B_{j+1} V_{j+1} = (\tilde{L}_{j+1}^\dagger \cdots \tilde{L}_{d/2}^\dagger) B (\tilde{L}_{d/2} \cdots \tilde{L}_{j+1}).
	\end{equation}
	From \Cref{lem:layer-by-layer-k}, we know that the operation performed in each iteration of \Cref{alg:gk} is $\hat{O}_j = \Pi_{j+1}(\tilde{L}_{j+1}^{\dagger\otimes 2} \hat{O}_{j+1} \tilde{L}_{j+1}^{\otimes 2})$.
	Recursing, we can see that
	\begin{equation}
		\hat{O}_j = \left(\prod_{l=j+1}^{d/2} \Pi_l\right)V_{j+1}^{\dagger\otimes 2} B_{j+1}^{\otimes 2} V_{j+1}^{\otimes 2}.
	\end{equation}
	Plugging this into the above and averaging over the implicit dependence of $\hat{O}_j$ on $\tilde{U}$, we have
	\begin{align}
		\E\norm{\Lambda_p(\mathrm{id} - \mathcal{F}_{3k})\hat{O}_j}_F^2 &\leq \sum_{S \in \{L,R\}} \E\norm{\mathrm{Proj}_S(\mathrm{id} - \mathrm{Proj}_{\mathrm{span}\{\I^{\otimes (p-3k)}\}}) \Lambda_p \left(\prod_{l=j+1}^{d/2}\Pi_l\right) V_{j+1}^{\dagger\otimes 2} B_{j+1}^{\otimes 2} V_{j+1}^{\otimes 2}}_F^2\\
        &= \sum_{S \in \{L,R\}} \E\norm{\left(\prod_{l=j+1}^{d/2}\Pi_l\right)\mathrm{Proj}_S(\mathrm{id} - \mathrm{Proj}_{\mathrm{span}\{\I^{\otimes (p-3k)}\}}) \Lambda_p V_{j+1}^{\dagger\otimes 2} B_{j+1}^{\otimes 2} V_{j+1}^{\otimes 2}}_F^2\\
		&\leq \sum_{S \in \{L,R\}} \E\norm{\mathrm{Proj}_S(\mathrm{id} - \mathrm{Proj}_{\mathrm{span}\{\I^{\otimes (p-3k)}\}}) \Lambda_p V_{j+1}^{\dagger\otimes 2} B_{j+1}^{\otimes 2} V_{j+1}^{\otimes 2}}_F^2\\
		&= \sum_{S \in \{L,R\}} \E\norm{\mathrm{Proj}_S(\mathrm{id} - \mathrm{Proj}_{\mathrm{span}\{\I^{\otimes (p-3k)}\}}) \Lambda_p B_{j+1}^{\otimes 2}}_F^2.
	\end{align}
	In the second line, we use that $\Pi_l$ projectors commute with these other projections.
    In the third line, we use contractivity of the product of the $\Pi_j$ projectors in $L2$ norm.
	In the last line, we use that the projectors commute with single-qubit two-copy rotations (because $\I$ and $\flip$ are invariant under single-qubit two-copy rotations) and unitary invariance of the Frobenius norm.
	Multiplying by $4^{-p}$, we have
	\begin{equation}
		\mathcal{E}_p((\mathrm{id} - \mathcal{F}_{3k})\hat{O}_j) \leq \sum_{S \in \{L,R\}} 4^{-p} \E\left[\norm{\mathrm{Proj}_S(\mathrm{id} - \mathrm{Proj}_{\mathrm{span}\{\I^{\otimes (p-3k)}\}}) \Lambda_p B_{j+1}^{\otimes 2}}_F^2\right].
	\end{equation}
	Then, we are done by the following claim because $|L| = |R| = k$.
	We prove the claim separately.
	\begin{claim}
		\label{claim:single-layer-w}
		Let $S \subseteq [p-3k+1,p]$.
		Then, with the notation above,
		\begin{equation}
			4^{-p} \E\left[\norm{\mathrm{Proj}_S(\mathrm{id} - \mathrm{Proj}_{\mathrm{span}\{\I^{\otimes (p-3k)}\}}) \Lambda_p B_{j+1}^{\otimes 2}}_F^2\right] \leq 2 \cdot 2^{-|S|}.
		\end{equation}
	\end{claim}

	\begin{proof}[Proof of \Cref{claim:single-layer-w}]
		Writing $B_{j+1} = \sum_{W\in \pauli_n} b_{W} W$, dropping dependence on $j$ for brevity.
		We split a Pauli $W \in \pauli_n$ into three parts: $P \in \pauli_{\overline{S}}$, $Q \in \pauli_{S},$ and $T \in \pauli_{[p+1,n]}$, where $\overline{S} = [p]\setminus S$.
		Then, expanding out the expression for $\Lambda_p B_{j+1}^{\otimes 2}$, we have
		\begin{equation}
			\Lambda_p B_{j+1}^{\otimes 2} = \frac{1}{2^{n-p}} \sum_{\substack{P,P' \in \pauli_{\overline{S}}\\Q,Q'\in\pauli_{S}}}\sum_{T \in \pauli_{[p+1,n]}} b_{P \otimes Q \otimes T} b_{P' \otimes Q' \otimes T}(P\otimes P')_{\overline{S}} \otimes (Q \otimes Q')_S \otimes \flip^{\otimes (n-p)},
		\end{equation}
		where the factor of $1/2^{n-p}$  comes from the orthogonal projection and $\norm{\flip^{\otimes(n-p)}}_F^2 = 4^{n-p}$.

		Applying the projection $(\mathrm{id} - \mathrm{Proj}_{\mathrm{span}\{\I^{\otimes (p-3k)}\}})$, we have
		\begin{align}
			\begin{split}
				&(\mathrm{id} - \mathrm{Proj}_{\mathrm{span}\{\I^{\otimes (p-3k)}\}})\Lambda_p B_{j+1}^{\otimes 2}\\
				&= \frac{1}{2^{n-p}} \sum_{\substack{P,P' \in \pauli_{\overline{S}}\\Q,Q'\in\pauli_{S}\\T \in \pauli_{[p+1,n]}}} \indicator\{P_{[1,p-3k]} \neq I \text{ or } P'_{[1,p-3k]} \neq I\} b_{P \otimes Q \otimes T}  b_{P' \otimes Q' \otimes T}(P\otimes P')_{\overline{S}} \otimes (Q \otimes Q')_{S} \otimes \flip^{\otimes (n-p)}.
			\end{split}
		\end{align}
		For $\mathrm{Proj}_S$, note that this is equal to the second Haar moment for each qubit in $S$.
		This is because second Haar moments project onto $\mathrm{span}\{\I,\flip\}$ by \Cref{lem:second-moment}.
		Using \Cref{lem:second-moment-pauli}, then for $Q = Q'$ with $S' \triangleq \supp(Q)$, we have
		\begin{equation}
			\mathrm{Proj}_S(Q \otimes Q) = \frac{1}{3^{|S'|}}\sum_{\substack{R \in \pauli_S\\\supp(R) = S'}} R \otimes R.
		\end{equation}
		For $Q \neq Q'$, then $\mathrm{Proj}_S(Q\otimes Q') = 0$.
		Regrouping terms, we can write
		\begin{equation}
			c_{P,P',S'} \triangleq \sum_{\substack{R \in \pauli_S\\\supp(R) = S'}} \sum_{T \in \pauli_{[p+1,n]}} b_{P\otimes R \otimes T} b_{P'\otimes R\otimes T},
		\end{equation}
		so regrouping the terms above, we have
		\begin{align}
            \mathrm{Proj}_S(\mathrm{id} - \mathrm{Proj}_{\mathrm{span}\{\I^{\otimes (p-3k)}\}})\Lambda_p B_{j+1}^{\otimes 2} 
            = 
            \frac{1}{2^{n-p}}\sum_{P,P' \in \pauli_{\overline{S}}} \indicator\{P_{[1,p-3k]} \neq I \text{ or } P_{[1,p-3k]}' \neq I\} \\ 
            \times \sum_{S' \subseteq S} \frac{1}{3^{|S'|}} c_{P,P',S'} \sum_{\substack{R \in \pauli_S\\\supp(R) = S'}} (P\otimes P')_{\overline{S}} \otimes (R\otimes R)_S \otimes \flip^{\otimes (n-p)}.
		\end{align}
		Multiplying by $4^{-p}$ and taking the Frobenius norm, we have
		\begin{align}
			&4^{-p}\norm{\mathrm{Proj}_S(\mathrm{id} - \mathrm{Proj}_{\mathrm{span}\{\I^{\otimes (p-3k)}\}})\Lambda_p B_{j+1}^{\otimes 2}}_F^2\\
			&= \sum_{P,P' \in \pauli_{\overline{S}}}\indicator\{P_{[1,p-3k]} \neq I \text{ or } P_{[1,p-3k]}' \neq I\} \sum_{S' \subseteq S} 3^{-|S'|} |c_{P,P',S'}|^2\\
			&= \sum_{S' \subseteq S} 3^{-|S'|} \sum_{\substack{P,P' \in \pauli_{\overline{S}}\\P_{[1,p-3k]} \neq I \text{ or } P'_{[1,p-3k]} \neq I}} \left|\sum_{\substack{R \in \pauli_S\\\supp(R) = S'}}\sum_{T \in \pauli_{[p+1,n]}} b_{P\otimes R \otimes T} b_{P'\otimes R \otimes T}\right|^2\\
			&\leq 2\sum_{S' \subseteq S} 3^{-|S'|} \sum_{\substack{P,P' \in \pauli_{\overline{S}}\\P_{[1,p-3k]} \neq I}} \left|\sum_{\substack{R \in \pauli_S\\\supp(R) = S'}}\sum_{T \in \pauli_{[p+1,n]}} b_{P\otimes R \otimes T} b_{P'\otimes R \otimes T}\right|^2\\
			&\leq 2\sum_{S' \subseteq S} 3^{-|S'|} \sum_{\substack{P,P' \in \pauli_{\overline{S}}\\P_{[1,p-3k]} \neq I}} \left(\sum_{\substack{R \in \pauli_S\\\supp(R) = S'\\T \in \pauli_{[p+1,n]}}} b_{P\otimes R\otimes T}^2\right)\left(\sum_{\substack{R \in \pauli_S\\\supp(R) = S'\\T \in \pauli_{[p+1,n]}}} b_{P'\otimes R\otimes T}^2\right)\\
			&\leq 2\sum_{S' \subseteq S} 3^{-|S'|} \sum_{\substack{P \in \pauli_{\overline{S}}\\P_{[1,p-3k]} \neq I}} \sum_{\substack{R \in \pauli_S\\\supp(R) = S'}}\sum_{T \in \pauli_{[p+1,n]}} b_{P\otimes R\otimes T}^2\\
			&= 2 \sum_{S' \subseteq S} 3^{-|S'|} \sum_{\substack{W \in \pauli_n\\W_{[1,p-3k]} \neq I\\\supp(W) \cap S = S'}} b^2_W.
		\end{align}
		where in the second line, we use that the terms are orthogonal.
		In the third line, we use the definition of $c_{P,P',S'}$.
		In the fourth line, we use symmetry between $P$ and $P'$.
		In the fifth line, we use Cauchy-Schwarz.
		In the sixth line, we use that
		\begin{equation}
			\sum_{\substack{P' \in \pauli_{\overline{S}}\\T \in \pauli_{[p+1,n]}}} \sum_{\substack{R \in \pauli_S\\\supp(R) = S'}} b^2_{P'\otimes R\otimes T} \leq \sum_{W \in \pauli_n} b_W^2 = 1.
		\end{equation}
		In the last line, we combine all of the Pauli labels into one.

		Finally, taking the expectation and reintroducing $b_W$'s dependence on $j$,
		\begin{align}
			4^{-p} \E\left[\norm{\mathrm{Proj}_S(\mathrm{id} - \mathrm{Proj}_{\mathrm{span}\{\I^{\otimes (p-3k)}\}}) \Lambda_p B_{j+1}^{\otimes 2}}_F^2\right] &\leq 2 \sum_{S' \subseteq S} 3^{-|S'|} \sum_{\substack{W \in \pauli_n\\W_{[1,p-3k]} \neq I\\\supp(W) \cap S = S'}} \E[b^2_{j+1,W}]\\
			&= 2 \sum_{S' \subseteq S} 3^{-|S'|} \sum_{\ell=1}^{p-3k} \sum_{\substack{W \in \pauli_n\\\ell(W) = \ell\\\supp(W) \cap S = S'}} \E[b^2_{j+1,W}]\\
			&= 2 \sum_{S' \subseteq S} 3^{-|S'|} \sum_{\ell=1}^{p-3k} \frac{p_{j+1,\ell}}{3\cdot 4^{n-\ell}} \cdot 3^{|S'| + 1} 4^{n-\ell-|S|}\\
			&= 2 \cdot 4^{-|S|} \sum_{S' \subseteq S} \sum_{\ell=1}^{p-3k} p_{j+1,\ell}\\
			&\leq 2 \cdot 2^{-|S|}.
		\end{align}
		In the second line, we rearrange the sum, defining $\ell(W)$ as the first non-identity site of $W$.
		Here, because $W_{[1,p-3k]} \neq I$, then $\ell(W) \leq p-3k$.
		In the third line, we use \Cref{lem:pauli-second-moment-B}.
		In the last line, we use that there are $2^{|S|}$ subsets of $S' \subseteq S$ and that $\sum_{\ell=1}^n p_{j+1,\ell} = 1$ by \Cref{lem:pauli-second-moment-B}.
		This completes the proof.
	\end{proof}
\end{proof}

\subsection{Computational time}
\label{sec:runtime}

The previous sections focused on proving the correctness of our algorithm.
In this section, we show it is also computationally efficient.

\begin{theorem}[Runtime guarantee]
    The runtime of the window propagation algorithm in~\Cref{alg:sim} is $\mathrm{poly}(n,d,1/\epsilon)$.
\end{theorem}

\begin{proof}
	For the purposes of this analysis, instead of the basis $\mathcal{A}$ used in the rest of the paper, we consider the orthonormal basis
	\begin{equation}
		\mathcal{A}^{\mathrm{ON}} \triangleq \left\{\frac{\I}{2}, \frac{2\flip - \I}{2\sqrt{3}},A_1,\dots, A_{14}\right\}^{\otimes n} \triangleq (\mathcal{A}_1^{\mathrm{ON}})^{\otimes n}
	\end{equation}
	and
	\begin{equation}
		\mathcal{A}_{\mathrm{front}}^{\mathrm{ON}}
		\triangleq
		\left\{
		  \left(\frac{\I}{2}\right)^{\otimes(\ell-1)} \otimes W_\ell \otimes \cdots \otimes W_r \otimes \left(\frac{\flip}{2}\right)^{\otimes(n-r)}
		  \;\middle|\;
		  \begin{gathered}
		    1 \leq \ell \leq n,\\
		    r = \min(n, \ell + 3k - 1),\\
		    W_j \in \mathcal{A}_1^{\mathrm{ON}} \quad (\ell \leq j \leq r)\\
            W_\ell \neq \I/2.
		  \end{gathered}
		\right\}.
	\end{equation}
   	We emphasize that this does not change the algorithm at all.
   	This is because $\mathrm{span}\,\mathcal{A}_{\mathrm{front}}^{\mathrm{ON}} = \mathrm{span}\,\mathcal{A}_{\mathrm{front}}$ and $\mathrm{span}\{\I,\flip\} = \mathrm{span}\left\{\frac{\I}{2},\frac{2\flip - \I}{2\sqrt{3}}\right\}$.
   	The only reason we consider these bases now is because it makes the time complexity analysis simpler, but the same result holds for our original basis.
   	We use the original basis in the rest of the paper for better exposition.

    First, we make formal the discussion following \Cref{alg:sim} regarding the number of coefficients one needs to keep track of in each iteration.
    In each iteration $j$, the observable $O_j$ lies in the range of the orthogonal projector $\mathcal{F}_{3k}$.
    The dimension of this space is
    \begin{align}
    	\dim \mathrm{ran}(\mathcal{F}_{3k}) \leq 1 + 15\sum_{\ell=1}^n 16^{\min\{n, \ell + 3k - 1\} - \ell} \leq 1 + 15n16^{3k-1} = \mathcal{O}(n16^{3k}).
    \end{align}
    Here, the $1$ comes from the identity operator.
    The other terms come from varying the leftmost non-identity site, indexed by $\ell$.
    At site $\ell$, there are $15$ possible non-identity choices for the basis element.
    For other sites within the window $W_\ell \otimes \cdots \otimes W_r$, there are $16$ possible basis element choices, and the window size is restricted to satisfy $r - \ell + 1 \leq 3k$.

    This is only the number of coefficients we need to store after applying both projections in each iteration.
    However, the number of coefficients in the intermediate observable $L_j^{\dagger \otimes 2} O_j L_j^{\otimes 2}$ may be larger.
    We can show that this is still on the order of $\mathcal{O}(n16^{3k})$, as follows.
    Consider a two-qubit gate $G_j(i,i+1)$ in layer $j$ acting on sites $i$ and $i+1$.
    Because $\I$ and $\flip$ are invariant under two-copy rotations (\Cref{fact:i-flip-invariant}), then if $i,i+1 \leq \ell-1$ or $i,i+1 \geq r+1$, then applying $G_j(i,i+1)^{\otimes 2}$ has no effect.
    Meanwhile, if $\ell \leq i,i+1 \leq r$, then only the coefficients change but the window does not change its size.
    Finally, if $i,i+1$ are crossing the boundary between the window and the identity/flip tails (e.g., $i = \ell-1$ or $i=r$), then the window size can increase by at most one.
    This means that one layer $U_j$ can increase the size of the window by at most two.
    In total, because $L_j = U_{2j} U_{2j-1}$, applying $L_j^{\otimes 2}$ increases the size of the window by at most four.
    Hence, by the same calculation as above, the intermediate operator $L_j^{\dagger\otimes 2} O_j L_j^{\otimes 2}$ can be stored using at most $\mathcal{O}(n16^{3k+4})$ coefficients.

    The above was a memory discussion, and we now turn to bounding the computational cost of computing the evolution $L_j^{\dagger\otimes 2} O_j L_j^{\otimes 2}$.
    Write $O_j = \sum_{A \in \mathcal{A}_{\mathrm{front}}^{\mathrm{ON}}} b_A A$.
    We first consider the cost of evolving under a two-qubit gate $G=G_j(i,i+1)$.
    Here, we have
    \begin{align}
    	G^{\dagger \otimes 2} O_j G^{\otimes 2} &= \sum_{A \in \mathcal{A}_{\mathrm{front}}^{\mathrm{ON}}} b_A G^{\dagger \otimes 2} A G^{\otimes 2}\\
    	&= \sum_{A \in \mathcal{A}_{\mathrm{front}}^{\mathrm{ON}}} b_A A_1 \otimes \cdots \otimes A_{i-1} \otimes (G^{\dagger\otimes 2} (A_i \otimes A_{i+1}) G^{\otimes 2}) \otimes A_{i+2} \otimes \cdots \otimes A_n.
    \end{align}
    Thus, the only term we have to evaluate is $G^{\dagger\otimes 2} (A_i \otimes A_{i+1}) G^{\otimes 2}$.
    These are constant-sized matrices, so this can be computed in constant time.
    Updating one basis element $A$ with this computation takes at most $\mathcal{O}(n)$ time.
    Recall that we only have $\mathcal{O}(n16^{3k})$ coefficients to keep track of.
    Thus, in total, for evolving under a two-qubit gate, the time is $\mathcal{O}(n^2 16^{3k})$.
    There are $\mathcal{O}(n)$ two-qubit gates in one layer, so this gives $\mathcal{O}(n^3 16^{3k})$ for the evolution step.

    Finally, we discuss the time to implement the two truncation rules.
    The diameter truncation can be implemented in time $\mathcal{O}(n)$ per coefficient $b_A$.
    This is because, to compute $S_A$, we can iterate through each factor of the corresponding basis element $A$.
    Then, we just need to check if $\mathrm{diam}(S_A) \leq k$.
    Thus, diameter truncation costs $\mathcal{O}(n^2 16^{3k})$ in total.

    Now, consider the window truncation.
    Writing $O_{j-1}' = \sum_A a_A A$, we want to compute $O_{j-1} = \mathcal{F}_{3k} (O_{j-1}') = \sum_{B \in \mathcal{A}_{\mathrm{front}}^{\mathrm{ON}}} b_B B$.
    To compute the coefficients $b_B$, because the basis $\mathcal{A}_{\mathrm{front}}^{\mathrm{ON}}$ is orthonormal under the Hilbert-Schmidt inner product, we can write
    \begin{equation}
    	O_{j-1} = \sum_{B \in \mathcal{A}_{\mathrm{front}}^{\mathrm{ON}}} \langle B, O_{j-1}' \rangle B.
    \end{equation}
    This implies that
    \begin{equation}
    	b_B = \langle B, O_{j-1}' \rangle = \sum_A a_A \tr(B^\dagger A).
    \end{equation}
    Computing $\tr(B^\dagger A)$ can be done in $\mathcal{O}(n)$ time because $\tr(B^\dagger A) = \prod_{i=1}^n \tr(B_i^\dagger A_i)$ and each of these products can be computed in constant time.
    On the right-hand side, there are $\mathcal{O}(n16^{3k})$ terms in the sum (as discussed earlier for intermediate observables).
    Also, we need to compute $\mathcal{O}(n 16^{3k})$ coefficients $b_B$.
    Thus, in total, this takes $\mathcal{O}(n^3 16^{6k})$ time.

    Overall, the dominant step per iteration costs $\mathcal{O}(n^3 16^{6k})$ time.
    There are $\mathcal{O}(d)$ iterations, so this gives a total of $\mathcal{O}(n^3 d 16^{6k})$ time.
    Using our choice of $k = \mathcal{O}(\log(d/\epsilon))$ gives the claim.
\end{proof}

\section{Details of numerical experiments}
\label{app:width-numerics}

In this section, we provide further details of our numerical simulations described in \Cref{sec:numerics-main} and include additional experiments.
We describe the simulations for Haar-random circuits in \Cref{app:haar-exact} and those for the Clifford-plus-$T$ circuits in \Cref{app:clifford-numerics}.
In both cases, our circuits follow the specifications in \Cref{sec:problem-main}, i.e., they act on a $1$D chain of qubits with open boundary conditions, and the butterfly operator $B= X_n$ and measurement operator $M = Z_1$ are placed on opposite ends of the chain.
At each tested system size, we choose the depth of the circuit such that the ensemble mean $\mathrm{OTOC}^{(1)}$ is approximately $0.5$, which is in the regime where we expect appreciable instance-to-instance fluctuations.

As noted in \Cref{sec:numerics-main}, for the numerics, we allow the truncation parameters $k,w$ of our window propagation algorithm (\Cref{alg:sim}) to vary independently, in contrast to our theoretical analysis, which requires $w = 3k$.
For each choice of the truncation parameters $k,w$, we compare our algorithm's predictions with the output of exact simulation.
For $m$ circuits $U_i$, the empirical mean-squared error (MSE) is
\begin{equation}
    \widehat M_{n,d}(k,w) \triangleq \frac{1}{m}\sum_{i=1}^{m}\left|h_{k,w}(U_i)-f^{(1)}(U_i)\right|^2,
    \label{eq:width-numerics-mse}
\end{equation}
where $h_{k,w}(U_i)$ denotes the output of our window propagation algorithm (\Cref{alg:sim}) applied to the circuit $U_i$ using truncation parameters $k,w$ and $f^{(1)}(U_i)$ denotes the reference calculation output for the circuit $U_i$, to which we compare our algorithm.
We use the same circuits for every choice of $k,w$.
Thus, changes in the measured error reflect only the effect of the truncation parameters rather than any fluctuations due to the choice of the random circuits.

\subsection{Haar-random circuits}
\label{app:haar-exact}

In this section, we consider Haar-random brickwork circuits, in which each two-qubit gate is drawn independently from Haar measure on $U(4)$ (see \Cref{def:brickwork}).
Our reference calculation for $f^{(1)}(U_i)$ uses truncation-free simulation\footnote{For system sizes up to $22$ qubits, the reference values are computed exactly in double precision. For $n=24,26$, we estimate the normalized trace using random-phase input states. Sampling the reference values can increase the measured MSE. We bound this average increase and include the allowance in the reported confidence intervals.}.

First, we describe the numerical implementation of our algorithm, which is summarized in~\Cref{alg:haar-numerical}.
The overall approach is similar to \Cref{alg:sim}.
At each iteration, the observable $O_j$ is stored as a linear combination of terms of the form
\begin{equation}
    \label{eq:window-contribution}
    \I^{\otimes(\ell-1)}
    \otimes W_{[\ell,r]}
    \otimes\flip^{\otimes(n-r)},
    \qquad r-\ell+1\leq w.
\end{equation}
For the numerics, we use the orthonormal basis
$\mathcal{A}_{\mathrm{front}}^{\mathrm{ON}}$ described in
\Cref{sec:runtime} with width $w$.
We write
\begin{equation}
    O_j =
    \sum_{A\in\mathcal{A}_{\mathrm{front}}^{\mathrm{ON}}}
    c_{j,A} A
\end{equation}
and group the basis terms by the starting site $\ell$ of
their window $[\ell,r]$, where $r=\min\{n,\ell+w-1\}$.
Each group gives an operator of the form
\Cref{eq:window-contribution}, which we call a
\emph{window term}.

For each window term, we store its starting site $\ell$
and the coefficients of $W_{[\ell,r]}$. The weights $c_{j,A}$
are absorbed into these coefficients, so no additional
weight needs to be stored for the whole window term.
We refer to this whole decomposition as the \emph{window
representation}.
The identity and flip factors outside the window are fixed
and need not be stored explicitly.
In this basis, the window projection $\mathcal{F}_w$ is computed by taking
Hilbert--Schmidt inner products with the retained basis
operators, as described in \Cref{sec:runtime}.

At each iteration, we Heisenberg evolve the observable $O_j$
through two layers of the circuit $L_j$.
Since both the evolution and the window projection $\mathcal{F}_w$ are linear,
we can apply them separately to each stored summand of $O_j$.
We express each projected summand in the same window basis
and add the coefficients of matching basis operators.
Thus, several input window terms can contribute to the same
output coefficient.

Gates acting entirely within an identity or flip region leave
it unchanged, so only gates inside a window or crossing its
boundaries require nontrivial updates.
We evaluate the projected coefficients directly by local
tensor contractions, without constructing the full evolved
operator.
\begin{algorithm}[htbp]
    \caption{Numerical implementation of window propagation}
    \label{alg:haar-numerical}

    \KwData{Circuit $U=L_{d/2}\cdots L_1$ on $n$ qubits,
    with two physical layers per block $L_j$;
    $B=X_n$, $M=Z_1$; truncation parameters
    $1\leq w\leq n$ and $0\leq k<w$.}

    \KwResult{The approximation $h_{k,w}(U)$.}

    Use the orthonormal window basis
    $\mathcal{A}_{\mathrm{front}}^{\mathrm{ON}}$
    of \Cref{sec:runtime}, with width $w$\;

    Initialize
    \[
        O_{d/2}
        = B\otimes B
        = \I^{\otimes(n-1)}\otimes(X\otimes X),
    \]
    and store its coefficients in this basis\;

    \For{$j=d/2,d/2-1,\ldots,1$}{
        Initialize all retained coefficients of $O_{j-1}$
        to zero\;
 
        \ForEach{stored summand $H$ of $O_j$}{
            Compute its contribution to the window-projected
            coefficients using local tensor contractions:
            \[
                \Delta c_A
                =
                \tr\!\left[
                    A^\dagger
                    (L_j^\dagger)^{\otimes2}
                    H L_j^{\otimes2}
                \right],
                \qquad
                A\in\mathcal{A}_{\mathrm{front}}^{\mathrm{ON}}.
            \]

            Add $\Delta c_A$ to the coefficient of $A$
            in $O_{j-1}$\;
        }

        Apply the diameter truncation: set the
        coefficients of basis operators $A$ for which
        $\mathrm{diam}(S_A)>k$ to zero\;
    }

    \Return{$h_{k,w}(U) \triangleq \displaystyle
        \frac{1}{2^n}\tr\!\left[
            O_0(M\otimes M)\flip_{\mathrm{all}}
        \right]$}\;
\end{algorithm}
The diameter and window projectors commute, so applying the diameter truncation after the window projection preserves the original algorithm.

In the remainder of this section, we set $k=w-1$, which gives the smallest empirical MSE among the tested diameter cutoffs at each fixed width, as discussed in \Cref{fig:numerics-main}(a) in the main text.
For this choice, the diameter truncation removes no terms, and the output is the window-only approximation $h_w(U)$.
Moreover, in \Cref{sec:numerics-main}, we have already discussed in detail how the mean-squared error decays with the window width, so in this appendix we focus on the scaling with system size.

To estimate the uncertainty in the MSE, we use bootstrap resampling~\cite{efron1993bootstrap}, with $10{,}000$ resamples for \Cref{fig:numerics-main}(b) and $50{,}000$ for the additional comparison across $14$--$26$ qubits. We report approximate $95\%$ confidence intervals for the ensemble MSE at each tested size and width.

\begin{figure}[htbp]
    \centering
    \includegraphics[width=\linewidth]
    {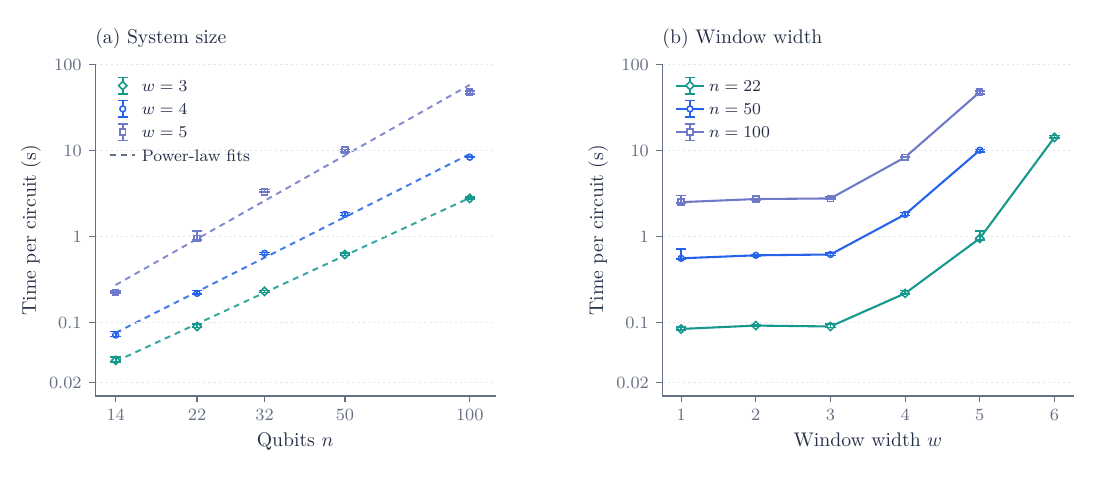}
    \caption{\textbf{Median runtime versus system size and window width.}
    In both subfigures, we evaluate $h_w=h_{w-1,w}$ on eight
    $n$-qubit Haar-random circuits with depth $d$ for
    $(n,d)\in\{(14,20), (22,32), (32,50), (50,80),(100,162)\}$, using the same circuits at every width.
    Symbols show median runtimes and bars show the full observed
    ranges of runtimes; the shortest ranges are comparable to
    the symbol size.
    Each evaluation uses a fresh process on one CPU thread
    of an Apple M4 Max.
    \textbf{(a)} Median runtime for fixed window widths
    $w\in\{3,4,5\}$. Dashed curves show the empirical fits
    described in \Cref{eq:haar-empirical-runtime}.
    \textbf{(b)} Median runtime at fixed system sizes
    $n\in\{22,50,100\}$. Solid lines connect the measured median
    runtimes and only guide the eye. Width $6$ is tested only
    at $22$ qubits.}
    \label{fig:haar-runtime-scaling}
\end{figure}

In \Cref{fig:haar-runtime-scaling}, we assess the computational cost of our algorithm.
All runtime estimates are quoted as the median runtime used to run our algorithm for a single circuit on one CPU thread of an Apple M4 Max, where the median is taken over eight randomly sampled circuits.
The timings include pre-processing, planning, and contraction but exclude circuit generation and reference simulations.
\Cref{fig:haar-runtime-scaling}(a) shows how the runtime of our algorithm scales with system size for fixed window widths.
We consider fixed window widths $w \in \{3,4,5\}$ and varying system sizes $n \in \{14,22,32,50,100\}$.
We note that this is only a discussion of the computational cost as the large-system-size regime is beyond the reach of exact simulation.

For each fixed width, we fit our timing data with a power-law relationship:
\begin{equation}
    T_w(n)\simeq a_w\left(\frac{n}{22}\right)^{p_w},
    \qquad 14\leq n\leq100.
    \label{eq:haar-empirical-runtime}
\end{equation}
We fit the logarithm of the median runtime linearly against $\log(n/22)$ using least squares.
The fitted coefficients $a_w$ and powers $p_w$ are
\begin{center}
\begin{tabular}{c c c}
    $w$ & $a_w$ (seconds) & $p_w$\\
    \hline
    $3$ & $0.0969$ & $2.23$\\
    $4$ & $0.229$  & $2.44$\\
    $5$ & $0.941$  & $2.73$
\end{tabular}
\end{center}

The largest relative deviations of these fits from the measured medians are approximately $7\%$, $11\%$, and $22\%$ for the window widths $w\in\{3,4,5\}$, respectively.
These power-law fits describe only the runtime scaling over the tested system sizes at our chosen depths; they do not establish the asymptotic complexity or separate the effects of system size and depth.

In \Cref{fig:haar-runtime-scaling}(b), we consider the runtime as a function of the window width for fixed system sizes.
We consider fixed system sizes $n \in \{22, 50,100\}$ and varying window widths $w \in \{1,\dots, 6\}$.
Again, we note that for $n \in \{50,100\}$, the accuracy of the algorithm remains untested; this is only a discussion of the runtime.
In the plot, we notice that the runtime changes very little for window width $w \in \{1,2,3\}$ but increases rapidly for $w > 3$.
At $22$ qubits, the measured median runtimes of our algorithm for $w\in\{4,5,6\}$ are $0.219$, $0.956$,
and $14.3$ seconds, respectively.
This sharp increase in the runtime confirms our theoretical prediction that the runtime scales exponentially in the window width $w$ (\Cref{sec:runtime}).

\subsection{Clifford-plus-$T$ circuits}
\label{app:clifford-numerics}

To test our algorithm at larger system sizes while still enabling comparison with exact simulation, we also consider simulating $\mathrm{OTOC}^{(1)}$ where the dynamics are given by Clifford-plus-$T$ circuits.
Our implementation builds on standard techniques for manipulating Clifford operations and Pauli rotations~~\cite{dehaene2003clifford,gosset2014tcount,begusic2025clifford}.
In the following, we explain how these tools can be leveraged to perform additional optimizations for both exact simulation and our algorithm.

We consider Clifford-plus-$T$ circuits of the following form.
We retain the same brickwork geometry as the previous section but replace each Haar-random gate by an independent, uniformly random two-qubit Clifford gate.
We also insert $3n/2$ $T$-gates, where $T=\operatorname{diag}(1,e^{i\pi/4})$, whose locations are chosen uniformly from the $nd$ possible qubit--layer positions.
Here, each $T$-gate is applied after the corresponding Clifford layer.
Thus, the number of $T$-gates is fixed, while their locations vary between circuits.
Throughout, we use $80$ randomly sampled $n$-qubit circuits with depth $d$ for $(n,d)\in\{(128,210),(256,422),(512,850)\}$, containing $192$, $384$, and $768$ $T$-gates, respectively.

\subsubsection{Exact simulation}

To assess the performance of our algorithm at such large system sizes, we exploit the Clifford structure to compute the $\mathrm{OTOC}^{(1)}$ value exactly and efficiently.
First, using that Clifford conjugation maps a Pauli string to another signed Pauli string, we can absorb the Clifford gates into the axes of the $T$ rotations, writing
\begin{equation}
    U=e^{i\phi}C\,R_t\cdots R_1,
    \qquad
    R_j=e^{-i\theta_jP_j/2},
    \label{eq:clifford-rotation-product}
\end{equation}
where $t=3n/2$ is the number of $T$ gates, $C$ is a Clifford
unitary, and each $P_j$ is a signed Pauli string
\cite{gosset2014tcount,begusic2025clifford}.
Indeed, $T=e^{i\pi/8}e^{-i\pi Z/8}$. The factor $e^{i\phi}$
is a global phase and cancels from $U^\dagger BU$. 

Defining $B'\triangleq C^\dagger BC$, the evolved butterfly operator becomes
\begin{equation}
    U^\dagger BU= R_1^\dagger\cdots R_t^\dagger\,B' \,R_t\cdots R_1.
\end{equation}
The remaining calculation thus involves only rotations
about Pauli strings.

For $\mathrm{OTOC}^{(1)}$, decomposing $U^\dagger BU$ in the Pauli basis, we have
\begin{equation}
    U^\dagger BU=\sum_{P\in\pauli_n}c_P P,
    \qquad
    \sum_{P\in\pauli_n}|c_P|^2=1.
\end{equation}
Plugging this into the expression for $f^{(1)}$ and using Pauli orthogonality, we have
\begin{equation}
    f^{(1)}(U) = 1-2\sum_{\substack{P\in\pauli_n\\\{P,M\}=0}} |c_P|^2.
    \label{eq:clifford-anticommuting-weight}
\end{equation}
Thus, in order to compute $\mathrm{OTOC}^{(1)}$, it suffices to compute the total squared Pauli weight that anticommutes with the measurement operator $M$.

Before evaluating this sum, we simplify the rotation sequence using
identities that preserve the $\mathrm{OTOC}^{(1)}$ value. We first
combine rotations generated by the same Pauli string, up to a sign,
whenever they commute with every rotation between them. Their angles
add when the Pauli strings have the same sign and subtract when
they have opposite signs.
We then remove rotations that cancel with their adjoints. A rotation
$R_j$ can be moved next to $B'$ if it commutes with all remaining
rotations lying between $R_j$ and $B'$ in the displayed Heisenberg
evolution. If $R_j$ also commutes with $B'$, it cancels with
$R_j^\dagger$. The same reasoning applies at the measurement end:
a rotation that commutes with $M$ and with all rotations between
it and $M$ cancels inside the trace. We repeat these two
cancellation checks until no further rotation is removed.

The simplified sequence has the same $\mathrm{OTOC}^{(1)}$ value,
so we can apply \Cref{eq:clifford-anticommuting-weight} to its Pauli
expansion.
For each of the $80$ sampled $512$-qubit circuits, at most $58$
rotations remain. We relabel these rotations as $R_j=e^{-i\theta_j P_j/2}$, where
$P_j$ is the Pauli string generator and $\theta_j$ is the angle
after simplification. Since combining rotations adds or subtracts
their original angles, $\theta_j$ can be any resulting integer
multiple of $\pi/4$.

After these simplifications, each remaining rotation either leaves a Pauli string unchanged or splits it into two terms:
\begin{equation}
    R_j^\dagger Q R_j =
    \begin{cases}
        Q, & [P_j,Q]=0,\\
        \cos\theta_j\,Q
        +i\sin\theta_j\,P_jQ,
        & \{P_j,Q\}=0.
    \end{cases}
\end{equation}
Iterating this identity generates paths through a
Pauli expansion \cite{begusic2025clifford}.
In general, different paths can reach the same final
Pauli string, in which case their amplitudes must be
added before squaring.

For our reference calculation of $f^{(1)}$, we check that the
remaining rotation axes are linearly independent in
their binary Pauli representation, where multiplication,
up to phase, corresponds to addition modulo two.
This guarantees that different paths end at different
Pauli strings. Their squared weights can therefore
be propagated directly: a split multiplies the
incoming weight by $\cos^2\theta_j$ and
$\sin^2\theta_j$ on its two branches.

We also use an efficient way to evaluate the resulting sum without enumerating every path. At a given step, after accounting for whether the current Pauli commutes or anticommutes
with $M$, its remaining contribution depends only
on its commutation relations with the rotation axes
still to be applied. Whenever this same pattern
reappears at the same step, the remaining calculation
is identical. We compute it once, store the result,
and reuse it.

Among the $80$ sampled $512$-qubit circuits, the largest exact simulation calculation stores $52{,}167$ such patterns.
This makes exact simulation tractable for the tested circuits, although the number of patterns can grow exponentially for other circuits.
We highlight that these exact calculations are deterministic. 

\subsubsection{Evaluating the window approximation}

To evaluate $h_w(U)$, we evolve $B\otimes B$ in the Heisenberg picture
and apply the window projection $\mathcal{F}_w$ after each two-layer
block, as in \Cref{alg:sim}. If $a_{P,Q}$ denotes the final coefficient
of the Pauli pair $P\otimes Q$, the OTOC estimate can be written as
\[
    h_w(U)=\frac{1}{2^n}\sum_{P,Q}a_{P,Q}\,\tr(PMQM).
\]
Since $MQM=\pm Q$, the trace equals $\pm\tr(PQ)$ and vanishes by
Pauli orthogonality unless $P=Q$. For an equal pair $P=Q$ and $M=Z_1$,
the normalized trace is $+1$ if $P$ carries $I$ or $Z$ on the first
qubit and $-1$ if it carries $X$ or $Y$.

The coefficient $a_{P,Q}$ collects contributions from all the ways
the evolution can reach $P\otimes Q$. A Clifford gate sends each
Pauli pair to a single signed pair, while a $T$ gate or a window
projection can produce several terms. Expanding these operations
in order gives sequences of Pauli pairs starting at $B\otimes B$.
The weight of a sequence is the product of the coefficients along it.

We group these sequences according to what happens at the $T$ gates,
numbered $j=1,\ldots,t$ in the order in which they act on $B\otimes B$.
At gate $j$, we record $\eta_j=1$ if exactly one string switches
between $X$ and $Y$ on the qubit where the gate acts, and $\eta_j=0$
if both strings switch or neither does.
We call the list
$\eta=(\eta_1,\ldots,\eta_t)$ the \emph{switch pattern} of the
sequence. Let $h_{w,\eta}(U)$ denote the total contribution from
sequences with pattern $\eta$, obtained by multiplying each sequence's
weight by its final normalized trace and summing. Since every sequence
has one switch pattern,
\[
    h_w(U)=\sum_{\eta\in\{0,1\}^t}h_{w,\eta}(U).
\]
We first determine which patterns allow the two strings to agree
at the measurement. We then work backwards from the trace to compute
the contribution of each such pattern.\\

\noindent\textbf{Finding the switch patterns.}
The final strings agree precisely when their product $PQ$ has the
identity label. To see how a switch affects this product, consider
the $j$th $T$ gate acting on $X\otimes X$. Since
$T^\dagger X T=(X-Y)/\sqrt{2}$, we obtain
\[
    X\otimes X \;\mapsto\;
    \tfrac12(X\otimes X+Y\otimes Y)
    -\tfrac12(X\otimes Y+Y\otimes X).
\]
In the first two pairs, both strings switch or neither does, so
$\eta_j=0$ and the product remains $I$. In the last two, exactly one
string switches, so $\eta_j=1$ and the product is $Z$ up to phase.
More generally, on qubit $i$ a switch between $X$ and $Y$ introduces
a factor $Z_i$ up to phase. Two switches at the same gate contribute
$Z_i^2=I$, so a $T$ gate changes the product label precisely when
$\eta_j=1$.

The window projection preserves this label on every pair it keeps.
For a pair other than $I\otimes I$, the window starts at the first
site where either string differs from $I$ and contains up to $w$
sites. Let $R_1$ and $R_2$ be the two strings on the $r$ sites to
its right. On those sites we project onto the span of the swap, giving
\[
    R_1\otimes R_2 \;\mapsto\;
    \frac{\indicator\{R_1=R_2\}}{4^r}
    \sum_{R\in\{I,X,Y,Z\}^{\otimes r}}R\otimes R.
\]
Thus unequal strings on those sites vanish. Equal strings are
replaced by an average over all $4^r$ choices of the same $R$ in
both copies. Their product is $I$ before the projection and in
every term of the average, while the factors inside the window
stay fixed.

It remains to follow the factors $Z_i$ through the Clifford gates.
Clifford conjugation acts on a product factor by factor, so we can
propagate each one separately. For $T$ gate $j$ on qubit $i$, let
$v_j$ be the binary Pauli label of $Z_i$ after conjugation by the
remaining Clifford gates in Heisenberg order. Since the initial
product is $B^2=I$ and Pauli multiplication adds labels modulo two,
the final label of $PQ$ is $\sum_j\eta_jv_j$. The identity has
label zero, so a pattern can contribute only if
\[
    \sum_{j=1}^t\eta_jv_j=0 \qquad\text{over }\mathbb F_2.
\]
We find all solutions by binary Gaussian elimination. Among the
$80$ sampled $512$-qubit circuits, each with $768$ $T$ gates,
$53$ had one such pattern, $25$ had two and $2$ had four.\\

\noindent\textbf{Summing the terms in one pattern.}
For each solution $\eta$ of the binary equations above, we now sum the
terms with that pattern. We start from the trace value of each final
pair and work backwards, keeping the outputs allowed by $\eta_j$ at
each $T$ gate. Consider the $T$ gate above with $\eta_j=0$.
For the incoming pair $X\otimes X$, the two allowed outputs are
$X\otimes X$ and $Y\otimes Y$, each with coefficient $1/2$.
If the rest of the circuit gives values $x$ and $y$ on these pairs,
the value before the gate is $(x+y)/2$.

To carry out this calculation for many pairs at once, we use the fact
that $\eta$ fixes the label of $PQ$ at each step. Specifying $P$
therefore determines $Q$. We write the remaining contribution as a
sum of functions of $P$. Each term has a coefficient $c$ and is
supported on a family of Pauli strings specified by binary equations.
Within that family, the term has value $+c$ or $-c$ according to
whether a specified sum of label bits is even or odd.

A two-qubit Clifford gate sends each individual pair to a single
pair, with a sign given by the product of the signs from the two
copies. For $P=Q$ this sign is always $+1$. For unequal pairs it
can vary across the family. We express this sign as a weighted sum
of parity functions and keep every resulting term within the same
switch pattern.

At a window projection, we keep the window fixed and average over
the $4^r$ possible Pauli strings $R$ on the $r$ sites to its right,
using the same $R$ in both copies. For each term, we only need to
count how many choices of $R$ give $+c$ and how many give $-c$.
Fixing the sign adds one binary equation to those defining the
family, so Gaussian elimination gives both counts. If they are
$N_+$ and $N_-$, respectively, the term contributes
$c(N_+-N_-)/4^r$ to the average.

To see why this gives $h_{w,\eta}(U)$ exactly, start at the
measurement, where the value of each pair is its normalized trace.
By linearity, the value of an incoming pair one step earlier is
the sum of the values of its allowed outputs, each multiplied by
its signed coefficient. Applying this identity backwards through
every gate and window projection gives the total contribution of
pattern $\eta$ when we reach $B\otimes B$. Every sequence with
nonzero final trace belongs to exactly one pattern satisfying
$\sum_j\eta_jv_j=0$. Summing over these patterns therefore gives
$h_w(U)$ exactly.\\

\begin{figure}[!htbp]
    \centering
    \includegraphics[width=0.90\linewidth,trim=0 10bp 0 10bp,clip]
        {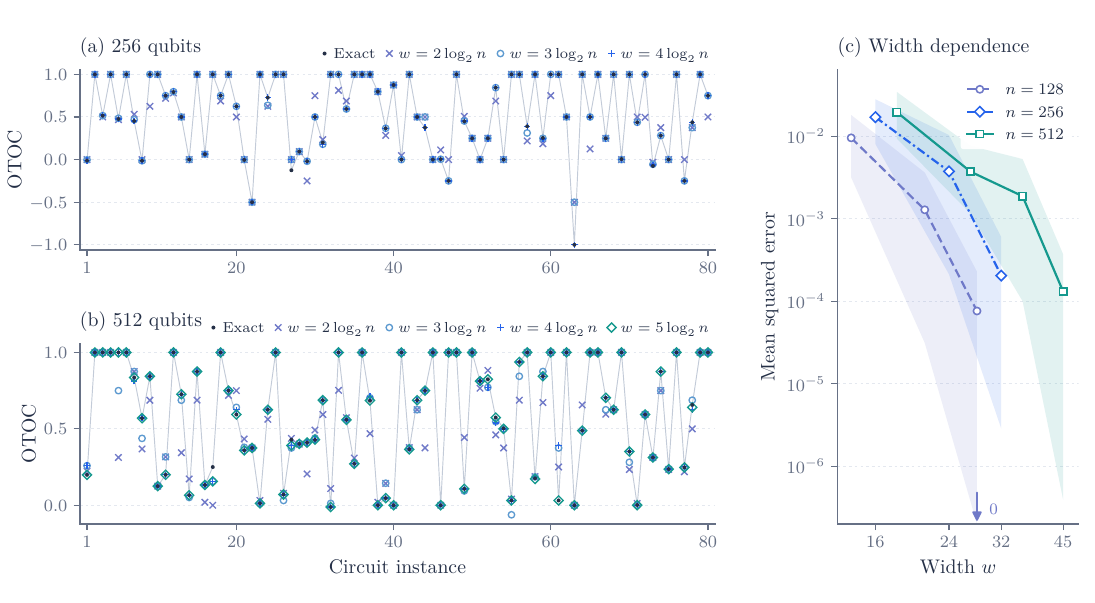}
    \caption{\textbf{Accuracy of our window propagation on
    Clifford-plus-$T$ circuits.}
    \textbf{(a,b)} Comparison of our algorithm with exact simulation
    for $80$ instances of Clifford-plus-$T$ circuits for
    $n\in\{256,512\}$. Black circles indicate exact simulation, and
    colored symbols show the results of our algorithm for window
    widths of $w\in\{2\log_2 n,\,3\log_2 n,\,4\log_2 n\}$ and
    $w=5\log_2 n$ also shown at $512$ qubits.
    \textbf{(c)} Mean squared error (MSE) as a function of window
    width for $n\in\{128,256,512\}$. The shaded area spans the middle
    $95\%$ of MSEs obtained at each window width by resampling the
    circuits with replacement~\cite{efron1993bootstrap}.
    At $128$ qubits and $w=4\log_2 n$, some resamples miss the only
    circuit with nonzero error. The lower bound of the confidence
    interval is zero, though it cannot be represented on the
    logarithmic axis and is instead indicated by an arrow.}
    \label{fig:clifford-t-width}
\end{figure}

\noindent\textbf{Comparison with exact OTOCs.}
\Cref{fig:clifford-t-width}(a,b) compares the output $h_w(U)$ of
our algorithm with the exact $\mathrm{OTOC}^{(1)}$ value on each
circuit for system sizes $n\in\{256,512\}$ and window widths
$w\in\{2\log_2 n,3\log_2 n,4\log_2 n\}$ and $w=5\log_2 n$ for
$n=512$. At $512$ qubits, they agree exactly on $29$ of $80$
circuits at $w=2\log_2 n$ and on $65$ of the circuits at
$w=5\log_2 n$.

In \Cref{fig:clifford-t-width}(c), we observe that the mean-squared
error decays with the window width $w$ for each fixed system size
$n\in\{128,256,512\}$. This is similar to the case of Haar-random
brickwork circuits in \Cref{fig:numerics-main}(b). Over all $80$
circuits, the MSE decreases from $1.97\times10^{-2}$ to
$1.31\times10^{-4}$ as we increase the window width from
$2\log_2 n$ to $5\log_2 n$.

\end{document}